\documentclass[11pt,a4paper]{article}
\pdfoutput=1
\usepackage{jheppub}

\usepackage{bold-extra}
\usepackage{enumerate}
\usepackage{fancyhdr}
\usepackage{lscape}
\usepackage{mathrsfs}
\usepackage{multirow}
\usepackage[final]{pdfpages}
\usepackage{rotating}
\usepackage{subfig}
\usepackage{textcomp}
\usepackage{url}
\usepackage{verbatim}
\usepackage{wrapfig}
\def\nn{\nonumber}

\title{One-Loop Corrections to the Perturbative Unitarity Bounds in the $CP$-Conserving Two-Higgs Doublet Model with a Softly Broken $\mathbb{Z}_2$ Symmetry}

\author[a]{Benjam\'{i}n Grinstein,}
\author[b]{Christopher W. Murphy,}
\author[c]{and Patipan Uttayarat}

\affiliation[a]{Department of Physics, University of California, San Diego, \\9500 Gilman Drive, La Jolla, CA 92093, USA}
\affiliation[b]{Scuola Normale Superiore, \\ Piazza dei Cavalieri 7, Pisa 56126, Italy}
\affiliation[c]{Department of Physics, Srinakharinwirot University, \\ Wattana, Bangkok 10110 Thailand}

\emailAdd{bgrinstein@ucsd.edu}
\emailAdd{christopher.murphy@sns.it}
\emailAdd{patipan@g.swu.ac.th}

\abstract{We compute all of the one-loop corrections that are enhanced, $O(\lambda_i \lambda_j / 16 \pi^2)$, in the limit $s \gg |\lambda_i | v^2 \gg M_{W}^2$, $s \gg m_{12}^2$ to all the $2 \to 2$ longitudinal vector boson and Higgs boson scattering amplitudes in the $CP$-conserving two-Higgs doublet model with a softly broken $\mathbb{Z}_2$ symmetry. In the two simplified scenarios we study, the typical bound we find is $|\lambda_i(s)| \lessapprox 4$.
}

\begin{document}
\maketitle

\section{Introduction}
Prior to the discovery of a Higgs boson at the LHC~\cite{Aad:2012tfa, Chatrchyan:2012xdj} with a mass $m_h \approx 125$~GeV~\cite{Aad:2015zhl}, theoretical considerations were used to place upper limits on how heavy the Higgs boson of the Standard Model (SM) could be. Lee, Quigg, and Thacker famously derived an upper bound of $\lambda \leq 16 \pi / 3$, or equivalently $m_h \leq \sqrt{16 \pi / 3}\, v \approx 1$~TeV with $v \approx 246$ GeV, by translating the unitarity of the $S$-matrix into an upper limit on the magnitude of partial wave amplitudes for $2 \to 2$ longitudinal vector boson and Higgs boson scattering~\cite{Lee:1977yc, Lee:1977eg}, see also~\cite{Dicus:1992vj}. This work was subsequently extended to the one-loop~\cite{Dawson:1988va, Dawson:1989up, Durand:1992wb} and two-loop~\cite{Maher:1993vj, Durand:1993vn} levels, which allowed for the study of bounds on $m_h$ due to perturbativity as well unitarity, and their interplay.

Despite the tremendous success of the Standard Model, there are good reasons to think that Nature might be described by an extended scalar sector. For starters, multiple generations of fermions are known to exist, so why shouldn't there be multiple Higgs doublets (or other multiplets) as well? More concretely, the couplings of the Higgs boson to vector bosons are consistent with the SM predictions, but the experimental uncertainties are currently in the tens of percents~\cite{Aad:2013wqa, Khachatryan:2014jba, Aad:2015gba}. Any deviation from the SM in these measurements would be a clear signal of additional Higgs bosons, and the uncertainties are currently large enough that this is a possibility.\footnote{Combining these experimental measurements of Higgs couplings with the bounds from perturbative unitarity is an efficient way to reduce the parameter space available in theories with extended Higgs sectors~\cite{Grinstein:2013fia}.} There are theoretical arguments which favor extended scalar sectors as well. The mass-squared parameter of the Higgs doublet of the SM is quadratically sensitive to the highest scale in the problem, rendering it unstable against quantum corrections. Beyond the SM solutions to this naturalness problem typically introduce new particles around TeV scale. The most well studied solution, the Minimal Supersymmetric Standard Model (MSSM), contains two Higgs doublets.

After the Standard Model, the theory with the next simplest scalar sector is the two-Higgs doublet model (2HDM)~\cite{Lee:1973iz, Deshpande:1977rw, Gunion:2002zf, Branco:2011iw}. Clearly, it is interesting to study the unitarity bounds in the 2HDM as the scale at which new particles are expected to appear is the same scale as the Lee, Quigg, Thacker upper limit on the Higgs mass in the SM. In fact, many authors have studied the tree level unitarity bounds on the quartic couplings and Higgs masses of the 2HDM~\cite{Casalbuoni:1986hy, Casalbuoni:1987eg, Maalampi:1991fb, Kanemura:1993hm, Akeroyd:2000wc, Ginzburg:2003fe, Ginzburg:2005dt, Horejsi:2005da, Haber:2010bw, Bhattacharyya:2013rya}. Extracting bounds on the masses of the Higgs bosons from the bounds on the quartic coupling is not as straightforward in the 2HDM as it was in the SM because in general the squares of the masses of the Higgs bosons of the 2HDM are not simply proportional to a linear combination of quartic couplings. In this work, we present the first one-loop level analysis of the perturbative unitarity bounds in the two-Higgs doublet model. Specifically, we compute all of the one-loop corrections that are enhanced, $O(\lambda_i \lambda_j / 16 \pi^2)$, in the limit $s \gg |\lambda_i | v^2 \gg M_{W}^2$, $s \gg m_{12}^2$ to all the $2 \to 2$ longitudinal vector boson and Higgs boson scattering amplitudes. As this is a first work, we do not consider the most general scalar potential in the 2HDM, but rather require that the potential is $CP$-conserving with a $\mathbb{Z}_2$ symmetry that is at most softly broken. Furthermore, we content ourselves with bounding the quartic couplings at the one-loop level, and save bounding the masses of the Higgs bosons for future studies. To this end, we study two simplified scenarios, and find that the typical bound on the quartic couplings is $|\lambda_i(s)| \lessapprox 4$.

The structure of the rest of the paper is as follows. We start in Sec.~\ref{sec:back} by giving the background necessary to understand the calculations and analysis we perform. After describing the 2HDM, which also gives us a chance to define our notation, we review the partial-wave analysis that is used to obtain upper limits on the quartic couplings. The details of the one-loop computation are discussed next in Sec.~\ref{sec:1L}. In particular, the computation is greatly simplified through use of the Goldstone boson equivalence theorem, which relates scattering amplitudes with external longitudinal vector bosons to amplitudes external Goldstone bosons. The conditions for the Goldstone boson equivalence theorem to hold at the one-loop level place restrictions on which renormalization schemes can be used to render the one-loop amplitudes finite. We then move on to the analysis of constraints on the quartic couplings due perturbative unitarity at the one-loop level, which is done in Sec.~\ref{sec:re}. After making some general considerations and reproducing the SM results, we analyze two simplified scenarios, the 2HDM in the limit where the longitudinal Goldstone boson scattering amplitudes possess an $SO(3)$ symmetry, as well as a scenario inspired by the form of the scalar potential in the MSSM. After that, our conclusions are given in Sec.~\ref{sec:con}. Finally, Appendix~\ref{sec:SE} contains our results for the self-energies, while Appendices~\ref{sec:RSA} and~\ref{sec:RSA2} contain our results for the scattering amplitudes.

\section{Background}
\label{sec:back}
\subsection{Two-Higgs Doublet Model}
\label{sec:back2HDM}
The two-Higgs doublet model contains two $SU(2)_L$ scalar doublets each with hypercharge $Y = 1 / 2$. We are using the convention $Q = T^3 + Y$, where $T^k = \tau^k / 2$ are the $SU(2)_L$ generators and $\tau^k$ are the Pauli matrices. The most general scalar potential consistent with $SU(2)_L \times U(1)_Y$ can be written as,
\begin{align}
\label{eq:genpot}
V &= m_{11}^2 (\Phi_1^{\dagger} \Phi_1) + m_{22}^2 (\Phi_2^{\dagger} \Phi_2) - [m_{12}^2 (\Phi_1^{\dagger} \Phi_2) + \text{h.c.}]  \\
&+ \tfrac{1}{2} \lambda_1 (\Phi_1^{\dagger} \Phi_1)^2 + \tfrac{1}{2} \lambda_2 (\Phi_2^{\dagger} \Phi_2)^2 + \lambda_3 (\Phi_1^{\dagger} \Phi_1)(\Phi_2^{\dagger} \Phi_2) + \lambda_4 (\Phi_1^{\dagger} \Phi_2)(\Phi_2^{\dagger} \Phi_1) \nn \\
&+ \{\tfrac{1}{2} \lambda_5 (\Phi_1^{\dagger} \Phi_2)^2 + [\lambda_6 (\Phi_1^{\dagger} \Phi_1) + \lambda_7 (\Phi_2^{\dagger} \Phi_2)] (\Phi_1^{\dagger} \Phi_2) + \text{h.c.} \}. \nn
\end{align}
The parameters $\lambda_{5},\, \lambda_{6},\, \lambda_{7}$, and $m_{12}^2$ are in general complex, while the rest of the parameters are always real. However, for the scalar potential to explicitly conserve $CP$-symmetry, there must exist a basis where the parameters $\lambda_{5},\, \lambda_{6},\, \lambda_{7}$, and $m_{12}^2$ are all simultaneously real. To avoid Higgs mediated flavor changing neutral currents (FCNCs) at tree level, we restrict the form of the potential by imposing a $\mathbb{Z}_2$ symmetry under which the scalar fields transform as $\Phi_1 \to \Phi_1$ and $\Phi_2 \to - \Phi_2$. We are agnostic about the $\mathbb{Z}_2$ charge assignments for the fermions in the theory. This symmetry forces $\lambda_{6},\, \lambda_{7} \to 0$, which then allows for $\lambda_5$ to be made real with a rephasing of $\Phi_1$~\cite{Haber:2015pua}. For the $\mathbb{Z}_2$ symmetry to be exact, $m_{12}^2$ must also be zero. However, we will allow for a soft breaking of the $\mathbb{Z}_2$ symmetry by keeping $m_{12}^2$ real to achieve a $CP$-conserving potential, but non-zero in general, as this scenario is more phenomenologically interesting. In any case, the bounds from perturbative unitarity on the quartic couplings are only very weakly dependent on $m_{12}^2$ at large $s$. This dependence is induced at one-loop due to terms of the form, for example, $\ln m_A^2 / m_h^2$ (with $m_A$ being the mass of the $CP$-odd Higgs). With these restrictions, the potential now has the form,
\begin{align}
\label{eq:Z2pot}
V &= m_{11}^2 (\Phi_1^{\dagger} \Phi_1) + m_{22}^2 (\Phi_2^{\dagger} \Phi_2) - m_{12}^2 [(\Phi_1^{\dagger} \Phi_2) + \text{h.c.}] + \tfrac{1}{2} \lambda_1 (\Phi_1^{\dagger} \Phi_1)^2  \\
&+ \tfrac{1}{2} \lambda_2 (\Phi_2^{\dagger} \Phi_2)^2 + \lambda_3 (\Phi_1^{\dagger} \Phi_1)(\Phi_2^{\dagger} \Phi_2) + \lambda_4 (\Phi_1^{\dagger} \Phi_2)(\Phi_2^{\dagger} \Phi_1) + \tfrac{1}{2} \lambda_5 [(\Phi_1^{\dagger} \Phi_2)^2 + \text{h.c.} ]. \nn
\end{align}
Requiring Eq.~\eqref{eq:Z2pot} to be bounded from below leads to the following tree level constraints on the parameters in the potential~\cite{Deshpande:1977rw},
\begin{equation}
\lambda_1 > 0,\quad \lambda_2 > 0 ,\quad \lambda_3 > - \sqrt{\lambda_1 \lambda_2},\quad \lambda_3 + \lambda_4 - |\lambda_5| > - \sqrt{\lambda_1 \lambda_2}.
\end{equation}
In what follows, it will be convenient to expand the fields in the basis where the $\mathbb{Z}_2$ is manifest as,
\begin{equation}
\label{eq:Z2phi}
\Phi_j = \begin{pmatrix}
w_j^{+} \\
n_j+ v_j / \sqrt{2}
\end{pmatrix}, \quad
n_j = \frac{h_j + i\, z_j}{\sqrt{2}}, \quad (j = 1,2).
\end{equation}
Here we have $\langle \Phi_j^{\dagger} \rangle = (0,\, v_j / \sqrt{2})$ with $v_1 = v\, c_{\beta}$, $v_2 = v\, s_{\beta}$, where we are using the notation $s_{\theta},\, c_{\theta},\, \text{and}\, t_{\theta}$ are the sine, cosine, and tangent of $\theta$ respectively. The minimization of scalar potential, which breaks $SU(2)_L \times U(1)_Y \to U(1)_{EM}$, is given by
\begin{align}
m_{11}^2 &= m_{12}^2 t_{\beta} - \tfrac{1}{2} v^2 (\lambda_1 c_{\beta}^2 + \lambda_{345} s_{\beta}^2), \quad m_{22}^2 = m_{12}^2 t_{\beta}^{-1} - \tfrac{1}{2} v^2 (\lambda_2 s_{\beta}^2 + \lambda_{345} c_{\beta}^2),
\end{align}
with $\lambda_{345} \equiv \lambda_3 + \lambda_4 + \lambda_5$. The relationships between the rest of the parameters in Eq.~\eqref{eq:Z2phi} and the Goldstone boson and physical Higgs states are: $h_1 = c_{\alpha} H - s_{\alpha} h$, $h_2 = s_{\alpha} H + c_{\alpha} h$, $z_1 = c_{\beta} z - s_{\beta} A$, $z_2 = s_{\beta} z + c_{\beta} A$, where $h$ and $H$ are the neutral, $CP$-even Higgs bosons, $A$ is the neutral, $CP$-odd Higgs boson, and $z$ is the would-be neutral Goldstone boson. The relations for the charged bosons are analogous to those for the $CP$-odd bosons.

In some instances, it will prove more convenient to use the Higgs basis rather than the $\mathbb{Z}_2$ basis. The Higgs basis can be obtained from the $\mathbb{Z}_2$ basis by making the following rotation, $H_1 =  c_{\beta} \Phi_1 + s_{\beta} \Phi_2$, $H_2 = - s_{\beta} \Phi_1 + c_{\beta} \Phi_2$, such that 
\begin{equation}
\label{eq:Hphi}
\sqrt{2} H_1 = \begin{pmatrix}
\sqrt{2} w^{+} \\
v + \phi_1 + i\, z
\end{pmatrix}, \quad
\sqrt{2} H_2 = \begin{pmatrix}
\sqrt{2} H^{+} \\
\phi_2 + i\, A
\end{pmatrix},
\end{equation}
with $\phi_1 = c_{\beta-\alpha} H + s_{\beta-\alpha} h$, $\phi_2 = - s_{\beta-\alpha} H + c_{\beta-\alpha} h$. In the notation of~\cite{Davidson:2005cw}, the potential in this basis is given by
\begin{align}
\label{eq:Hpot}
V &= M_{11}^2 (H_1^{\dagger} H_1) + M_{22}^2 (H_2^{\dagger} H_2) - M_{12}^2 [(H_1^{\dagger} H_2) + \text{h.c.}]  \\
&+ \tfrac{1}{2} \Lambda_1 (H_1^{\dagger} H_1)^2 + \tfrac{1}{2} \Lambda_2 (H_2^{\dagger} H_2)^2 + \Lambda_3 (H_1^{\dagger} H_1)(H_2^{\dagger} H_2) + \Lambda_4 (H_1^{\dagger} H_2)(H_2^{\dagger} H_1) \nn \\
&+ \{\tfrac{1}{2} \Lambda_5 (H_1^{\dagger} H_2)^2 + [\Lambda_6 (H_1^{\dagger} H_1) + \Lambda_7 (H_2^{\dagger} H_2)] (H_1^{\dagger} H_2) + \text{h.c.} \}, \nn
\end{align}
where all the parameters are real due to the $CP$-symmetry. Since there are only five quartic couplings in the $\mathbb{Z}_2$-basis, two of the seven quartic couplings in the Higgs basis are dependent upon the other five. The minimization conditions are simpler in the Higgs basis, and are given by
\begin{equation}
\label{eq:Hvevs}
M_{11}^2 = - \tfrac{1}{2} \Lambda_1 v^2, \quad M_{12}^2 = \tfrac{1}{2} \Lambda_6 v^2.
\end{equation}

We will occasionally refer to the alignment limit of the 2HDM, where $s_{\beta-\alpha} \to 1$ for $m_h < m_H$ or $c_{\beta-\alpha} \to 1$ for $m_H < m_h$, and the couplings of the aligned Higgs boson approach those of the SM, see~\cite{Gunion:2002zf} or more recently~\cite{Dev:2014yca, Bernon:2015qea, Bernon:2015wef}. Results from Run-1 of the LHC have pushed the parameter space of the 2HDM towards this limit~\cite{Grinstein:2013npa}. There are two ways to achieve the alignment limit, decoupling, or alignment without decoupling (or both simultaneously). In the Higgs basis, decoupling occurs when $M_{22}^2 \gg v^2$, while alignment without decoupling can achieved by taking $\Lambda_6 \ll 1$.

\subsection{Partial-Wave Analysis}
We are interested in seeing how large the parameters of the 2HDM can be. To this end, we perform a partial-wave analysis. Partial wave amplitudes are bounded by the unitarity of the $S$-matrix, $\mathbf{S}^{\dagger} \mathbf{S} = \mathbf{1}$, which requires 
\begin{equation}
\label{eq:SmatboundA}
\left| a_{\ell}^{2 \to 2} - \tfrac{1}{2} i \right|^2 + \sum_{n > 2} \left| a_{\ell}^{2 \to n} \right|^2 = \tfrac{1}{4}.
\end{equation}
Here, $a_{\ell}^{2 \to 2}$ are the eigenvalues of the matrix of $2 \to 2$ $\ell$-th partial-wave amplitudes, $\mathbf{a}_{\ell}^{2 \to 2}$. In this work, we do not compute any of the inelastic scattering amplitudes that appear in Eq.~\eqref{eq:SmatboundA}.\footnote{Note that because the $a_{\ell}$'s are eigenvalues, all of the $2 \to 2$ processes in Eq.~\eqref{eq:SmatboundA} are elastic, and similarly, all of the inelastic processes in Eq.~\eqref{eq:SmatboundA} are the $2 \to n$ amplitudes. This is course not true in general, e.g. $w^+ w^- \to z z$ is inelastic $2 \to 2$ scattering.} We do however make a few comments about the $2 \to n$ amplitudes before continuing with the analysis of the $2 \to 2$ amplitudes. The inelastic scattering amplitudes in Eq.~\eqref{eq:SmatboundA} are computed in a basis where $\mathbf{a}_{\ell}^{2 \to 2}$ is diagonal, and in each term in the sum contains an implicit integral over the $n$-body phase space.  The scattering amplitudes that enter the $2 \to 3$ partial-wave amplitudes scale as $\mathcal{M}^{2 \to 3} \sim \lambda_i^2 v / s$, leading to $| a_{\ell}^{2 \to 3}|^2 \sim \lambda_i^4 v^2 / s$ after the phase space integration is performed. Thus, in the energy limit under consideration, $s \gg |\lambda_i| v^2$, the $2 \to 3$ partial-wave amplitudes can be neglected. The leading inelastic amplitudes that persist in the energy regime we are considering are the $2 \to 4$ scatterings, which have the following scalings, $\mathcal{M}^{2 \to 4} \sim \lambda_i^2 / s$ and $| a_{\ell}^{2 \to 4}|^2 \sim \lambda_i^4$. In the SM, the $2 \to 4$ amplitudes are a few percent of the total contribution to the partial-wave amplitudes for moderate values of the quartic coupling~\cite{Durand:1992wb}.

Henceforth we will drop the superscripts from $\mathbf{a}_{\ell}$ and only consider elastic scattering, unless explicitly stated otherwise. In this case, the unitarity of the $S$-matrix puts an upper limit on the magnitude of the eigenvalues of $\mathbf{a}_{\ell}$,
\begin{equation}
\label{eq:Smatbound}
\left| a_{\ell} - \tfrac{1}{2} i \right|^2 \leq \tfrac{1}{4}.
\end{equation}
As can be seen from Eq.~\eqref{eq:SmatboundA}, the equality is satisfied if and only if all of the inelastic scattering processes vanish. From~\eqref{eq:Smatbound}, two perhaps slightly more familiar, but in general weaker bounds can be derived,
\begin{equation}
\label{eq:Smatbound2}
\left| a_{\ell} \right| \leq 1, \quad \left| \text{Re}(a_{\ell}) \right| \leq \tfrac{1}{2}.
\end{equation}
At tree level, and for the energy regime of interest, $s \gg |\lambda_i | v^2 \gg M_{W}^2$, $s \gg m_{12}^2$, the only non-zero partial wave is the $\ell = 0$ wave, so it will be the only partial-wave we will consider. 

To compute $\mathbf{a}_0$, we adapt the approach of~\cite{Ginzburg:2003fe, Ginzburg:2005dt} to the one-loop level. Refs.~\cite{Ginzburg:2003fe, Ginzburg:2005dt} showed that the tree level derivation simplifies considerably in the $\mathbb{Z}_2$ basis with non-physical Higgs fields $w_j^{\pm}$, $n_j$, and $n_j^{*}$. At high energies, the $SU(2)_L \times U(1)_Y$ symmetry is manifest, and weak isospin $(\tau$) and hypercharge $(Y)$ are conserved by the $2 \to 2$ scattering processes at tree level. Thus, $\mathbf{a}_0$ is block diagonal at leading order, with blocks of definite isospin and hypercharge. These blocks can themselves be broken down into smaller blocks by noting that, at tree level, $\mathbb{Z}_2$-even and -odd states do not mix. 

The full set of initial states, and their representations under these symmetries is given in Table~\ref{tab:reps}. For the $\tau = 1$ states, $k = \{3, +, -\}$. We have defined $\tilde{\Phi}_j = (i \tau^2 \Phi_j)^T$. The states with $\mathbb{Z}_2$-even, $Y = 1$, $\tau = 0$, i.e. $\tfrac{1}{2} (\tilde{\Phi}_1 \Phi_1)$, are identically zero since they are proportional to $\varepsilon^{ij} \delta_{ij}$. States with hypercharge $-1$ can be obtained from the states with $Y = 1$ by charge conjugation.
\begin{table}
\centering
 \begin{tabular}{c | c | c c }
 $Y$ & $\tau$ & $\mathbb{Z}_2$-even & $\mathbb{Z}_2$-odd  \\ \hline
\multirow{2}{*} 0 & \multirow{2}{*} 0 & $\tfrac{1}{\sqrt{2}} (\Phi_1^{\dagger} \Phi_1)$ & $\tfrac{1}{\sqrt{2}} (\Phi_1^{\dagger} \Phi_2)$ \\
 &  & $\tfrac{1}{\sqrt{2}} (\Phi_2^{\dagger} \Phi_2)$ & $\tfrac{1}{\sqrt{2}} (\Phi_2^{\dagger} \Phi_1)$ \\ \hline
 \multirow{2}{*} 0 & \multirow{2}{*} 1 & $\tfrac{1}{\sqrt{2}} (\Phi_1^{\dagger} \tau^k \Phi_1)$ & $\tfrac{1}{\sqrt{2}} (\Phi_1^{\dagger} \tau^k \Phi_2)$ \\
 &  & $\tfrac{1}{\sqrt{2}} (\Phi_2^{\dagger} \tau^k \Phi_2)$ & $\tfrac{1}{\sqrt{2}} (\Phi_2^{\dagger} \tau^k \Phi_1)$ \\ \hline
 1 & 0 & --- & $\tfrac{1}{\sqrt{2}} (\tilde{\Phi}_1 \Phi_2)$ \\ \hline
 \multirow{2}{*} 1 & \multirow{2}{*} 1 & $\tfrac{1}{2} (\tilde{\Phi}_1 \tau^k \Phi_1)$ & $\tfrac{1}{\sqrt{2}} (\tilde{\Phi}_1 \tau^k \Phi_2)$ \\
  & & $\tfrac{1}{2} (\tilde{\Phi}_2 \tau^k \Phi_2)$ & \\
  \end{tabular}
  \caption{Initial states for $2 \to 2$ scattering broken down by total hypercharge, total weak isospin, and transformation under $\mathbb{Z}_2$. The $\mathbb{Z}_2$-even, $Y = 1$, $\tau = 0$ states are identically zero. We have omitted the $Y = - 1$ states, which can be obtained from the $Y = 1$ states by charge conjugation.}
  \label{tab:reps}
\end{table}

For a given initial state $i$ and final state $f$, the corresponding element of $\mathbf{a}_0$ is given by,
\begin{equation}
(\mathbf{a}_0)_{i, f} = \frac{1}{16 \pi s} \int_{-s}^{0} \! dt \, \mathcal{M}_{i \otimes f}(s, t),
\end{equation}
where we have assumed the states can be treated as massless. Here, $\mathcal{M}_{i \otimes f}$ represents the sum of all possible amplitudes involving $w_j^{\pm}$, $n_j$, and $n_j^{*}$ (with the appropriate weights) that can be formed from the initial and final states. For example, suppressing the explicit dependence on $s$ and $t$,
\begin{align}
\label{eq:Mex}
\mathcal{M}_{\tfrac{1}{\sqrt{2}} (\Phi_1^{\dagger} \Phi_1) \otimes \tfrac{1}{\sqrt{2}} (\Phi_2^{\dagger} \tau^3 \Phi_2)} = \frac{1}{2} &\left( \mathcal{M}_{w_1^+ w_1^- \to w_2^+ w_2^-} - \mathcal{M}_{w_1^+ w_1^- \to n_2 n_2^*} \right. \\
&\left. + \mathcal{M}_{n_1 n_1^* \to w_2^+ w_2^-} - \mathcal{M}_{n_1 n_1^* \to n_2 n_2^*} \right). \nn
\end{align}
The amplitude in Eq.~\eqref{eq:Mex} is actually zero at tree level (it's non-zero at one-loop), but was chosen as it is a simple example of the combinatoric exercise. 

The block diagonal structure of $\mathbf{a}_0$ does not hold beyond tree level. However, it still has an important consequence for the analysis at the one-loop level. For all tree level blocks whose eigenvalues are unique (for a given net electric charge in the scattering process), because the block diagonal elements start at tree level and the off-block diagonal elements start at one-loop, the off-block diagonal elements do not affect these eigenvalues until the two-loop level. Thus, they can be ignored for the purposes of the one-loop analysis. For neutral initial states, eight of the 14 eigenvalues are unique, with three additional eigenvalues appearing twice. On the other hand, all of the eigenvalues for the charged initial states are unique. This difference occurs because, for example, both $\tfrac{1}{2} (\tilde{\Phi}_i \tau^+ \Phi_j)$ and $\tfrac{1}{2} (\tilde{\Phi}_i^{*} \tau^- \Phi_j^{*})$ are neutral initial states (that lead to the same block of scattering amplitudes), whereas $\tfrac{1}{2} (\tilde{\Phi}_i \tau^3 \Phi_j)$ and $\tfrac{1}{2} (\tilde{\Phi}_i^{*} \tau^3 \Phi_j^{*})$ have opposite electric charges.

At one-loop, the approach of~\cite{Ginzburg:2003fe, Ginzburg:2005dt} works for all diagrams where the particles can all be treated as massless. In the high energy limit under consideration, this corresponds to all the 1PI one-loop diagrams. The only diagrams that can not be computed using this strategy are the external wavefunction corrections, as they are independent of $s$ (and $t$).

\section{One-Loop Calculation}
\label{sec:1L}
\subsection{Equivalence Theorem}
\label{sec:Gbet}
We are interested in the full set of one-loop amplitudes for longitudinal vector boson and Higgs boson scattering in the energy regime, $s \gg |\lambda_i | v^2 \gg M_{W}^2$, $s \gg m_{12}^2$.\footnote{Note that this energy regime does not imply the decoupling limit discussed in Sec.~\ref{sec:back2HDM} as this set of conditions does not generally require that one of the Higgs bosons be parametrically lighter than the rest. Furthermore, it is worth mentioning that partial-wave amplitudes satisfy unitarity bounds uniformly for all $s$ and $t$ that are sufficiently far away from the resonances of the theory~\cite{Dawson:1988va, Dawson:1989up, Murphy:2015kag}. We have chosen to examine the bounds from perturbative unitarity in this energy regime as the computation in simpler in this case.} The computation of these amplitudes can be greatly simplified through use of the Goldstone boson equivalence theorem~\cite{Lee:1977yc, Lee:1977eg, Chanowitz:1985hj, Bagger:1989fc, Denner:1996gb}. At the one-loop level, the theorem states that an amplitude involving $n$ external, longitudinally polarized vector bosons is related to an amplitude with $n$ external Goldstone bosons as,
\begin{equation}
\label{eq:equiv}
\mathcal{M}(W^{\pm}_L,\, Z_L,\, h, \ldots) = (i C)^n \mathcal{M}(w^{\pm},\, z,\, h, \ldots) + O(M_W / \sqrt{s}).
\end{equation}
To make the computation of scattering amplitudes involving longitudinal vector bosons as simple as possible, we will use Eq.~\eqref{eq:equiv} and choose our renormalization scheme, to be discuss in Sec.~\ref{sec:renorm}, such that $C = 1$.
As was just alluded to, the constant $C$ depends on the choice of renormalization scheme~\cite{Bagger:1989fc},
\begin{equation}
C = \frac{M_W^0}{M_W} \frac{Z_{W^+W^-}^{1/2}}{Z_{w^+w^-}^{1/2}} [1 + O(g_2^2)],
\end{equation}
where $M_W^0$ and $M_W$ are the bare and renormalized mass of $W^{\pm}$ respectively. In general, we denote the bare value of a parameter $X$, as $X^0$, and its counterterm is defined by $\delta X = X^0 - X$. $Z_{W^+W^-}$ and $Z_{w^+w^-}$ are the wavefunction renormalization constants of the physical $W^{\pm}$ bosons and the charged Goldstone bosons, $w^{\pm}$, respectively.  

Ref.~\cite{Bagger:1989fc} showed that $C = 1 + O(g_2^2)$ when the Goldstone bosons are renormalized using a momentum subtraction scheme with subtraction scale $m^2 \ll \lambda_i v^2$, where $g_2$ is the gauge coupling of $SU(2)_L$. Since $M_W^2 = g_2^2 v^2 / 4$ at tree level, the $O(g_2^2)$ terms are small in the parameter regime of interest, $g_2^2 \ll \lambda_i$. In addition, this hierarchy in parameters further simplifies that calculation by allowing us to consider only scalar particles in the loop diagrams. Furthermore, since $Z_{W^+W^-}^{1/2} = 1 + O(g_2^2)$, then it follows that $M_W^0 / M_W = 1 + \delta M_W / M_W =  Z_{w^+w^-}^{1/2} [1 + O(g_2^2)]$. This relation implies 
\begin{equation}
\label{eq:dv}
\frac{\delta v^2}{v^2} = Z_w - 1,
\end{equation}
with $Z_w \equiv Z_{w^+w^-}$.

\subsection{Renormalization}
\label{sec:renorm}
The renormalization of the two-Higgs doublet model is discussed in depth in~\cite{Santos:1996vt}. In contrast with that work, and the loop level SM perturbative unitarity analyses~\cite{Dawson:1988va, Dawson:1989up, Durand:1992wb, Maher:1993vj, Durand:1993vn}, we use the $\overline{\text{MS}}$ renormalization scheme with two exceptions, which are necessary to satisfy the Goldstone boson equivalence theorem. The first exception is the finite renormalization of $v$, Eq.~\eqref{eq:dv}. In addition, instead of $\overline{\text{MS}}$, we exactly cancel the tadpole diagrams by subtracting the appropriate combination of Goldstone boson self-energy and Goldstone-Higgs mixing at zero momentum from all the scalar self-energies and mixings~\cite{Taylor:1976ru}. The relevant part of the bare Lagrangian in the Higgs basis is,
\begin{align}
\label{eq:tad}
\mathcal{L} \supset &- ((M_{11}^2)^0 + \tfrac{1}{2} \Lambda_1^0 (v^2)^0)(w^+ w^- \tfrac{1}{2} z^2 + \tfrac{1}{2} \phi_1^2 + v^0 \phi_1) \\
&- ((M_{12}^2)^0 + \tfrac{1}{2} \Lambda_6^0 (v^2)^0)(w^+ H^- + H^+ w^- + z A + \phi_1 \phi_2 + v^0 \phi_2).\nn
\end{align}
At tree level, the right hand side of~\eqref{eq:tad} is zero due to Eqs.~\eqref{eq:Hvevs}, but this cancellation does not hold in general at the loop level. More to the point,~\eqref{eq:tad} shows that the tadpole counterterms are related to the self-energies of the Goldstone bosons and the Goldstone-Higgs mixing at zero momentum. The particular combinations are,
\begin{align}
\delta T_h &= - v^0 [s_{\beta - \alpha} \Pi_{z z}(0) + c_{\beta - \alpha} \Pi_{z A}(0)], \\ 
\delta T_H &= - v^0 [c_{\beta - \alpha} \Pi_{z z}(0) - s_{\beta - \alpha} \Pi_{z A}(0)]. \nn
\end{align}
Note that $\Pi_{z z}(0)  = \Pi_{w^+ w^-}(0)$ and $\Pi_{z A}(0)  = \Pi_{w^+ H^-}(0)$, and $\Pi_{ij}(p^2) = \Pi_{ji}(p^2)$. All the tadpole diagrams can then be ignored provided the scalar self-energies are modified as follows,
\begin{align}
\widetilde{\Pi}_{w^+ w^-}(p^2) &= \Pi_{w^+ w^-}(p^2) - \Pi_{z z}(0), \\
\widetilde{\Pi}_{z z}(p^2) &= \Pi_{z z}(p^2) - \Pi_{z z}(0), \nn \\
\widetilde{\Pi}_{h h}(p^2) &= \Pi_{h h}(p^2) - s_{\beta - \alpha}^2 \Pi_{z z}(0) - 2 s_{\beta - \alpha} c_{\beta - \alpha} \Pi_{z A}(0), \nn \\
\widetilde{\Pi}_{H H}(p^2) &= \Pi_{H H}(p^2) - c_{\beta - \alpha}^2 \Pi_{z z}(0) + 2 s_{\beta - \alpha} c_{\beta - \alpha} \Pi_{z A}(0), \nn 
\end{align}
with $\Pi_{H^+ H^-}$ and $\Pi_{A A}$ unchanged. The mixing between the Goldstone bosons and the physical Higgs bosons must also be modified,
\begin{align}
\widetilde{\Pi}_{w^+ H^-}(p^2) &= \Pi_{w^+ H^-}(p^2) - \Pi_{z A}(0),\\
\widetilde{\Pi}_{z A}(p^2) &= \Pi_{z A}(p^2) - \Pi_{z A}(0), \nn \\
\widetilde{\Pi}_{h H}(p^2) &= \Pi_{hH}(p^2) - s_{\beta - \alpha} c_{\beta - \alpha} \Pi_{z z}(0) - (c_{\beta - \alpha}^2 - s_{\beta - \alpha}^2) \Pi_{z A}(0). \nn
\end{align}
Explicit expressions for the self-energies can be found in Appendix~\ref{sec:SE}. The wavefunction renormalization then depends on the shifted self-energies as well,
\begin{align}
Z_{i i}^{1/2} &= 1 + \frac{1}{2} \left. \frac{d \widetilde{\Pi}_{i i}(p^2)}{dp^2} \right|_{p^2 \to m_i^2}, \\ 
Z_{i j}^{1/2} &= \frac{\widetilde{\Pi}_{i j}(m_i^2)}{m_i^2 - m_j^2}. \nn
\end{align}
Note that $Z_{i j}^{1/2}$ is not symmetric, e.g. $Z_{w^+ H^-}^{1/2} = 0$, but $Z_{H^+ w^-}^{1/2} \neq 0$. For later convenience, we define a reduced wavefunction renormalization,
\begin{equation}
\label{eq:reduce}
z_{ij}^{1/2} = 16 \pi^2 (Z_{ij}^{1/2} - \delta_{ij}).
\end{equation}
Importantly, in addition to exactly canceling the tadpoles diagrams, this scheme renormalizes the Goldstone bosons on-shell, which satisfies the condition for the Goldstone boson equivalence theorem to hold at one-loop as discussed in Sec.~\ref{sec:Gbet}.

The quartic couplings and the soft $\mathbb{Z}_2$ breaking parameter are renormalized using the $\overline{\text{MS}}$ scheme. The renormalized parameters are defined in terms of the bare parameters as,
\begin{equation}
\lambda_i^0 = \lambda_i + \delta \lambda_i, \quad (m_{12}^2)^0 = m_{12}^2 + \delta m_{12}^2,
\end{equation}
where, as previously stated, $X^0$ are bare parameters and $X$ are renormalized parameters. In $D = 4 - 2 \epsilon$ dimensions, after making the following replacements in the Lagrangian, $\lambda_i \to \lambda_i \tilde{\mu}^{2\epsilon}$ with $\mu^2 = 4 \pi e^{- \gamma} \tilde{\mu}^2$, the counterterms can be written as
\begin{equation}
\label{eq:dC}
\delta X = \frac{1}{16 \pi^2 \epsilon} \beta_X.
\end{equation}
Our findings for the beta functions in Eq.~\eqref{eq:dC} agree with the well known results in the literature, see e.g.~\cite{Hill:1985tg},\footnote{Recently, the complete two-loop beta functions in the $CP$-conserving 2HDM with a softly broken $\mathbb{Z}_2$ symmetry have been determined~\cite{Chowdhury:2015yja}.}
\begin{align}
\label{eq:beta}
\beta_{\lambda_1} &= 6 \lambda_1^2 + 2 \lambda_3^2 +2 \lambda_3 \lambda_4 + \lambda_4^2 + \lambda_5^2, \\
\beta_{\lambda_2} &= 6 \lambda_2^2 + 2 \lambda_3^2 +2 \lambda_3 \lambda_4 + \lambda_4^2 + \lambda_5^2, \nn \\
\beta_{\lambda_3} &= 2 \lambda_3^2 + \lambda_4^2 + \left(\lambda_1 + \lambda_2\right)\left(3 \lambda_3 + \lambda_4\right) + \lambda_5^2, \nn \\
\beta_{\lambda_4} &= \left(\lambda_1 + \lambda_2 +4 \lambda_3\right) \lambda_4 + 2 \lambda_4^2 + 4 \lambda_5^2, \nn \\
\beta_{\lambda_5} &= \left(\lambda_1 + \lambda_2 + 4 \lambda_3 + 6 \lambda_4\right) \lambda_5, \nn \\
\beta_{m_{12}^2} &= m_{12}^2 \left(\lambda_3 + 2 \lambda_4 + 3 \lambda_5\right). \nn
\end{align}
For a given parameter $X$ in Eq.~\eqref{eq:beta},
\begin{equation}
\beta_{X} = 16 \pi^2 \mu^2  \frac{d X}{d \mu^2}.
\end{equation}
From Eqs.~\eqref{eq:dv} and~\eqref{eq:dC}, it is straightforward to derive counterterms for the mass parameters,
\begin{align}
\delta m_h^2 &= (\delta m_h^2)_{\overline{\text{MS}}} + (Z_w - 1) (m_h^2 + m_{Z_w}^2 c_{\beta-\alpha}^2), \\
\delta m_H^2 &= (\delta m_H^2)_{\overline{\text{MS}}} + (Z_w - 1) (m_H^2 + m_{Z_w}^2 s_{\beta-\alpha}^2), \nn \\
\delta (m_h m_H) &= (\delta (m_h m_H))_{\overline{\text{MS}}} - (Z_w - 1) m_{Z_w}^2 s_{\beta-\alpha} c_{\beta-\alpha}, \nn \\
\delta m_{H^+}^2 &= (\delta m_{H^+}^2)_{\overline{\text{MS}}} + (Z_w - 1) (m_{H^+}^2 + m_{Z_w}^2), \nn \\
\delta m_A^2 &= (\delta m_A^2)_{\overline{\text{MS}}} + (Z_w - 1) (m_A^2 + m_{Z_w}^2), \nn
\end{align}
where we have defined,
\begin{equation}
2 m_{Z_w}^2 = m_h^2 s_{\beta-\alpha}^2 + m_H^2 c_{\beta-\alpha}^2 + (m_h^2 - m_H^2) s_{2\beta-2\alpha} t_{2\beta}^{-1} - 2 m_{12}^2 s_{\beta}^{-1} c_{\beta}^{-1} .
\end{equation}
These counterterms render the Higgs self-energies finite, which in turn modify the tree level relations between the physical Higgs masses and the parameters in Eq.~\eqref{eq:Z2pot}. The loop level relations can be written in a form analogous to the tree level relations,
\begin{align}
\label{eq:treereps}
\lambda_1 v^2 c_{\beta}^2 &= \bar{m}_H^2 c_{\alpha}^2 + \bar{m}_h^2 s_{\alpha}^2 - m_{12}^2 t_{\beta}, \\
\lambda_2 v^2 s_{\beta}^2 &= \bar{m}_H^2 s_{\alpha}^2 + \bar{m}_h^2 c_{\alpha}^2 - m_{12}^2 t_{\beta}^{-1}, \nn \\
\lambda_3 v^2 s_{2\beta} &= (\bar{m}_H^2 - \bar{m}_h^2) s_{2\alpha} - 2 (m_{12}^2 - \bar{m}_{H^+}^2 s_{2\beta}), \nn \\
\lambda_4 v^2 &= m_{12}^2 s_{\beta}^{-1} c_{\beta}^{-1} + \bar{m}_A^2 - 2 \bar{m}_{H^+}^2, \nn \\
\lambda_5 v^2 &= m_{12}^2 s_{\beta}^{-1} c_{\beta}^{-1} - \bar{m}_A^2, \nn
\end{align}
with $\bar{m}_i^2 \equiv m_i^2 - \text{Re}[\widetilde{\Pi}_{ii}(m_i^2)]$ (and $\widetilde{\Pi}$ being the renormalized self-energy). We have chosen not to rediagonalize the mass matrix for the neutral, $CP$-even Higgs bosons, which would have induced a dependence of Eq.~\eqref{eq:treereps} on $\widetilde{\Pi}_{hH}$ and a redefinition of $\alpha$. 

\subsection{$2 \to 2$ Scattering Amplitudes}
The only one-loop diagrams that survive in the limit $s \gg |\lambda_i | v^2 \gg M_{W}^2$, $s \gg m_{12}^2$ are the 1PI diagrams with two internal lines, i.e. 1PI bubble diagrams, and the external wavefunction renormalization diagrams.\footnote{The 1PI diagrams with three and four internal lines scale as $v^2 / s$ and $v^4 / s^2$ respectively. The contribution of these diagrams to longitudinal vector boson scattering in the SM is IR-finite~\cite{Dawson:1989up}. In the 2HDM, there are no new topologies, and the presence of extra masses in the loops can only serve to improve the regulation of the IR behavior of these diagrams, such that they can indeed be neglected in the limit $s \gg v^2$.}  In this limit, the masses of the internal particles can be neglected in the bubble diagrams. This allows us to use the non-physical Higgs fields $w_j^{\pm}$, $n_j$, and $n_j^{*}$ in computing the bubble diagram contribution to $\mathbf{a}_0$. Furthermore, in this limit, the bubble diagrams preserve the block diagonal form of $\mathbf{a}_0$. Up to symmetry factors, all of the bubble diagrams have the form,
\begin{equation}
\frac{1}{16 \pi^2} \lambda_i \lambda_j \left( \frac{1}{\epsilon} - \ln\left(\frac{- p^2 - i 0^+}{\mu^2}\right) + 2 \right).
\end{equation}
For $p^2 > 0$, the branch cut in the log yields $ \ln(- p^2) \to \ln(p^2) - i \pi$. 

Unfortunately, this trick of using non-physical Higgs fields will not work when computing the one-loop corrections to the external legs of the amplitudes because the masses of the Higgs bosons can not be neglected for those diagrams. Instead we calculate and renormalize the external wavefunction corrections in the Higgs basis with the physical Higgs fields, as the expressions are simpler in this basis. The results are then converted back to the parameters of the $\mathbb{Z}_2$-basis such that the parameterization of the scattering amplitudes is consistent amongst all of its contributions.

All of the energy dependence of $\mathbf{a}_0$ in this limit can be subsumed into running couplings through standard renormalization group (RG) methods. The running couplings, $\lambda_i(\mu^2)$, are the solutions to Eq.~\eqref{eq:beta} with initial conditions at the scale $\mu_0$ given by Eq.~\eqref{eq:treereps}. By setting $\mu^2 = s$ in the fixed order scattering amplitudes, we remove all of the explicit energy dependence from (the high energy limit of) the amplitudes. Then the couplings appearing in the scattering amplitudes should be interpreted as the running couplings evaluated at $\mu^2 = s$, i.e. $\lambda_i(s)$. 

Consider a generic block of one-loop scattering amplitudes in $\mathbf{a}_0$,
\begin{equation}
256 \pi^3 \mathbf{a}_0^{Q Y \tau \mathbb{Z}_2} = 
\begin{pmatrix}
- 16 \pi^2 b_0 + b_1 & - 16 \pi^2 c_0 + c_1 \\[2mm]
- 16 \pi^2 c_0 + c_1 & - 16 \pi^2 d_0 + d_1
\end{pmatrix}.
\end{equation}
We label blocks of $\mathbf{a}_0$ and their eigenvalues by the electric charge ($Q$), hypercharge ($Y$), weak isospin ($\tau$), and transformation under $\mathbb{Z}_2$ of their initial state. For a given $Q$, if the tree level eigenvalues for this block are unique (with respect to all of the eigenvalues of $\mathbf{a}_0$ for that $Q$), the corresponding one-loop level eigenvalues are
\begin{align}
\label{eq:EVmod}
256 \pi^3 a_{0\pm}^{Q Y \tau \mathbb{Z}_2} = &- 8 \pi^2 \left(b_0 + d_0 \pm \sqrt{\left(b_0 - d_0\right)^2 + 4 c_0^2}\right) \\
&+ \frac{1}{2} \left(b_1 + d_1 \pm \frac{\left(b_0 - d_0\right) \left(b_1 - d_1\right) + 4 c_0 c_1}{\sqrt{\left(b_0 - d_0\right)^2 + 4 c_0^2}} \right). \nn
\end{align}
Explicit expressions for the scattering amplitudes that form the block diagonal and off-block diagonal elements of $\mathbf{a}_0$ are given in Appendices~\ref{sec:RSA} and~\ref{sec:RSA2} respectively. With this organization, for eigenvalues that are unique at tree level, the corresponding one-loop eigenvalues only depend on the results of Appendix~\ref{sec:RSA}. On the other hand, for degenerate tree level eigenvalues, the corresponding one-loop eigenvalues depend on the results of both Appendices~\ref{sec:RSA} and~\ref{sec:RSA2}. Our results for the tree level eigenvalues agree with those of~\cite{Kanemura:1993hm, Akeroyd:2000wc, Ginzburg:2003fe, Ginzburg:2005dt},
\begin{align}
\label{eq:EVtree}
- 16 \pi a_{0\pm}^{0 1 \text{even}} &= \frac{1}{2} \left(\lambda_1 + \lambda_2 \pm \sqrt{\left(\lambda_1 - \lambda_2\right)^2 + 4 \lambda_4^2} \right), \quad - 16 \pi a_{0\pm}^{0 1 \text{odd}} = \lambda_3 \pm \lambda_5, \\
- 16 \pi a_{0\pm}^{1 1 \text{even}} &= \frac{1}{2} \left(\lambda_1 + \lambda_2 \pm \sqrt{\left(\lambda_1 - \lambda_2\right)^2 + 4 \lambda_5^2} \right), \quad - 16 \pi a_{0}^{1 1 \text{odd}} = \lambda_3 + \lambda_4, \nn \\
- 16 \pi a_{0\pm}^{0 0 \text{even}} &= \frac{1}{2} \left(3 \lambda_1 + 3 \lambda_2 \pm \sqrt{\left(3 \lambda_1 - 3 \lambda_2\right)^2 + 4 \left(2 \lambda_3 + \lambda_4\right)^2}\right), \nn \\
- 16 \pi a_{0\pm}^{0 0 \text{odd}} &= \lambda_3 + 2 \lambda_4 \pm 3 \lambda_5, \quad - 16 \pi a_{0}^{1 0 \text{odd}} = \lambda_3 - \lambda_4. \nn 
\end{align}
In Eq.~\eqref{eq:EVtree}, the tree level eigenvalues are not labeled with $Q$ as, unlike the one-loop eigenvalues, they are degenerate with respect to the electric charge of the initial state.

\section{Analysis of One-Loop Perturbative Unitarity Constraints}
\label{sec:re}
In this Section, we analyze the one-loop level unitarity constraints on the 2HDM. In addition to reproducing the SM results with our methods, we consider two simplified scenarios for the 2HDM: the case where the Goldstone boson scattering amplitudes have an SO(3) symmetry, and a 2HDM whose parameters are inspired by the form of the Higgs potential in the MSSM. It should be noted however that the results in Appendices~\ref{sec:SE}, ~\ref{sec:RSA}, and~\ref{sec:RSA2} can be used to analyze the $CP$-conserving 2HDM with a softly-broken $\mathbb{Z}_2$ symmetry, which is more general than any of the scenarios considered in this Section.

\subsection{General Considerations}
Before getting into specific examples, we make some general considerations regarding the bounds on one-loop amplitudes from perturbative unitarity. Consider the case of when the tree level eigenvalue does not contain a square root, e.g. all of the $\mathbb{Z}_2$-odd eigenvalues in Eqs.~\eqref{eq:EVtree}. At one-loop, an eigenvalue of this type can be parameterized as
\begin{equation}
\label{eq:singEV}
256 \pi^3 a_0 = - 16 \pi^2 b_0 + b_1.
\end{equation}
We will explicitly break $b_1$ up into its real and imaginary parts in what follows, $b_1 = b_R + i b_I$. The two constraints that are commonly considered in tree level analyses are~\eqref{eq:Smatbound2}, $1 \geq |a_0|$ and $\tfrac{1}{2} \geq |\text{Re}(a_0)|$. At one-loop, these bounds become
\begin{equation}
8 \pi \geq |b_0 - \tfrac{1}{16 \pi^2} b_R|,\, 16 \pi \geq \sqrt{(b_0 - \tfrac{1}{16 \pi^2} b_R)^2 + (\tfrac{1}{16 \pi^2} b_I)^2}.
\end{equation}
From this, we see the usual interplay between perturbativity and unitarity. The more interesting bound is~\eqref{eq:Smatbound}, $\tfrac{1}{2} \geq |a_0 - i / 2|$, which first becomes non-trivial at the one-loop order. Expanding this unitarity constraint yields
\begin{equation}
\label{eq:sol1}
\frac{1}{4} \geq \frac{1}{4} + \frac{b_0^2}{256 \pi^2} - \frac{b_I}{256 \pi^3} - \frac{b_0 b_R}{2048 \pi^4} + \frac{b_R^2}{65536 \pi^6} + \frac{b_I^2}{65536 \pi^6}.
\end{equation}
The leading order bound from~\eqref{eq:sol1} is
\begin{equation}
\label{eq:sol1a}
b_I = \text{Im}(b_1) \geq \pi b_0^2.
\end{equation}
Assuming~\eqref{eq:sol1a} is saturated, or that perturbation theory holds, leads to a constraint on the real part of $b_1$,
\begin{align}
\label{eq:sol2}
b_0 b_R &\geq \frac{b_R^2}{32 \pi^2} + \frac{b_I^2}{32 \pi^2}, \\
b_0 b_R &\geq \frac{b_R^2}{32 \pi^2}, \nn \\
b_0 &\geq \frac{b_R}{32 \pi^2},\quad (b_R > 0) . \nn
\end{align}
Neglecting the wavefunction renormalization contribution to the scattering amplitude,~\eqref{eq:sol2} leads to bounds on the beta functions of the theory,
\begin{equation}
\label{eq:sol4}
b_0 \left(16 \pi^2 + b_0 + \pi \sqrt{256 \pi^2 - b_0^2}\right) \geq 3 \beta_{b_0} \geq b_0 \left(16 \pi^2 + b_0 - \pi \sqrt{256 \pi^2 - b_0^2}\right),
\end{equation}
where $16 \pi \geq |b_0|$ and $\beta_{b_0}$ is the linear combination of beta functions appearing in the scattering amplitude. For example, if $b_0 = \lambda_3 + 2 \lambda_4 + 3 \lambda_5$, then $\beta_{b_0} = \beta_{\lambda_3} + 2 \beta_{\lambda_4} + 3 \beta_{\lambda_5}$. Expanding~\eqref{eq:sol4} to leading order in $b_0$ yields,
\begin{equation}
\label{eq:sol5}
32 \pi^2 \left|b_0\right| \gtrapprox 3 \left|\beta_{b_0}\right| \gtrapprox b_0^2.
\end{equation}

Now consider the more general case where the eigenvalue has the form of Eq.~\eqref{eq:EVmod}. We will again expand the one-loop parts of the eigenvalues into the real and imaginary parts. The leading order bound from~\eqref{eq:Smatbound} is,
\begin{align}
\label{eq:sol6a}
&(b_I - d_I) (b_0 - d_0) + 4 c_I c_0 + (b_I + d_I) \sqrt{(b_0 - d_0)^2 + 4 c_0^2}  \\
&\geq \pi \left[(b_0^2 - d_0^2) (b_0 - d_0) + 4 (b_0 + d_0) c_0^2 + (b_0^2 + 2 c_0^2 + d_0^2) \sqrt{(b_0 - d_0)^2 + 4 c_0^2}\right] . \nn
\end{align}
The constraint~\eqref{eq:sol6a} is saturated when,
\begin{equation}
\label{eq:sol6}
b_I = \pi (b_0^2 + c_0^2),\quad c_I = \pi (b_0 + d_0) c_0,\quad d_I = \pi (d_0^2 + c_0^2).
\end{equation}
For all of the scattering amplitudes in Appendix~\ref{sec:RSA}, the 1PI contribution to the amplitudes satisfies Eqs.~\eqref{eq:sol6}. This property of the scattering amplitudes is perfectly consistent with the statement that the equality in~\eqref{eq:Smatbound} (or~\eqref{eq:sol6a}) is satisfied only when all of the $2 \to n$ scattering processes vanish. When the wavefunction renormalization contribution to the scattering amplitudes contains an imaginary part, it is due to there being open decay channels. Clearly, decays are inelastic, and so the equality in~\eqref{eq:Smatbound} cannot be satisfied in this case. Neglecting the imaginary parts of the one-loop amplitudes, the generalization of~\eqref{eq:sol2} to eigenvalues with the form of Eq.~\eqref{eq:EVmod} is,
\begin{align}
&32 \pi^2 [\left(b_0 b_R + 2 c_0 c_R + d_0 d_R\right) \sqrt{\left(b_0 - d_0\right)^2 + 4 c_0^2} \\
& + \left(b_0^2 +2 c_0^2\right) b_R + \left(d_0^2 +2 c_0^2\right) d_R + 2 c_0 \left(b_0 + d_0\right) c_R - b_0 d_0 \left(b_R + d_R\right) ] \nn \\
&\geq \left(b_R^2 + 2 c_R^2 + d_R^2\right) \sqrt{\left(b_0 - d_0\right)^2 + 4 c_0^2} + b_0 \left(b_R^2 - d_R^2\right) + d_0 \left(d_R^2 - b_R^2\right) + 2 \left(b_0 + d_0\right) c_R^2  \nn \\
& \frac{2 c_0}{\left(b_0 - d_0\right)^2 + 4 c_0^2} \left[b_0 b_R^2 c_0 + 8 b_R c_0^2 c_R - 4 b_0 c_0 c_R^2 + b_R^2 c_0 d_0 - 4 b_0 b_R c_R d_0 - 4 c_0 c_R^2 d_0  \right. \nn \\
&\left. + 4 b_R c_R d_0^2 + 2 d_R \left(2 b_0^2 c_R + c_0\left(4 c_0 c_R - b_R d_0\right) - b_0 \left(b_R c_0 + 2 c_R d_0\right)\right) + c_0 \left(b_0 + d_0 \right) d_R^2\right]. \nn
\end{align}

\subsection{Reproduction of the Standard Model Results}
Since neither this renormalization scheme nor this basis for computing $\mathbf{a}_0$ has been used for the Standard Model, we begin our analysis of specific models by reproducing the results of the SM. The matrix of scattering amplitudes for neutral initial states is,
\begin{align}
\label{eq:matSM}
&256 \pi^3 \mathbf{a}_{0}^{Q=0} / \lambda = \\ 
&\begin{pmatrix}
\tfrac{3 (- 64 \pi^2 + (51 + 12 i \pi + 2 \sqrt{3} \pi) \lambda)}{4} & \tfrac{(13 - 2 \sqrt{3} \pi) \lambda}{2} & \tfrac{(- 13 + 2 \sqrt{3} \pi) \lambda}{2} & \tfrac{(- 13 + 2 \sqrt{3} \pi) \lambda}{2} \\
\tfrac{(13 - 2 \sqrt{3} \pi) \lambda}{2} & \tfrac{- 64 \pi^2 + (59 + 4 i \pi + 2 \sqrt{3} \pi) \lambda}{4} & \tfrac{(13 - 2 \sqrt{3} \pi) \lambda}{4} & \tfrac{(13 - 2 \sqrt{3} \pi) \lambda}{4} \\
\tfrac{(- 13 + 2 \sqrt{3} \pi) \lambda}{2} & \tfrac{(13 - 2 \sqrt{3} \pi) \lambda}{4} & \tfrac{- 32 \pi^2 + (23 + 2 i \pi + 2 \sqrt{3} \pi) \lambda}{2} & 0 \\
\tfrac{(- 13 + 2 \sqrt{3} \pi) \lambda}{2} & \tfrac{(13 - 2 \sqrt{3} \pi) \lambda}{4} & 0 & \tfrac{- 32 \pi^2 + (23 + 2 i \pi + 2 \sqrt{3} \pi) \lambda}{2}
\end{pmatrix}, \nn
\end{align}
where the initial (final) states of the columns (rows) are,
\begin{equation}
\mathbf{a}_{0}^{Q=0} = \bordermatrix{~ & \tfrac{1}{\sqrt{2}} \Phi^{\dagger} \Phi &  \tfrac{1}{\sqrt{2}} \Phi^{\dagger} \tau^3 \Phi & \tfrac{1}{2} \tilde{\Phi} \tau^+ \Phi & \tfrac{1}{2} \tilde{\Phi}^{*} \tau^+ \Phi^{*} \cr
\tfrac{1}{\sqrt{2}} \Phi^{\dagger} \Phi & & & \cr
\tfrac{1}{\sqrt{2}} \Phi^{\dagger} \tau^3 \Phi & & & \cr
\tfrac{1}{2} \tilde{\Phi} \tau^+ \Phi & & & \cr
\tfrac{1}{2} \tilde{\Phi}^{*} \tau^+ \Phi^{*} & & & \cr} ,
\end{equation}
and $\Phi$ is the Higgs doublet of the SM with the Higgs mass at tree level given by $m_h^2 = \lambda v^2$. Note that at tree level, $\mathbf{a}_0$ is diagonal in the SM, as opposed to the block diagonal structure of the 2HDM. The eigenvalues of Eq.~\eqref{eq:matSM} are
\begin{align}
\label{eq:evSM}
a_{0,1}^{Q=0} &= - 6 \bar{\lambda} + \frac{(153 + 36 i \pi + 6 \sqrt{3} \pi)\bar{\lambda}^2}{\pi}, \quad a_{0,2}^{Q=0} = - 2 \bar{\lambda} + \frac{(72 + 4 i \pi)\bar{\lambda}^2}{\pi}, \\
a_{0,3}^{Q=0} &= - 2 \bar{\lambda} + \frac{(46 + 4 i \pi + 4 \sqrt{3} \pi)\bar{\lambda}^2}{\pi}, \quad a_{0,4}^{Q=0} = - 2 \bar{\lambda} + \frac{(33 + 4 i \pi + 6 \sqrt{3} \pi)\bar{\lambda}^2}{\pi}, \nn
\end{align}
where $\bar{\lambda} = \lambda / 32 \pi$. Eq.~\eqref{eq:evSM} is in agreement with Ref.~\cite{Durand:1992wb}. Notice that $a_{0,1}^{Q=0}$ is unaffected by the diagonalization of Eq.~\eqref{eq:matSM}. This is due to the fact at tree level, $a_{0,1}^{Q=0}$ is different from the other three diagonal elements of Eq.~\eqref{eq:matSM}. Along the same lines, because $a_{0,2}^{Q=0} = a_{0,3}^{Q=0} = a_{0,4}^{Q=0}$ at tree level, all three of these eigenvalues are affected by the off-diagonal elements of $\mathbf{a}_0$.

Another check of our SM result is to look at the fixed order expressions for $a_{0}^{Q=0}$ in terms of the physical Higgs mass. Expanding the running coupling to the one-loop order,
\begin{align}
\lambda \left(s\right) &= \lambda \left(\mu_0^2\right) \left(1 - \frac{3}{8 \pi^2} \lambda \left(\mu_0^2\right) \ln \frac{s}{\mu_0^2}\right)^{-1}, \\
 &\approx \lambda \left(\mu_0^2\right) \left(1 + \frac{3}{8 \pi^2} \lambda \left(\mu_0^2\right) \ln \frac{s}{\mu_0^2}\right), \nn 
\end{align}
and eliminating $\lambda$ through 
\begin{equation}
\lambda \left(\mu_0^2\right) = \frac{m_h^2}{v^2} - \frac{m_h^4}{32 \pi^2 v^4} \left(3 \sqrt{3} \pi - 25 + 12 \ln \frac{m_h^2}{\mu_0^2}\right),
\end{equation}
we find
\begin{equation}
\label{eq:a1mh}
a_{0,1}^{Q=0} = - \frac{3 m_h^2}{16 \pi v^2} \left[1 - \frac{m_h^2}{16 \pi^2 v^2} \left(\frac{1}{4} + 3 i \pi + 2 \sqrt{3} \pi - 6 \ln \frac{s}{m_h^2}\right) + O\left(\frac{m_h}{4 \pi v}\right)^4 \right]  . 
\end{equation}
Eq.~\eqref{eq:a1mh} is also in agreement with the results of~\cite{Durand:1992wb}. 

\begin{figure}
  \centering
  \includegraphics[width=0.7\textwidth]{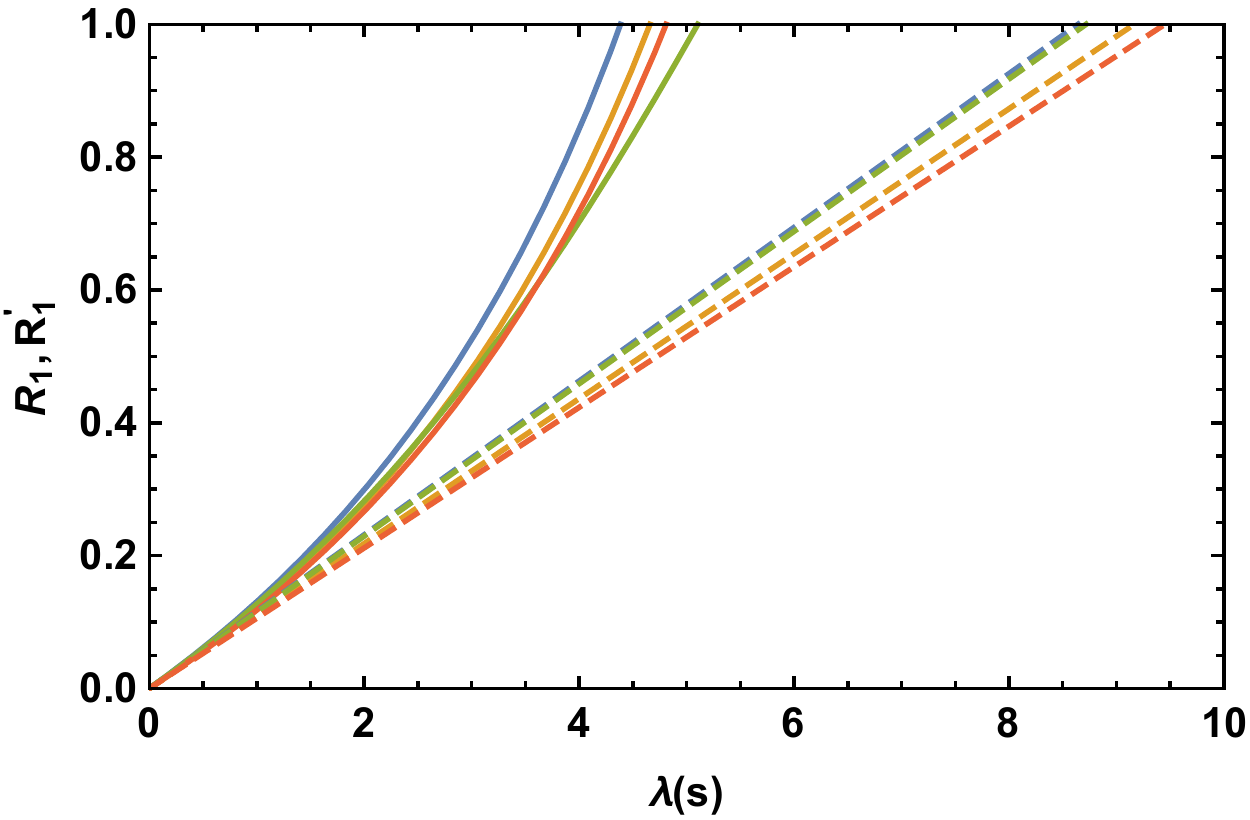}
  \caption{The validity of perturbation theory in the SM. The solid curves and dashed lines correspond to $R_1$ and $R_1^{\prime}$ respectively, which are defined by~\eqref{eq:R1}. The green, blue, orange, and red curves/lines correspond to $a_{0,1}^{Q=0}$, $a_{0,2}^{Q=0}$, $a_{0,3}^{Q=0}$, and $a_{0,4}^{Q=0}$ respectively.}
  \label{fig:SMper}
\end{figure}

The unitarity constraint $| a_{0,1}^{Q=0} - i / 2| \leq 1 /2$ yields the bound $\lambda(s) \leq 15.5$. It's interesting to note that a numerically similar bound is obtained when only the 1PI diagrams are included in the analysis, $|(a_{0,1}^{Q=0})_{\text{1PI}} - i / 2| \leq 1 /2 \to \lambda(s) \leq 15.1$. While unitarity can in principle hold up to $\lambda(s) \approx 15$, perturbativity does not hold for such large couplings. To see this consider the following quantities,
\begin{equation}
\label{eq:R1}
R_1 \equiv \frac{|a_0^{(1)}|}{|a_0^{(0)} + a_0^{(1)}|}, \quad R_1^{\prime} \equiv \frac{|a_0^{(1)}|}{|a_0^{(0)}|},
\end{equation}
where $a_0^{(0)}$ and $a_0^{(1)}$ are the tree level and one-loop contributions to the eigenvalue $a_0$ respectively. Minimal requirements for perturbation theory to hold are that the next-to-leading order contribution to an amplitude should be smaller in magnitude than  both the leading order contribution and the total amplitude. Thus, perturbativity is violated when $R_1 = 1$ or $R_1^{\prime} = 1$. Based on the criterion, perturbativity is violated when $\lambda(s) \sim 4.3 - 5.1$, as can be seen from Figure~\ref{fig:SMper}. The solid curves and dashed lines in Fig.~\ref{fig:SMper} correspond to $R_1$ and $R_1^{\prime}$ respectively. The eigenvalues entering into $R_1$ and $R_1^{\prime}$ in the green, blue, orange, and red curves in Fig.~\ref{fig:SMper} are $a_{0,1}^{Q=0}$, $a_{0,2}^{Q=0}$, $a_{0,3}^{Q=0}$, and $a_{0,4}^{Q=0}$ respectively. Ref.~\cite{Durand:1992wb} states that the range of $R_1$ for $\lambda(s) = 5$ (in our notation) is $1.08 - 1.31$.\footnote{Note that $\lambda_{\text{this work}} = 2 \lambda_{\text{Ref.~\cite{Durand:1992wb}}}$.} Whereas we find that $R_1 = \{0.97,\, 1.31,\, 1.15,\, 1.08\}$ for $a_{0,1-4}^{Q = 0}$ with $\lambda(s) = 5$.

\subsection{$SO(3)$ Symmetric Limit}
In the SM, the Goldstone boson scattering amplitudes possess an $SO(3)$ symmetry, analogous to the strong isospin symmetry of the pions. We start our analysis of the 2HDM by considering the highly simplified scenario where the Goldstone boson scattering amplitudes in the 2HDM retain the $SO(3)$ symmetry they had in the SM.

In the Higgs basis, it's clear that if the Goldstone boson scattering amplitudes are to have an $SO(3)$ symmetry, at least at high energies, then only $\Lambda_1$, $\Lambda_2$, and $\Lambda_3$ can be non-zero. This choice brings about the alignment limit, and forces $m_H = m_{H^+} = m_A$. In the $\mathbb{Z}_2$-basis, these choices can only be achieved if $\lambda_1 = \lambda_2 = \lambda_3$, and $\lambda_4 = \lambda_5 = 0$. Thus, we have the further simplification, $\Lambda_1 = \Lambda_2 = \Lambda_3$. For definiteness, the potential in this case is,
\begin{equation}
V = m_{12}^2 \left(\Phi_1^{\dagger} \Phi_1 t_{\beta} + \Phi_2^{\dagger} \Phi_2 t_{\beta}^{-1} - \Phi_1^{\dagger} \Phi_2 - \Phi_2^{\dagger} \Phi_1\right) + \tfrac{1}{2} \lambda_1 \left(\Phi_1^{\dagger} \Phi_1 + \Phi_2^{\dagger} \Phi_2 - \tfrac{v^2}{2} \right)^2.
\end{equation}
The masses of the Higgs bosons are,
\begin{equation}
\label{eq:so3mass}
m_h^2 = \lambda_1 v^2,\quad m_A^2 = m_H^2 = m_{H^+}^2 = m_{12}^2 s_{\beta}^{-1} c_{\beta}^{-1}.
\end{equation}
There are other symmetry considerations that lead to the mass spectrum in Eqs.~\eqref{eq:so3mass} as well, such as the Maximally Symmetric 2HDM potential based on $SO(5)$~\cite{Dev:2014yca}. Alternatively, this mass spectrum can also be obtained by demanding the stability of the scalar potential up to the Planck scale~\cite{Das:2015mwa}.

The reason for considering such a simple scenario is that it isolates one of the differences between one-loop scattering amplitudes in the SM and the 2HDM. The main difference between the tree level scattering amplitudes in the SM and the 2HDM is that there are five parameters in the 2HDM versus only one parameter in the SM. This scenario allows us to eliminate that difference. In doing so, we are able to isolate another difference, in the 2HDM the external wavefunction corrections contain terms of the form $\ln m_A^2 / m_h^2$.

Due to this being such a simplified scenario, there are only two unique tree level eigenvalues, $a_{0+}^{00\text{even}} = - 5 \lambda_1 / 16 \pi$, while all the rest are $- \lambda_1 / 16 \pi$. As a result, we will focus on the $\mathbf{a}_0^{000\text{even}}$ block of $\mathbf{a}_0$,
\begin{align}
\label{eq:a000even}
&256 \pi^3 \mathbf{a}_0^{000\text{even}} = \\
&- 16 \pi^2 \lambda_1 \begin{pmatrix}
3 & 2  \\
2  & 3 
\end{pmatrix} 
+ \frac{\lambda_1^2}{8} \begin{pmatrix}
C_1 + 24 c_{\beta}^2 f(x) + 48 s_{\beta}^2 g(x) & C_2 + 8 f(x) + 16 g(x) \\
C_2 + 8 f(x) + 16 g(x) & C_1 + 24 s_{\beta}^2 f(x) + 48 c_{\beta}^2 g(x)
\end{pmatrix} . \nn
\end{align}
The definitions in Eq.~\eqref{eq:a000even} are
\begin{align}
x &= m_A^2 / m_h^2, \\
C_1 &= 445 + 104 i \pi + 6 \sqrt{3} \pi + 3 (2 \sqrt{3} \pi - 9) c_{2\beta}, \nn \\
C_2 &= 270 + 4 \pi (24 i + \sqrt{3}), \nn \\
f(x) &= \frac{1 - 4 x - 2 x \sqrt{1 - 4 x}\, \ln\left(\tfrac{- 1 + \sqrt{1 - 4 x} + 2 x}{2 x}\right)}{- 1 + 4 x}, \nn \\
g(x) &= \frac{[4 x - h(x)][2 x + (1 - x) \ln(x)] - 2 (1 - 3 x) h(x) \ln\left(\tfrac{h(x)}{2 \sqrt{x}}\right)}{2 x^2 [h(x) - 4 x]}, \nn \\
h(x) &= 1 + \sqrt{1 - 4 x}. \nn
\end{align}
The functions $f$ and $g$ have similar limiting behavior,
\begin{equation}
f(1) = g(1) = - 1 + \frac{2 \sqrt{3} \pi}{9}, \quad f(x \gg 1),\, g(x \gg 1) \propto \frac{1}{x}.
\end{equation}
At one-loop, the eigenvalue of interest is
\begin{equation}
\label{eq:5L1}
a_{0+}^{000\text{even}} = - \frac{5 \lambda_1}{16 \pi} + \frac{5 \lambda_1^2}{2048 \pi^3} \left(143 + 40 i \pi +2 \sqrt{3} \pi + 4 f(x) + 8 g(x)\right).
\end{equation}

\begin{figure}
  \centering
\includegraphics[width=0.7\textwidth]{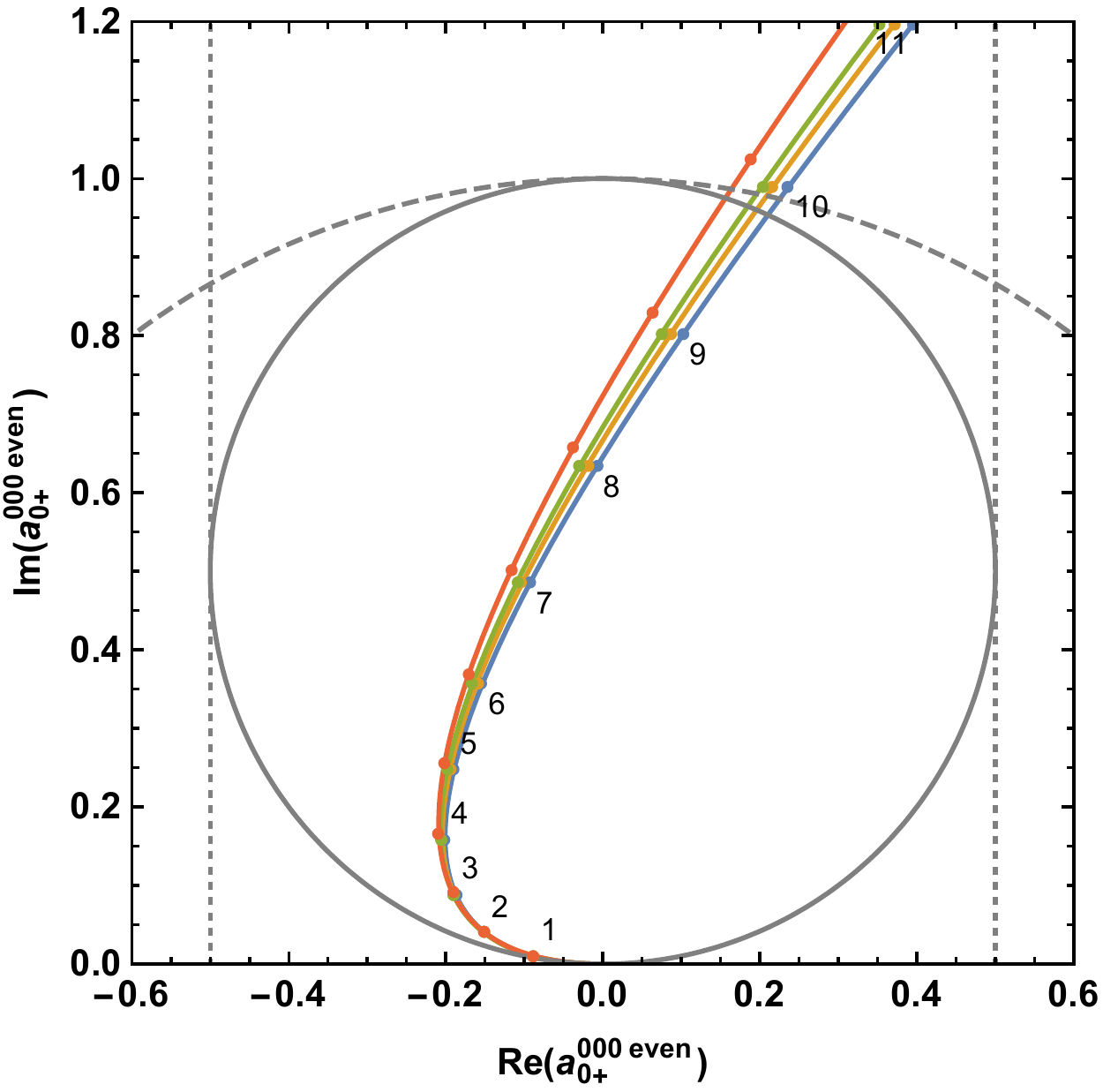}
 \caption{Argand diagram of $a_{0+}^{000\text{even}}$ in the $SO(3)$ symmetric 2HDM as a function of the running coupling $\lambda_1(s)$. The blue and red curves corresponds to $m_h^2 = m_A^2$ and $m_h^2 = 8 m_A^2$ respectively, and are labeled with various values of $\lambda_1(s)$. The orange curve is the limit $s \gg m_A^2 \gg m_h^2$. Finally, the green corresponds to the contribution of the 1PI diagrams alone. The solid circle is the bound $|a_0 - i / 2| \leq 1 / 2$, whereas the dashed arc and the dotted vertical lines represent the bounds $|a_0| \leq 1$ and $|\text{Re}(a_0)| \leq 1 / 2$ respectively.}
  \label{fig:argandSO3}
\end{figure}

The Argand diagram of $a_{0+}^{000\text{even}}$ is shown in Figure~\ref{fig:argandSO3} as a function of the running coupling $\lambda_1(s)$. The solid circle is the bound $|a_0 - i / 2| \leq 1 / 2$, whereas the dashed arc and the dotted vertical lines represent the bounds $|a_0| \leq 1$ and $|\text{Re}(a_0)| \leq 1 / 2$ respectively. The blue curve corresponds to $m_A = m_h$, and is labeled with various values of $\lambda_1(s)$. The orange curve instead corresponds to the limit $s \gg m_A^2 \gg m_h^2$. In practice, this limit amounts to setting $f(x) = g(x) = 0$ in Eq.~\eqref{eq:5L1}, but the heavy Higgses do not decouple completely ($s \gg m_A^2$) as there is an $O(1)$ difference between the orange curve and the SM value, $a_{0+\text{SM}}^{00\text{even}} = - 3 \lambda_1 / 16 \pi$. Figure~\ref{fig:argandSO3} shows that the effect of the real parts of the $\ln m_A^2 / m_h^2$ terms are numerically unimportant, at least in the limit of an $SO(3)$ symmetry. The green curve emphasizes a similar point, as it corresponds to neglecting the external wavefunction corrections completely. This shows that the contribution of the external wavefunction renormalization diagrams are typically small with respect to the 1PI diagrams (tree $+$ one-loop), again at least in the case of an $SO(3)$ symmetry. On the other hand, the red curve corresponds to $m_h^2 = 8 m_A^2$, leading to an imaginary part for $Z_h$ because the decay $h \to A A$ is now allowed. As can be seen from Figure~\ref{fig:argandSO3}, the imaginary part of $Z_h$ is positive. However, even though this curve is further away from the other three curves in the Argand plane, it still doesn't cause a significant change to the bound on the quartic coupling; the orange curve yields $\lambda_1(s) \leq 9.85$, whereas the red curve yields $\lambda_1(s) \leq 9.76$.

\begin{figure}
  \centering
  \includegraphics[width=0.7\textwidth]{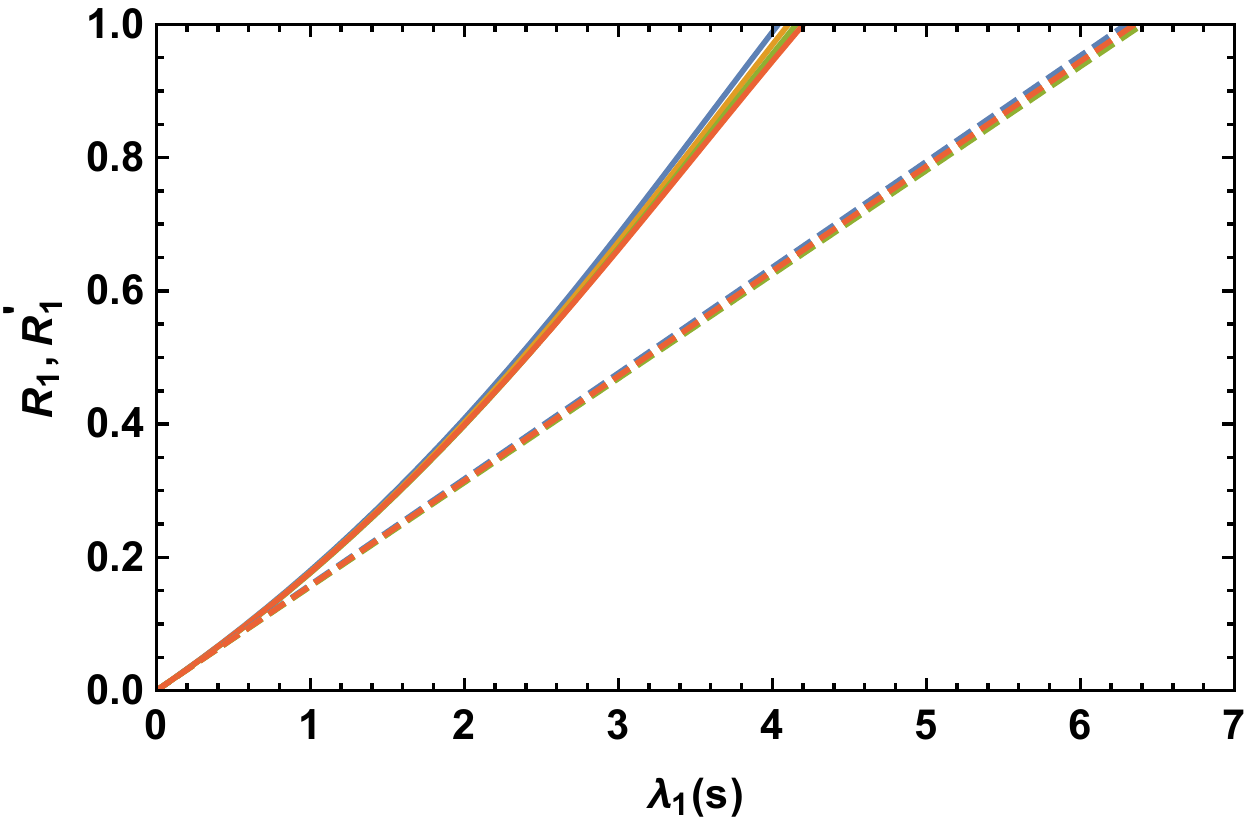}
  \caption{The validity of perturbation theory in $SO(3)$ symmetric limit of the 2HDM. The solid curves and dashed lines correspond to $R_1$ and $R_1^{\prime}$ for the eigenvalue $a_{0+}^{000\text{even}}$ respectively. The blue, orange, green, and red curves have the same parameterizations as they do in Fig.~\ref{fig:argandSO3}.}
  \label{fig:SO3per}
\end{figure}

As we have just shown, unitarity can in principle hold up to $\lambda_1(s) \approx 9.8$ in the $SO(3)$ symmetric limit. However, just as in the SM, perturbativity does not hold for such large couplings. Based on the criterion $R_1 < 1$, perturbativity is violated when $\lambda_1(s) \sim 4.0 - 4.2$, which can be seen from Figure~\ref{fig:SO3per}. Similarly, based on $R_1^{\prime} < 1$, perturbativity is violated when $\lambda_1(s) \sim 6.3 - 6.4$. The solid curves and dashed lines in Fig.~\ref{fig:SO3per} correspond to $R_1$ and $R_1^{\prime}$ for the eigevnalue $a_{0+}^{000\text{even}}$ respectively. The blue, orange, green, and red curves/lines in Fig.~\ref{fig:SO3per} have the same parameterizations as the curves in Fig.~\ref{fig:argandSO3}. 

\subsection{MSSM-like 2HDM}
\label{sec:mssm}
In the MSSM, the Higgs quartic couplings are related to the gauge couplings,
\begin{equation}
\label{eq:Lmssm}
\lambda_1 = \lambda_2 = \frac{g_2^2 + g_1^2}{4},\, \lambda_3 = \frac{g_2^2 - g_1^2}{4},\, \lambda_4 = - \frac{g_2^2}{2},\, \lambda_5 = 0,
\end{equation}
where $g_1$ is the gauge coupling associated with $U(1)_Y$, and again, $g_2$ is the gauge coupling of $SU(2)_L$. The soft $\mathbb{Z}_2$-breaking parameter is given by $m_{12}^2 = m_A^2 s_{\beta} c_{\beta}$. Loop corrections to the MSSM potential are important, as at tree level the MSSM predicts $\text{min}\{m_h,\, m_H\} < M_Z$, which is incompatible with the LHC measurements of a Higgs boson at 125 GeV. Clearly, the quartic coupling of the MSSM satisfy the tree level unitarity bounds, as we have assumed $\lambda_i \gg g_{2,1}$ in everything that proceeded Eq.~\eqref{eq:Lmssm}.

However, by considering a scenario inspired by Eq.~\eqref{eq:Lmssm}, we can get a feel for the impact of the one-loop corrections without having to deal with the full complexity of the 2HDM parameter space. Specifically, we will take $\lambda_1$, $\lambda_3$, $m_A$, and $\tan \beta$ to be free parameters, and enforce at tree level
\begin{equation}
\label{eq:LmssmIn}
\lambda_1 = \lambda_2,\quad \lambda_4 = - (\lambda_1 + \lambda_3),\quad \lambda_5 = 0 .
\end{equation}
It should noted however that the relations in Eqs.~\eqref{eq:Lmssm} are RG-invariant in the MSSM, whereas the analogous relations, Eqs.~\eqref{eq:LmssmIn}, are not RG-invariant if supersymmetry is not imposed on the 2HDM~\cite{Branco:2011iw}. In this analysis, we impose the relations in Eqs.~\eqref{eq:LmssmIn} at $\mu = \sqrt{s}$.

\begin{figure}
  \centering
\includegraphics[width=0.7\textwidth]{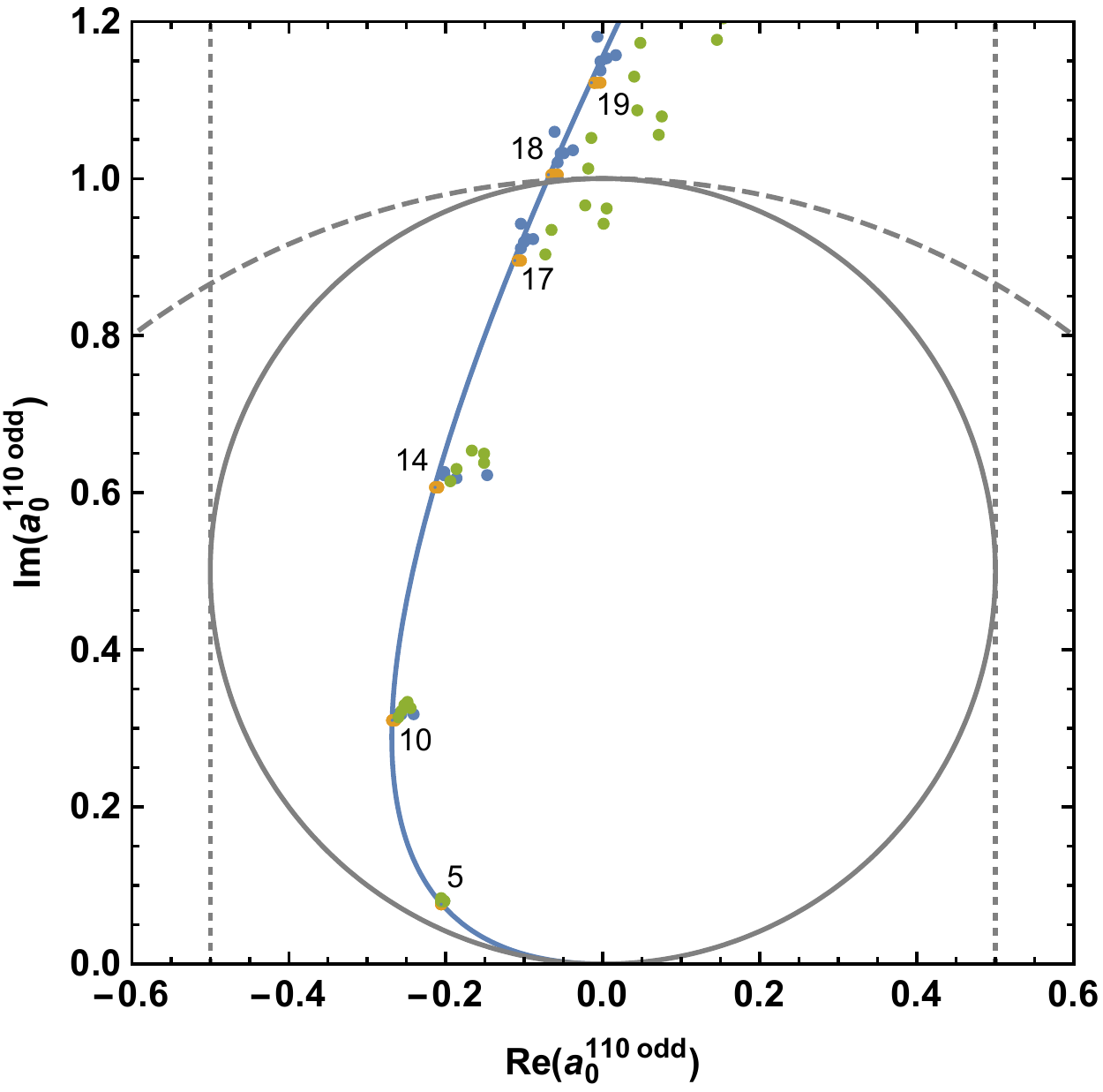}
 \caption{Argand diagram of $a_{0}^{110\text{odd}}$ in the MSSM-like 2HDM as a function of the running coupling $\lambda_1(s)$ with $\lambda_3(s) = 9 \lambda_1(s) / 10$. The blue curve corresponds to the contribution of the 1PI diagrams alone, and is labeled with various values of $\lambda_1(s)$. The blue, orange, and green points are the complete one-loop calculations for $a_{0}^{110\text{odd}}$ at each of the labeled values of $\lambda_1(s)$. The choices for $m_A$ and $\tan \beta$ for each point are given in the text. The unitarity bounds are the same as those in Fig.~\ref{fig:argandSO3}.}
  \label{fig:MSSM3}
\end{figure}

As was the case for the $SO(3)$ symmetric 2HDM, because of the relative simplicity of the MSSM-like 2HDM, there are more degenerate tree level eigenvalues of $\mathbf{a}_0$ than there are in the general case of a $CP$-conserving 2HDM with a softly broken $\mathbb{Z}_2$ symmetry. Due to this fact, we will first focus on $a_{0}^{10\text{odd}} = - (\lambda_1 + 2 \lambda_3)$, which is unique at tree level. Neglecting the external wavefunction corrections, the one-loop eigenvalue in the MSSM-like 2HDM is
\begin{equation}
256 \pi^3 a_{0}^{110\text{odd}} = - 16 \pi^2 (\lambda_1 + 2 \lambda_3) - (2 \lambda_1 - 11 \lambda_3)(2 \lambda_1 + \lambda_3) + i \pi (\lambda_1 + 2 \lambda_3)^2.
\end{equation}
The full expression for $a_{0}^{110\text{odd}}$, which valid for the more general case of the $CP$-conserving 2HDM with a softly broken $\mathbb{Z}_2$ symmetry, is given in Eq.~\eqref{eq:110}. A typical result of our investigation in shown in Figure~\ref{fig:MSSM3}, which is an Argand diagram of the eigenvalue $a_{0}^{100\text{odd}}$ for $\lambda_3(s) = 9 \lambda_1(s) / 10$. The various unitarity bounds in gray are the same as they were in Fig.~\ref{fig:argandSO3}. The blue curve corresponds to neglecting the external wavefunction corrections. Several values of $\lambda_1(s)$ are labeled along the curve. For each labeled value of $\lambda_1(s)$ we plotted the complete one-loop prediction for $a_{0}^{100\text{odd}}$ for three choices of $m_A$. The blue, orange, and green points respectively correspond to $m_A = \{1\,\text{TeV},\, 14\,\text{TeV},\, 400\,\text{GeV}\}$. Five choices for $\tan \beta$ are plotted for each value of $m_A$, $\tan \beta = \{1.1,\, 1.6,\, 2.5,\, 5.0,\, 60\}$. The scalar integrals entering into the wavefunction renormalization terms are computed using \texttt{LoopTools-2.12}~\cite{Hahn:1998yk}. It's clear from Fig.~\ref{fig:MSSM3} that the approximation of neglecting the external wavefunction corrections becomes worse as the theory becomes more strongly coupled. However, the overall change in the bound extracted on $\lambda_1(s)$ does not change much despite this modest spread in predictions for $a_{0}^{100\text{odd}}$ near the unitarity circle, as can be seen by inspecting Fig.~\ref{fig:MSSM3}.

\begin{figure}
  \centering
 \subfloat{\includegraphics[width=0.48\textwidth]{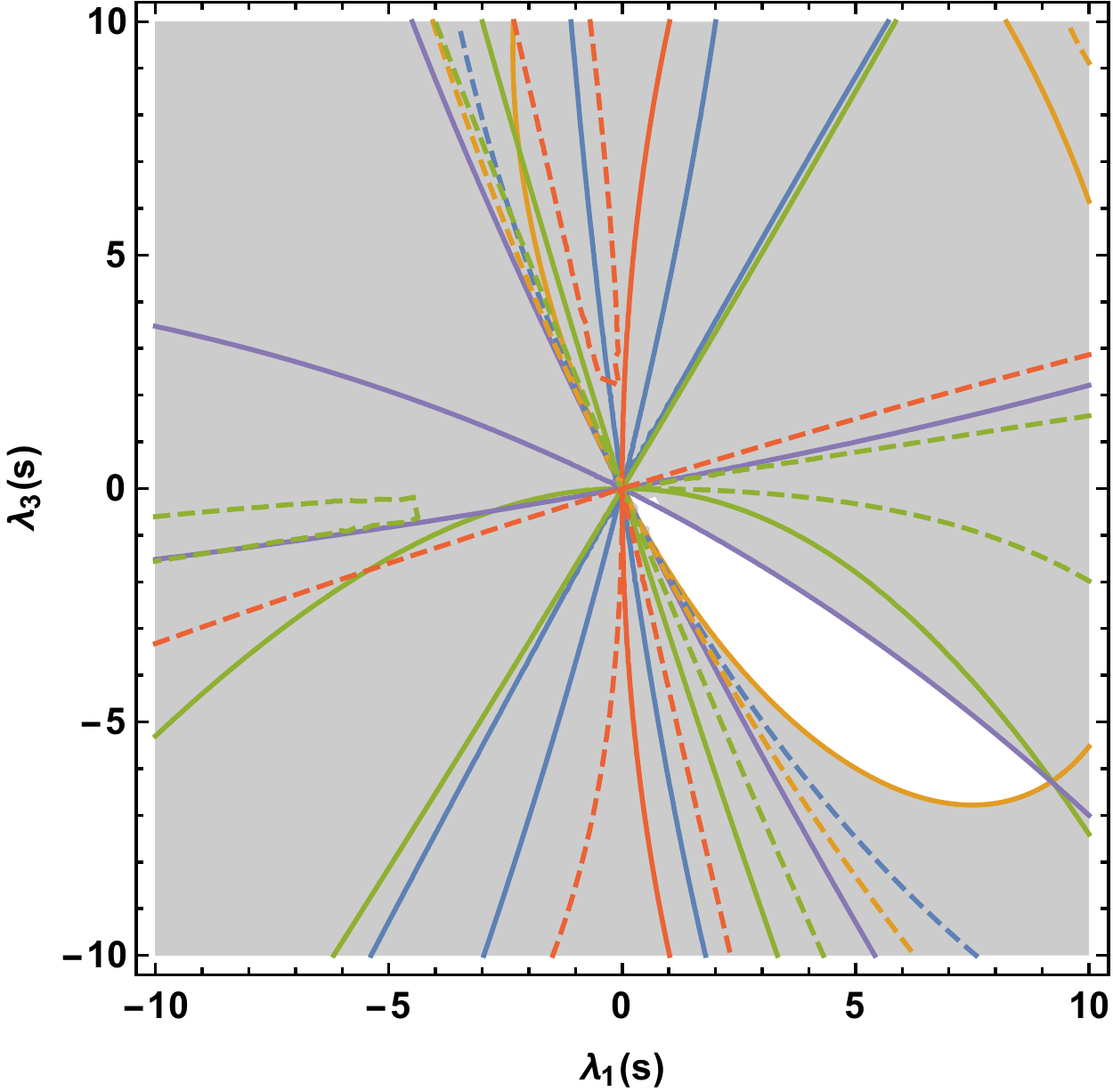}}\,
\subfloat{\includegraphics[width=0.48\textwidth]{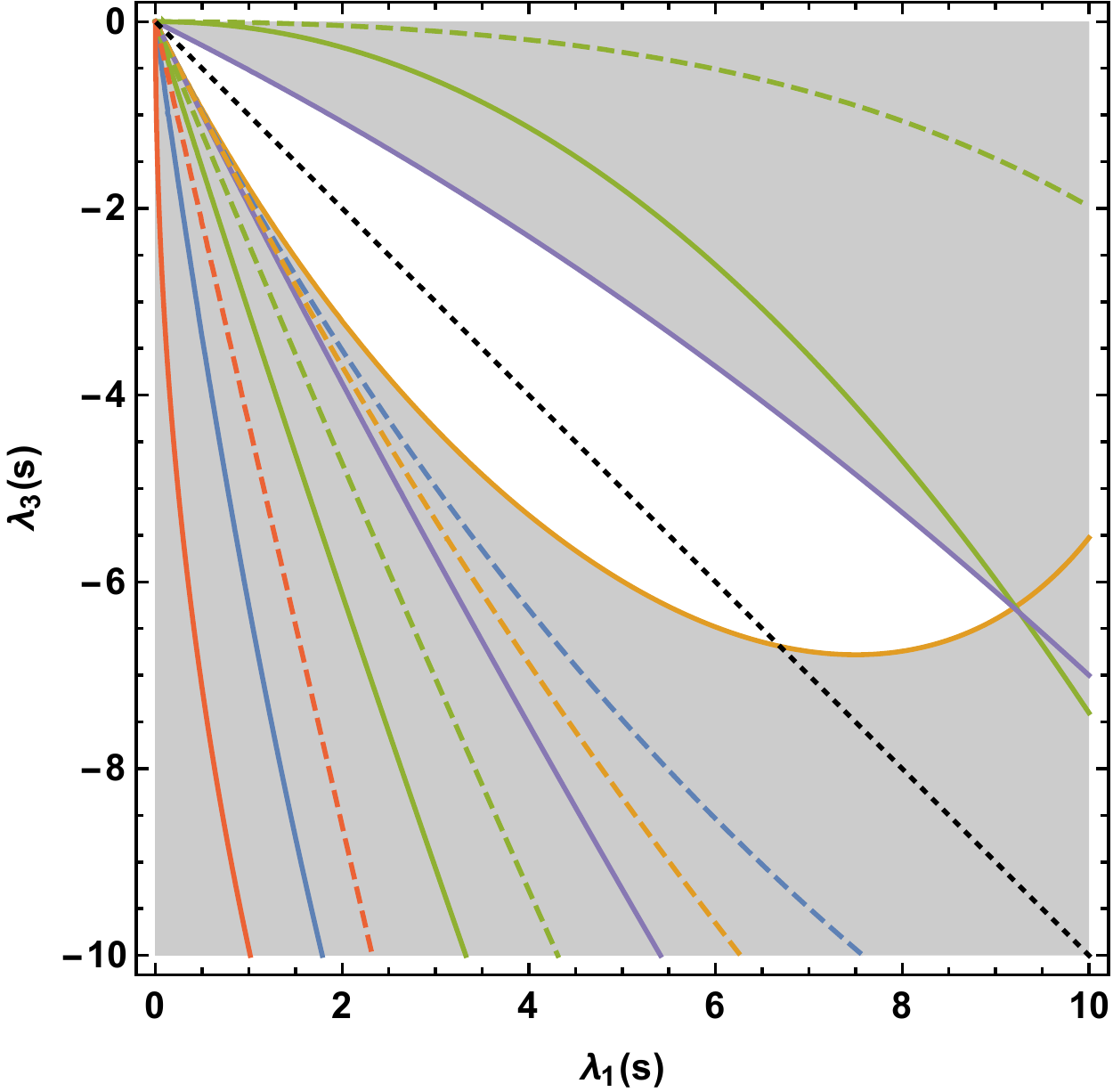}}
 \caption{(left) Limits on $\lambda_1(s)$ and $\lambda_3(s)$ in the MSSM-like 2HDM due to unitarity, $| a_{0} - i / 2| \leq 1 /2$. See the text for which eigenvalue corresponds to which curve. (right) Zoomed in version of the left panel, which also includes the tree level stability bound, $\lambda_3 > - \lambda_1$, given by the dotted,  black line.}
  \label{fig:MSSM4a}
\end{figure}

Stronger limits can be obtained by combining the bounds for multiple channels. Fig.~\ref{fig:MSSM4a} shows the upper limits on $\lambda_1(s)$ and $\lambda_3(s)$ obtained by combining the constraints in the nine unique channels, neglecting the external wavefunction corrections, from the unitarity bound $| a_{0} - i / 2| \leq 1 /2$. The solid blue, orange, green, red, and purple curves correspond to $a_{0+}^{000\text{even}}$, $a_{0-}^{000\text{even}}$, $a_{0-}^{001\text{even}}$, $a_{0}^{011\text{even}}$, and $a_{0}^{110\text{odd}}$ respectively. While the dashed blue, orange, green, and red curves correspond to $a_{0+}^{001\text{even}}$, $a_{0}^{000\text{odd}}$, $a_{0}^{001\text{odd}}$, and $a_{0}^{011\text{odd}}$ respectively. Note that the subscript $+$ or $-$ has been dropped in some cases because those eigenvalues become degenerate when the wavefunction corrections are omitted. Lastly, the gray parameter space is ruled out due to at least one of the eigenvalue exceeding the bound $| a_{0} - i / 2| \leq 1 /2$. Much of the parameter space in Fig.~\ref{fig:MSSM4a} that is viable with respect to unitarity can be eliminated by enforcing the tree level stability bounds, which in the MSSM-like 2HDM take the form $\lambda_1 > 0$, $\lambda_3 > - \lambda_1$. The black, dotted line in the right panel of Fig.~\ref{fig:MSSM4a} indicates the tree level stability bound, and the parameter space to the left of this line is ruled out this bound.

Figure~\ref{fig:MSSM4a} requires $\lambda_3$ to be negative at the high scale, $\mu = \sqrt{s}$. Curiously, in the MSSM $\lambda_3 = (g_2^2 - g_1^2) / 4$ is positive at the low scale, say $\mu = M_Z$. However, one should not rush to conclusions as there are important differences between the MSSM-like 2HDM and the actual MSSM, as noted at the beginning of this Subsection,~\ref{sec:mssm}.

\begin{figure}
  \centering
\subfloat{\includegraphics[width=0.48\textwidth]{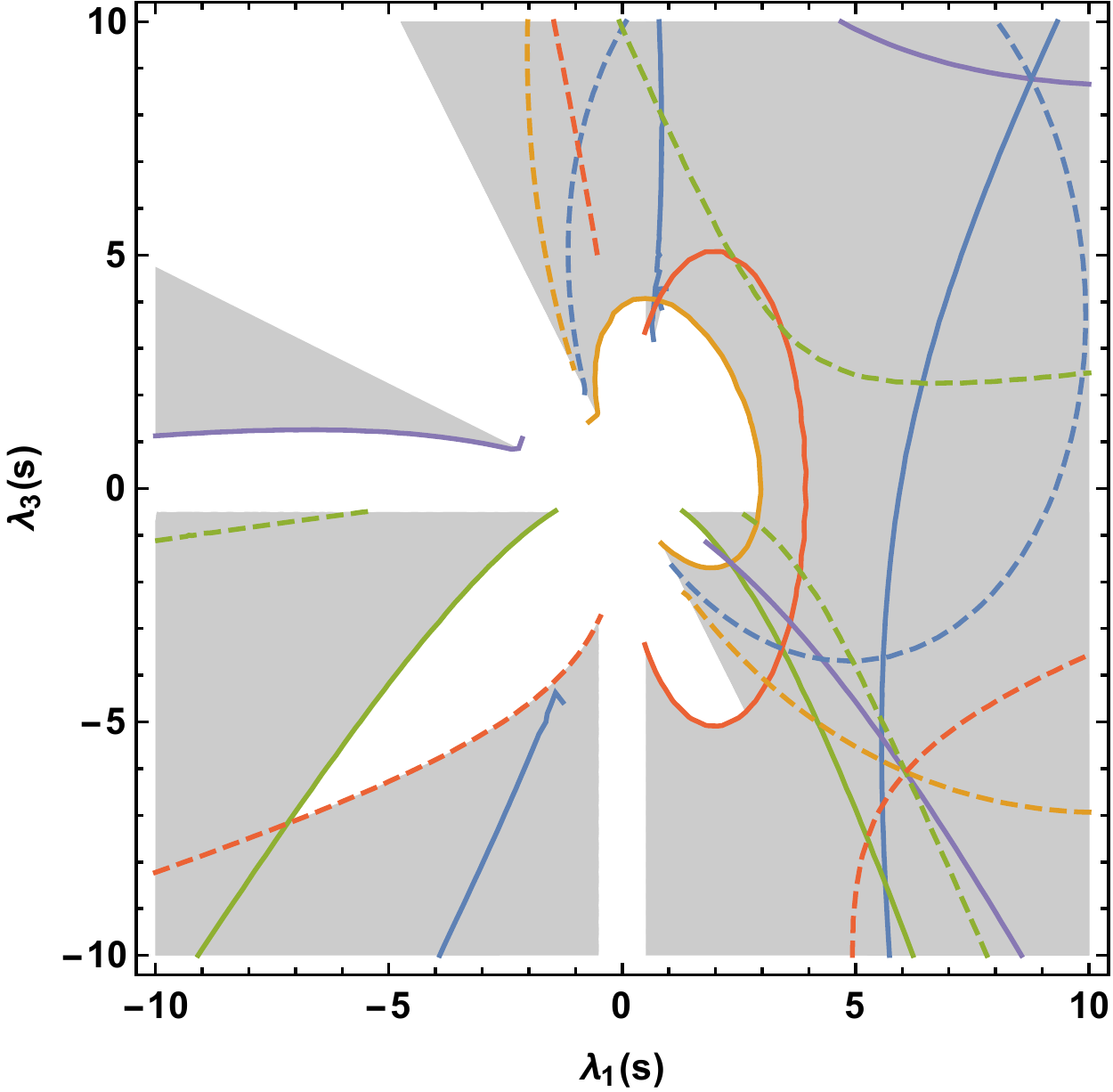}}\,
\subfloat{\includegraphics[width=0.48\textwidth]{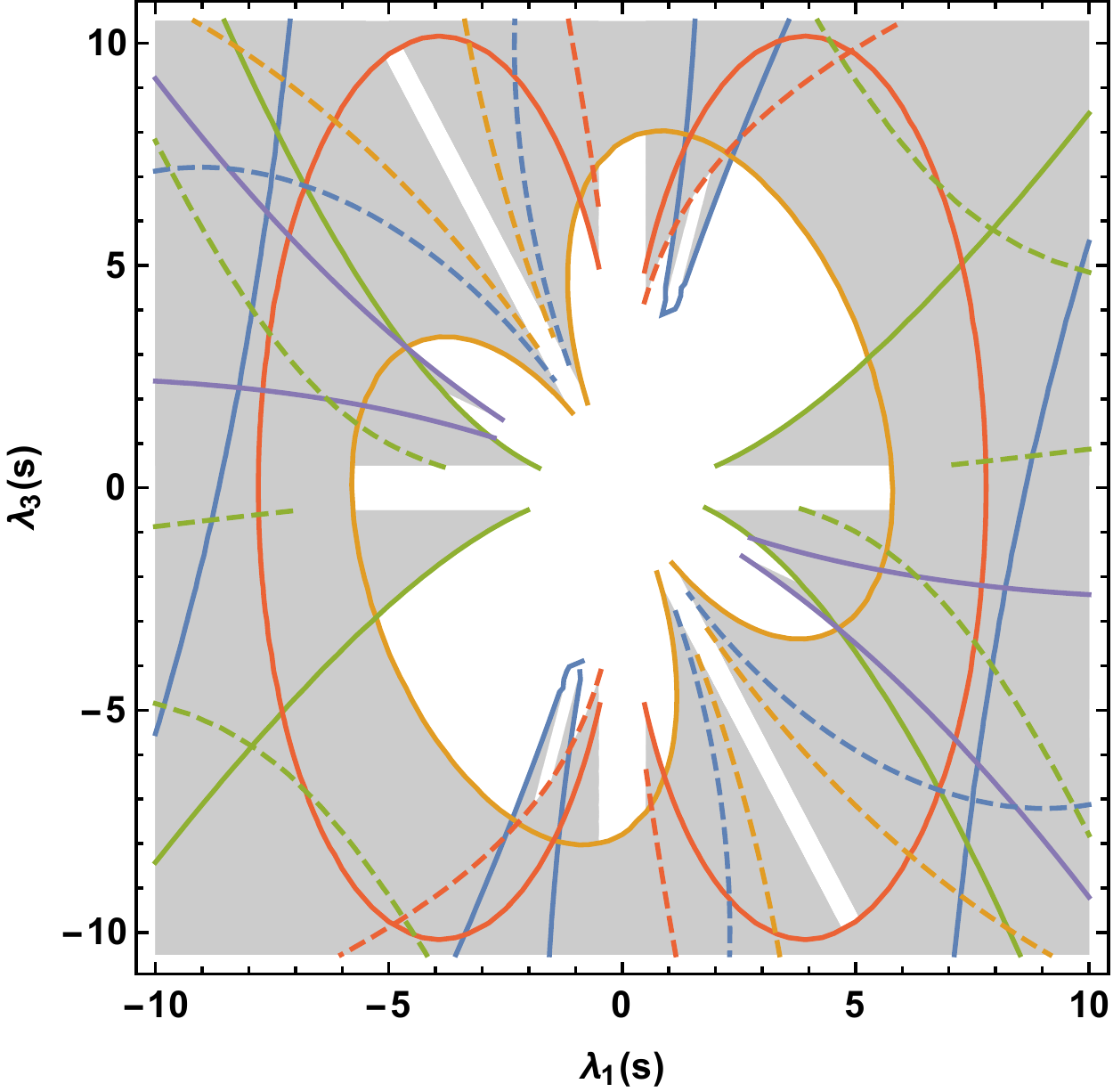}}
 \caption{Limits on $\lambda_1(s)$ and $\lambda_3(s)$ in the MSSM-like 2HDM due to perturbativity from requiring (left) $R_1 < 1$ (right) $R_1^{\prime} < 1$. The cut $|a_0^{(0)}| > 0.01$ is imposed to prevent the tree level amplitudes from accidentally becoming small. See the text more details on this cut, and for which eigenvalue corresponds to which curve.}
  \label{fig:MSSM4b}
\end{figure}

As was the case for the SM and the $SO(3)$ symmetric limit of the 2HDM, we also consider the limits obtained from perturbativity. Figure~\ref{fig:MSSM4b} shows the limits on $\lambda_1(s)$ and $\lambda_3(s)$ in the MSSM-like 2HDM due to perturbativity from requiring $R_1 < 1$ in the left panel, and $R_1^{\prime} < 1$ in the right panel. The various solid and dashed curves correspond to the same eigenvalues as they did in Fig.~\ref{fig:MSSM4a}. As was the case for Fig.~\ref{fig:MSSM4a}, the gray parameter space is ruled out.  Unlike the cases of the SM and the $SO(3)$ symmetric limit of the 2HDM, there are two unique quartic couplings in the MSSM-like 2HDM. This can lead to accidental cancellations in the tree level partial-wave amplitudes, which may fail the perturbativity tests $R_1^{(\prime)}$ even for reasonable values of $\lambda_1$ and $\lambda_3$. To prevent this from happening, we imposed a cut, $|a_0^{(0)}| > a_{0\text{cut}}^{(0)}$, to prevent the tree level amplitudes from accidentally becoming small. We choose $a_{0\text{cut}}^{(0)} = 0.01$, as this roughly corresponds to $|\lambda_i| > 1 /2$ assuming $a_0^{(0)} \sim \lambda_i / (16 \pi)$.

Combining the bounds from Figs.~\ref{fig:MSSM4a} and~\ref{fig:MSSM4b}, the limits on the quartic couplings are $|\lambda_{1,3}(s)| \lessapprox 4$, at least for the regions of parameter space that satisfy the tree level stability bounds. In the MSSM-like 2HDM, the neglect of the external wavefunction corrections is justified \textit{a posteriori} by comparing Fig.~\ref{fig:MSSM3} against Figs.~\ref{fig:MSSM4a} and~\ref{fig:MSSM4b}. Interestingly, both unitarity and perturbativity dominate the bounds on $\lambda_{1,3}(s)$ in certain regions of parameter space, whereas perturbativity was always the more dominant constraint in the SM and $SO(3)$ symmetric limit of the 2HDM.

\section{Conclusions}
\label{sec:con}
In this work, we computed all of the one-loop corrections that are enhanced in the limit $s \gg |\lambda_i | v^2 \gg M_{W}^2$, $s \gg m_{12}^2$ to all the $2 \to 2$ longitudinal vector boson and Higgs boson scattering amplitudes in the $CP$-conserving two-Higgs doublet model with a softly broken $\mathbb{Z}_2$ symmetry. We found that the external wavefunction corrections are generally numerically subdominant with respect to the 1PI one-loop corrections, and that they can often be neglected to a good approximation. In the two simplified scenarios we studied, it was shown that combining perturbativity and unitarity places bounds on the magnitude of the quartic couplings of $|\lambda_i(s)| \lessapprox 4$. It would be interesting to compute the tree level $2 \to 4$ scattering amplitudes in the 2HDM, which should be the leading contribution to the $2 \to n$ partial-wave amplitudes. Then the equality Eq.~\eqref{eq:SmatboundA} could be used to bound the quartic couplings rather than the inequality~\eqref{eq:Smatbound}.

\begin{acknowledgments}
We thank Debtosh Chowdhury, Otto Eberhardt, and Howard Haber for helpful discussions. This work was supported in part by the U.S. Department of Energy under grant no.~DE-SC0009919 (BG), the MIUR-FIRB under grant no.~RBFR12H1MW (CM), and the Thailand Research Fund under the contract no.~TRG5880061 (PU). BG and PU also acknowledge the hospitality of the Munich Institute for Astro- and Particle Physics (MIAPP) of the DFG cluster of excellence ``Origin and Structure of the Universe" where part of this work was completed.
\end{acknowledgments}

\appendix

\section{Results for Self-Energies}
\label{sec:SE}
Our results for the self-energies in the 2HDM, which enter into the external wavefunction renormalization of the scattering amplitudes given in Appendices~\ref{sec:RSA} and~\ref{sec:RSA2}, as well as the threshold corrections to the parameters of the 2HDM, are given in~\ref{sec:SEse}. The cubic and quartic couplings that enter into the self-energies are given in~\ref{sec:SE3} and~\ref{sec:SE4} respectively. The particles that interact through a given coupling are listed to the left of the formula for the coupling, and the corresponding Feynman rules for these three- and four-point interactions are $i m_n$ and $i g_n$ respectively. The self-energies are given in terms of the finite parts of the usual one-point and two-point scalar integrals, in $D = 4 - 2 \epsilon$ dimensions,
\begin{equation}
A_0[m^2] = \frac{m^2}{\epsilon} + \bar{A}_0[m^2], \quad B_0[p^2, m_1^2, m_2^2] = \frac{1}{\epsilon} + \bar{B}_0[p^2, m_1^2, m_2^2].
\end{equation}
Explicitly, the finite pieces of the scalar integrals are,
\begin{align}
\bar{A}_0[m^2] &= m^2 \left(- \ln \frac{m^2}{\mu^2} + 1\right), \\
\bar{B}_0[p^2, m_1^2, m_2^2] &= - \int_0^1 \! dx\, \ln \frac{p^2 x^2 - x\left(p^2 - m_1^2 + m_2^2\right) + m_2^2 - i 0^+}{\mu^2}. \nn
\end{align}

\subsection{Self-Energies}
\label{sec:SEse}
\begin{align}
&\widetilde{\Pi}_{w^+w^-}(p^2) = - \tfrac{1}{16\pi^2}(m_1^2 (\bar{B}_0[p^2, 0, m_h^2] - \bar{B}_0[0, 0, m_h^2]) \\
&+ m_3^2 (\bar{B}_0[p^2, 0, m_H^2] - \bar{B}_0[0, 0, m_H^2]) + m_5^2 (\bar{B}_0[p^2, m_{H^+}^2, m_h^2] - \bar{B}_0[0, m_{H^+}^2, m_h^2]) \nn \\
&+ m_7^2 (\bar{B}_0[p^2, m_{H^+}^2, m_H^2] - \bar{B}_0[0, m_{H^+}^2, m_H^2]) + |m_9|^2 (\bar{B}_0[p^2, m_{H^+}^2, m_A^2] - \bar{B}_0[0, m_{H^+}^2, m_A^2]))  \nn
\end{align}
\begin{align}
&\widetilde{\Pi}_{zz}(p^2) = - \tfrac{1}{16\pi^2}(m_2^2 (\bar{B}_0[p^2, 0, m_h^2] -\bar{B}_0[0, 0, m_h^2]) \\
&+ m_4^2 (\bar{B}_0[p^2, 0, m_H^2] - \bar{B}_0[0, 0, m_H^2]) + m_6^2 (\bar{B}_0[p^2, m_A^2, m_h^2] - \bar{B}_0[0, m_A^2, m_h^2])  \nn \\
&+ m_8^2 (\bar{B}_0[p^2, m_A^2, m_H^2] - \bar{B}_0[0, m_A^2, m_H^2])) \nn
\end{align}
\begin{align}
&\widetilde{\Pi}_{w^+H^-}(p^2) = - \tfrac{1}{16\pi^2}(m_1 m_5 (\bar{B}_0[p^2, 0, m_h^2] - \bar{B}_0[0, 0, m_h^2])  \\
&+ m_3 m_7 (\bar{B}_0[p^2, 0, m_H^2] - \bar{B}_0[0, 0, m_H^2]) + m_5 m_{10} (\bar{B}_0[p^2, m_{H^+}^2, m_h^2] - \bar{B}_0[0, m_{H^+}^2, m_h^2]) \nn \\
&+ m_7 m_{12} (\bar{B}_0[p^2, m_{H^+}^2, m_H^2] - \bar{B}_0[0, m_{H^+}^2, m_H^2])) \nn
\end{align}
\begin{align}
&\widetilde{\Pi}_{zA}(p^2) = - \tfrac{1}{16\pi^2}(m_2 m_6 (\bar{B}_0[p^2, 0, m_h^2] - \bar{B}_0[0, 0, m_h^2]) \\
&+ m_4 m_8 (\bar{B}_0[p^2, 0, m_H^2] - \bar{B}_0[0, 0, m_H^2])  + m_6 m_{11} (\bar{B}_0[p^2, m_A^2, m_h^2] - \bar{B}_0[0, m_A^2, m_h^2]) \nn \\
&+ m_8 m_{13} (\bar{B}_0[p^2, m_A^2, m_H^2] - \bar{B}_0[0, m_A^2, m_H^2]))  \nn
\end{align}
\begin{align}
\Pi_{zz}(0) &= \Pi_{w^+w^-}(0) = \tfrac{1}{32\pi^2}(g_1 \bar{A}_0[m_h^2] + g_2 \bar{A}_0[m_H^2] + 2 g_3 \bar{A}_0[m_{H^+}^2] + g_4 \bar{A}_0[m_A^2]) \\
&- \tfrac{1}{16\pi^2}(m_2^2 \bar{B}_0[0, 0, m_h^2] + m_4^2 \bar{B}_0[0, 0, m_H^2] + m_6^2 \bar{B}_0[0, m_A^2, m_h^2] + m_8^2 \bar{B}_0[0, m_A^2, m_H^2]) \nn
\end{align}
\begin{align}
\Pi_{zA}(0) &= \Pi_{w^+H^-}(0) = \tfrac{1}{32\pi^2}(g_5 \bar{A}_0[m_h^2] + g_6 \bar{A}_0[m_H^2] + 2 g_7 \bar{A}_0[m_{H^+}^2] + g_8 \bar{A}_0[m_A^2]) \\
&- \tfrac{1}{16\pi^2}(m_2 m_6 \bar{B}_0[0, 0, m_h^2] + m_4 m_8 \bar{B}_0[0, 0, m_H^2]  + m_6 m_{11} \bar{B}_0[0, m_A^2, m_h^2] \nn \\
&+ m_8 m_{13} \bar{B}_0[0, m_A^2, m_H^2])  \nn
\end{align}
\begin{align}
\widetilde{\Pi}_{hh}(p^2) &= \tfrac{1}{32\pi^2}(2 g_{9} \bar{A}_0[m_{H^+}^2] +  g_{10} \bar{A}_0[m_A^2] + g_{15} \bar{A}_0[m_h^2] + g_{17} \bar{A}_0[m_H^2]) \\
&- \tfrac{1}{32\pi^2}((2 m_1^2 + m_2^2) \bar{B}_0[p^2, 0, 0] + 4 m_5^2 \bar{B}_0[p^2, 0, m_{H^+}^2] + 2 m_6^2 \bar{B}_0[p^2, 0, m_A^2] \nn \\
&+ 2 m_{10}^2 \bar{B}_0[p^2, m_{H^+}^2, m_{H^+}^2] + m_{11}^2 \bar{B}_0[p^2, m_A^2, m_A^2] +m_{14}^2 \bar{B}_0[p^2, m_h^2, m_h^2]  \nn \\
&+ 2 m_{15}^2 \bar{B}_0[p^2, m_H^2, m_h^2] + m_{16}^2 \bar{B}_0[p^2, m_H^2, m_H^2]) - s_{\beta-\alpha}^2 \Pi_{zz}(0) \nn \\
&- 2 s_{\beta-\alpha} c_{\beta-\alpha} \Pi_{zA}(0) + (Z_w - 1) (m_h^2 + m_{Z_w}^2 c_{\beta-\alpha}^2) \nn
\end{align}
\begin{align}
\widetilde{\Pi}_{HH}(p^2) &= \tfrac{1}{32\pi^2}(2 g_{11} \bar{A}_0[m_{H^+}^2] +  g_{12} \bar{A}_0[m_A^2] + g_{17} \bar{A}_0[m_h^2] + g_{19} \bar{A}_0[m_H^2]) \\
&- \tfrac{1}{32\pi^2}((2 m_3^2 + m_4^2) \bar{B}_0[p^2, 0, 0] + 4 m_7^2 \bar{B}_0[p^2, 0, m_{H^+}^2] + 2 m_8^2 \bar{B}_0[p^2, 0, m_A^2] \nn \\
&+ 2 m_{12}^2 \bar{B}_0[p^2, m_{H^+}^2, m_{H^+}^2] + m_{13}^2 \bar{B}_0[p^2, m_A^2, m_A^2] +m_{15}^2 \bar{B}_0[p^2, m_h^2, m_h^2]  \nn \\
&+ 2 m_{16}^2 \bar{B}_0[p^2, m_H^2, m_h^2] + m_{17}^2 \bar{B}_0[p^2, m_H^2, m_H^2]) - c_{\beta-\alpha}^2 \Pi_{zz}(0)  \nn \\
&+ 2 s_{\beta-\alpha} c_{\beta-\alpha} \Pi_{zA}(0) + (Z_w - 1) (m_H^2 + m_{Z_w}^2 s_{\beta-\alpha}^2) \nn
\end{align}
\begin{align}
\widetilde{\Pi}_{hH}(p^2) &= \tfrac{1}{32\pi^2}(2 g_{13} \bar{A}_0[m_{H^+}^2] +  g_{14} \bar{A}_0[m_A^2] + g_{16} \bar{A}_0[m_h^2] + g_{18} \bar{A}_0[m_H^2]) \\
&- \tfrac{1}{32\pi^2}((2 m_1 m_3 + m_2 m_4) \bar{B}_0[p^2, 0, 0] + 4 m_5 m_7 \bar{B}_0[p^2, 0, m_{H^+}^2] \nn \\
&+ 2 m_6 m_8 \bar{B}_0[p^2, 0, m_A^2] + 2 m_{10} m_{12} \bar{B}_0[p^2, m_{H^+}^2, m_{H^+}^2] + m_{11} m_{13} \bar{B}_0[p^2, m_A^2, m_A^2] \nn \\
&+ m_{14} m_{15} \bar{B}_0[p^2, m_h^2, m_h^2] + 2 m_{15} m_{16} \bar{B}_0[p^2, m_H^2, m_h^2] + m_{16} m_{17} \bar{B}_0[p^2, m_H^2, m_H^2]) \nn \\
&- s_{\beta-\alpha} c_{\beta-\alpha} \Pi_{zz}(0) - (c_{\beta-\alpha}^2 - s_{\beta-\alpha}^2) \Pi_{zA}(0) - (Z_w - 1) m_{Z_w}^2 s_{\beta-\alpha} c_{\beta-\alpha} \nn 
\end{align}
\begin{align}
\Pi_{H^+H^-}(p^2) &= \tfrac{1}{32\pi^2}(g_{9} \bar{A}_0[m_h^2] + g_{11} \bar{A}_0[m_H^2] + 2 g_{20} \bar{A}_0[m_{H^+}^2] + g_{21} \bar{A}_0[m_A^2]) \\
&- \tfrac{1}{16\pi^2}(m_5^2 \bar{B}_0[p^2, 0, m_h^2] + m_7^2 \bar{B}_0[p^2, 0, m_H^2] + |m_9|^2 \bar{B}_0[p^2, 0, m_A^2] \nn \\
&+ m_{10}^2 \bar{B}_0[p^2, m_{H^+}^2, m_h^2] + m_{12}^2 \bar{B}_0[p^2, m_{H^+}^2, m_H^2] ) + (Z_w - 1) (m_{H^+}^2 + m_{Z_w}^2) \nn
\end{align}
\begin{align}
\Pi_{AA}(p^2) &= \tfrac{1}{32\pi^2}(g_{10} \bar{A}_0[m_h^2] + g_{12} \bar{A}_0[m_H^2] + 2 g_{21} \bar{A}_0[m_{H^+}^2] + g_{22} \bar{A}_0[m_A^2]) \\
&- \tfrac{1}{16\pi^2}(m_6^2 \bar{B}_0[p^2, 0, m_h^2] + m_8^2 \bar{B}_0[p^2, 0, m_H^2] + 2 |m_9|^2 \bar{B}_0[p^2, 0, m_{H^+}^2] \nn \\
&+ m_{11}^2 \bar{B}_0[p^2, m_A^2, m_h^2] + m_{13}^2 \bar{B}_0[p^2, m_A^2, m_H^2] ) + (Z_w - 1) (m_A^2 + m_{Z_w}^2) \nn
\end{align}

\subsection{Cubic Couplings}
\label{sec:SE3}
\begin{equation}
h w^+ w^-: \quad m_1 v = - m_h^2 s_{\beta-\alpha}
\end{equation}
\begin{equation}
h z z: \quad m_2 = m_1
\end{equation}
\begin{equation}
H w^+ w^-: \quad m_3 v = - m_H^2 c_{\beta-\alpha}
\end{equation}
\begin{equation}
H z z: \quad m_4 = m_3
\end{equation}
\begin{equation}
h w^+ H^-: \quad m_5 v = - (m_h^2 - m_{H^+}^2) c_{\beta-\alpha}
\end{equation}
\begin{equation}
h z A: \quad m_6 v = - (m_h^2 - m_A^2) c_{\beta-\alpha}
\end{equation}
\begin{equation}
H w^+ H^-: \quad m_7 v = (m_H^2 - m_{H^+}^2) s_{\beta-\alpha}
\end{equation}
\begin{equation}
H z A: \quad m_8 v = (m_H^2 - m_A^2) s_{\beta-\alpha}
\end{equation}
\begin{equation}
A w^+ H^-: \quad m_9 v = - i (m_A^2 - m_{H^+}^2)
\end{equation}
\begin{equation}
h H^+ H^-: \quad m_{10} v = s_{\beta}^{-1} c_{\beta}^{-1} (m_{12}^2 c_{\beta+\alpha} s_{\beta}^{-1} c_{\beta}^{-1} - m_h^2 c_{2\beta} c_{\beta-\alpha}) - (m_h^2 +2 m_{H^+}^2) s_{\beta-\alpha}
\end{equation}
\begin{equation}
h A A: \quad m_{11} v = s_{\beta}^{-1} c_{\beta}^{-1} (m_{12}^2 c_{\beta+\alpha} s_{\beta}^{-1} c_{\beta}^{-1} - m_h^2 c_{2\beta} c_{\beta-\alpha}) - (m_h^2 +2 m_A^2) s_{\beta-\alpha}
\end{equation}
\begin{equation}
H H^+ H^-: \quad m_{12} v = s_{\beta}^{-1} c_{\beta}^{-1} (m_{12}^2 s_{\beta+\alpha} s_{\beta}^{-1} c_{\beta}^{-1} + m_H^2 c_{2\beta} s_{\beta-\alpha}) - (m_H^2 +2 m_{H^+}^2) c_{\beta-\alpha}
\end{equation}
\begin{equation}
H A A: \quad m_{13} v = s_{\beta}^{-1} c_{\beta}^{-1} (m_{12}^2 s_{\beta+\alpha} s_{\beta}^{-1} c_{\beta}^{-1} + m_H^2 c_{2\beta} s_{\beta-\alpha}) - (m_H^2 +2 m_A^2) c_{\beta-\alpha}
\end{equation}
\begin{equation}
h h h: \quad 4 m_{14} v s_{2\beta}^2 / 3 = 16 m_{12}^2 c_{\beta+\alpha} c_{\beta-\alpha}^2 - m_h^2 (3 s_{3\beta+\alpha} + 3 s_{\beta-\alpha} + s_{3\beta-3\alpha} + s_{\beta+3\alpha}) 
\end{equation}
\begin{equation}
h h H: \quad m_{15} v s_{2\beta} = - c_{\beta-\alpha} [2 m_{12}^2 + (m_H^2 + 2 m_h^2 - 3 m_{12} s_{\beta}^{-1} c_{\beta}^{-1}) s_{2\alpha}]
\end{equation}
\begin{equation}
h H H: \quad m_{16} v s_{2\beta} = s_{\beta-\alpha} [- 2 m_{12}^2 + (m_h^2 + 2 m_H^2 - 3 m_{12} s_{\beta}^{-1} c_{\beta}^{-1}) s_{2\alpha}]
\end{equation}
\begin{equation}
H H H: \quad 4 m_{17} v s_{2\beta}^2 / 3 = 16 m_{12}^2 s_{\beta+\alpha} s_{\beta-\alpha}^2 + m_H^2 (3 c_{3\beta+\alpha} - 3 c_{\beta-\alpha} + c_{3\beta-3\alpha} - c_{\beta+3\alpha}) 
\end{equation}

\subsection{Quartic Couplings}
\label{sec:SE4}
\begin{align}
h h z z: \quad - g_1 v^2 &= m_H^2 c_{\beta-\alpha}^4 + 2 (m_h^2 - m_H^2) c_{\beta-\alpha}^3 s_{\beta-\alpha} t_{2\beta}^{-1} + m_h^2 s_{\beta-\alpha}^4 \\
&+ c_{\beta-\alpha}^2 [2 m_A^2 - 2 m_{12} s_{\beta}^{-1} c_{\beta}^{-1} + (3 m_h^2 - m_H^2) s_{\beta-\alpha}^2] \nn
\end{align}
\begin{align}
H H z z: \quad - g_2 v^2 &= m_H^2 c_{\beta-\alpha}^4 + 2 (m_h^2 - m_H^2) c_{\beta-\alpha} s_{\beta-\alpha}^3 t_{2\beta}^{-1} + m_h^2 s_{\beta-\alpha}^4 \\
&+ s_{\beta-\alpha}^2 [2 m_A^2 - 2 m_{12} s_{\beta}^{-1} c_{\beta}^{-1} + (3 m_H^2 - m_h^2) c_{\beta-\alpha}^2] \nn
\end{align}
\begin{align}
H^+ H^- z z: \quad g_3 v^2 &= 2 m_{12}^2 s_{\beta}^{-1} c_{\beta}^{-1} - 2 m_{H^+}^2 - m_H^2 c_{\beta-\alpha}^2 - m_h^2 s_{\beta-\alpha}^2 \\
&+ (m_H^2 - m_h^2) t_{2\beta}^{-1} s_{2\beta-2\alpha} \nn
\end{align}
\begin{align}
A A z z: \quad g_4 v^2 &= 2 m_{12}^2 s_{\beta}^{-1} c_{\beta}^{-1} - (m_H^2 + 2 m_h^2) c_{\beta-\alpha}^2 - (m_h^2 + 2 m_H^2) s_{\beta-\alpha}^{2} \\
&+ (m_H^2 - m_h^2) t_{2\beta}^{-1} s_{2\beta-2\alpha} \nn
\end{align}
\begin{align}
h h z A: \quad 2 g_5 v^2 s_{2\beta} &= m_H^2 s_{2\beta-2\alpha} s_{2\alpha} - 2 m_A^2 s_{2\beta} s_{2\alpha-2\beta} \\
&+c_{\beta-\alpha} [4 m_{12}^2 c_{\beta-\alpha} s_{\beta}^{-1} c_{\beta}^{-1} c_{2\beta} - m_h^2 (c_{-\beta+3\alpha} + 3 c_{\beta+\alpha})]  \nn
\end{align}
\begin{align}
H H z A: \quad 2 g_6 v^2 s_{2\beta} &= m_h^2 s_{2\beta-2\alpha} s_{2\alpha} + 2 m_A^2 s_{2\beta} s_{2\alpha-2\beta}  \\
&+s_{\beta-\alpha} [4 m_{12}^2 s_{\beta-\alpha} s_{\beta}^{-1} c_{\beta}^{-1} c_{2\beta} - m_H^2 (s_{-\beta+3\alpha} - 3 s_{\beta+\alpha})]  \nn
\end{align}
\begin{align}
H^+ H^- z A: \quad 8 g_7 v^2 s_{2\beta}^2 &= 32 m_{12}^2 c_{2\beta} \\
&+ 2 (m_H^2 - m_h^2) (3 c_{2\alpha} + c_{4\beta-2\alpha}) s_{2\beta} - 4 (m_h^2 + m_H^2) s_{4\beta} \nn
\end{align}
\begin{align}
A A z A: \quad g_8 &= 3 g_7
\end{align}
\begin{align}
h h H^+ H^-: \quad 16 g_9 v^2 s_{2\beta} &= 2 s_{\beta}^{-2} c_{\beta}^{-2} [c_{2\alpha-6\beta} + 2(3 + c_{2\alpha-2\beta} + c_{4\beta}) + 5 c_{2\alpha+2\beta}] m_{12}^2 \\
&- s_{\beta}^{-1} c_{\beta}^{-1} (9 + 3 c_{4\alpha} + 6 c_{2\alpha-2\beta} + c_{4\alpha-4\beta} + 3 c_{4\beta} + 10 c_{2\alpha+2\beta}) m_h^2 \nn \\
&- 2 s_{\beta}^{-1} c_{\beta}^{-1} s_{2\alpha} (3 s_{2\alpha} + s_{2\alpha-4\beta} + 2 s_{2\beta}) m_H^2 - 32 s_{\alpha-\beta}^2 s_{2\beta} m_{H^+}^2 \nn
\end{align}
\begin{align}
h h A A: \quad 16 g_{10} v^2 s_{2\beta} &= 2 s_{\beta}^{-2} c_{\beta}^{-2} [c_{2\alpha-6\beta} + 2(3 + c_{2\alpha-2\beta} + c_{4\beta}) + 5 c_{2\alpha+2\beta}] m_{12}^2 \\
&- s_{\beta}^{-1} c_{\beta}^{-1} (9 + 3 c_{4\alpha} + 6 c_{2\alpha-2\beta} + c_{4\alpha-4\beta} + 3 c_{4\beta} + 10 c_{2\alpha+2\beta}) m_h^2 \nn \\
&- 2 s_{\beta}^{-1} c_{\beta}^{-1} s_{2\alpha} (3 s_{2\alpha} + s_{2\alpha-4\beta} + 2 s_{2\beta}) m_H^2 - 32 s_{\alpha-\beta}^2 s_{2\beta} m_A^2 \nn
\end{align}
\begin{align}
H H H^+ H^-: \quad 16 g_{11} v^2 s_{2\beta} &= 2 s_{\beta}^{-2} c_{\beta}^{-2} [ 2(3 - c_{2\alpha-2\beta} + c_{4\beta}) - c_{2\alpha-6\beta} - 5 c_{2\alpha+2\beta}] m_{12}^2 \\
&- s_{\beta}^{-1} c_{\beta}^{-1} (9 + 3 c_{4\alpha} - 6 c_{2\alpha-2\beta} + c_{4\alpha-4\beta} + 3 c_{4\beta} - 10 c_{2\alpha+2\beta}) m_H^2 \nn \\
&- 2 s_{\beta}^{-1} c_{\beta}^{-1} s_{2\alpha} (3 s_{2\alpha} + s_{2\alpha-4\beta} - 2 s_{2\beta}) m_h^2 - 32 c_{\alpha-\beta}^2 s_{2\beta} m_{H^+}^2 \nn
\end{align}
\begin{align}
H H A A: \quad 16 g_{12} v^2 s_{2\beta} &= 2 s_{\beta}^{-2} c_{\beta}^{-2} [ 2(3 - c_{2\alpha-2\beta} + c_{4\beta}) - c_{2\alpha-6\beta} - 5 c_{2\alpha+2\beta}] m_{12}^2 \\
&- s_{\beta}^{-1} c_{\beta}^{-1} (9 + 3 c_{4\alpha} - 6 c_{2\alpha-2\beta} + c_{4\alpha-4\beta} + 3 c_{4\beta} - 10 c_{2\alpha+2\beta}) m_H^2 \nn \\
&- 2 s_{\beta}^{-1} c_{\beta}^{-1} s_{2\alpha} (3 s_{2\alpha} + s_{2\alpha-4\beta} - 2 s_{2\beta}) m_h^2 - 32 c_{\alpha-\beta}^2 s_{2\beta} m_A^2 \nn
\end{align}
\begin{align}
h H H^+ H^-: \quad 8 g_{13} v^2 s_{2\beta} &= c_{\beta-\alpha} s_{\beta}^{-1} c_{\beta}^{-1} (3 s_{\beta-\alpha} + s_{3\beta-3\alpha} - 3 s_{\beta+3\alpha} - s_{3\beta+\alpha}) m_h^2 \\
& + s_{\beta-\alpha} s_{\beta}^{-1} c_{\beta}^{-1} (3 c_{\beta-\alpha} - c_{3\beta-3\alpha} - 3 c_{\beta+3\alpha} + c_{3\beta+\alpha}) m_H^2 \nn \\
& - 8 s_{2\beta-2\alpha} s_{2\beta} m_{H^+}^2 - 4 s_{2\beta}^{-2} (2(1 + 3 c_{4\beta}) s_{2\beta-2\alpha} - 4 c_{2\beta-2\alpha} s_{4\beta}) m_{12}^2 \nn
\end{align}
\begin{align}
h H A A: \quad 8 g_{14} v^2 s_{2\beta} &= c_{\beta-\alpha} s_{\beta}^{-1} c_{\beta}^{-1} (3 s_{\beta-\alpha} + s_{3\beta-3\alpha} - 3 s_{\beta+3\alpha} - s_{3\beta+\alpha}) m_h^2 \\
& + s_{\beta-\alpha} s_{\beta}^{-1} c_{\beta}^{-1} (3 c_{\beta-\alpha} - c_{3\beta-3\alpha} - 3 c_{\beta+3\alpha} + c_{3\beta+\alpha}) m_H^2 \nn \\
& - 8 s_{2\beta-2\alpha} s_{2\beta} m_A^2 - 4 s_{2\beta}^{-2} (2(1 + 3 c_{4\beta}) s_{2\beta-2\alpha} - 4 c_{2\beta-2\alpha} s_{4\beta}) m_{12}^2 \nn
\end{align}
\begin{align}
h h h h: \quad 4 g_{15} v^2 s_{2\beta}^2 / 3 &= 4 c_{\beta-\alpha}^2 (4 m_{12}^2 s_{\beta}^{-1} c_{\beta}^{-1} c_{\beta+\alpha}^2 - m_H^2 s_{2\alpha}^2) \\
&- m_h^2 (c_{-\beta+3\alpha} + 3 c_{\beta+\alpha})^2 \nn
\end{align}
\begin{align}
h h h H: \quad 2 g_{16} v^2 s_{2\beta}^2 &= 3 s_{2\alpha} [m_H^2 s_{2\alpha} s_{2\beta-2\alpha} - m_h^2  c_{\beta-\alpha} (c_{-\beta+3\alpha} + 3 c_{\beta+\alpha})] \\
&+ 12 m_{12}^2 s_{2\beta}^{-1} c_{\beta-\alpha} (s_{\beta+3\alpha} - s_{\beta-\alpha}) \nn
\end{align}
\begin{align}
h h H H: \quad 8 g_{17} v^2 s_{2\beta}^2 &= 4 s_{\beta}^{-1} c_{\beta}^{-1} (2 + c_{4\beta} - 3 c_{4\alpha}) m_{12}^2 + 6 (c_{4\alpha} - 1) (m_h^2 + m_H^2) \\
&+ (3 c_{-2\beta+6\alpha} - c_{2\beta+2\alpha} - 2 c_{2\beta-2\alpha}) (m_h^2 - m_H^2) \nn
\end{align}
\begin{align}
h H H H: \quad 2 g_{18} v^2 s_{2\beta}^2 &= 3 s_{2\alpha} [m_h^2 s_{2\alpha} s_{2\beta-2\alpha} - m_H^2 s_{\beta-\alpha} (s_{-\beta+3\alpha} - 3 s_{\beta+\alpha})] \\
&+ 12 m_{12}^2 s_{2\beta}^{-1} s_{\beta-\alpha} (c_{\beta+3\alpha} - c_{\beta-\alpha}) \nn
\end{align}
\begin{align}
H H H H: \quad 4 g_{19} v^2 s_{2\beta}^2 / 3 &= 4 s_{\beta-\alpha}^2 (4 m_{12}^2 s_{\beta}^{-1} c_{\beta}^{-1} s_{\beta+\alpha}^2 - m_h^2 s_{2\alpha}^2) \\
&- m_H^2 (s_{-\beta+3\alpha} - 3 s_{\beta+\alpha})^2 \nn\end{align}
\begin{align}
H^+ H^- H^+ H^-: \quad g_{20} v^2 / 2 &= (m_H^2 - m_h^2) c_{2\beta} s_{\beta}^{-1} c_{\beta}^{-1} s_{2\beta-2\alpha} - m_h^2 s_{\beta-\alpha}^2 \\
&- c_{\beta-\alpha}^2 (m_H^2 + 4 m_h^2 t_{2\beta}^{-2}) - 4 t_{2\beta}^{-2} (m_H^2 s_{\beta-\alpha}^2 - m_{12}^2 s_{\beta}^{-1} c_{\beta}^{-1}) \nn
\end{align}
\begin{align}
A A H^+ H^-: \quad g_{21} = g_{20} / 2
\end{align}
\begin{align}
A A A A: \quad g_{22} = 3 g_{20} / 2
\end{align}

\section{Results for Scattering Amplitudes I: Block Diagonal Elements of $\mathbf{a}_0$}
\label{sec:RSA}
Each amplitude, $ i \to f$, given in Appendices~\ref{sec:RSA} and~\ref{sec:RSA2} corresponds to $256 \pi^3 (\mathbf{a}_0)_{i, f}$. The reduced wavefunction renormalization, Eq.~\eqref{eq:reduce}, is used heavily is these expressions. All of the scattering amplitudes appearing in Appendix~\ref{sec:RSA} are part of the diagonal blocks of $\mathbf{a}_0$. Off-block diagonal elements of $\mathbf{a}_0$ have been relegated to Appendix~\ref{sec:RSA2}, as they do not contribute to the eigenvalues that are unique at tree level until the two-loop level.

\subsection{$Y = 0$, $\tau = 0$, $\mathbb{Z}_2$-even}
\begin{align}
&\tfrac{1}{\sqrt{2}} (\Phi_1^{\dagger} \Phi_1) \to \tfrac{1}{\sqrt{2}} (\Phi_1^{\dagger} \Phi_1) = - 48 \pi^2 \lambda_1 + 9 \beta_{\lambda_1} + \left(i \pi - 1\right) \left(9 \lambda_1^2 + \left(2 \lambda_3 + \lambda_4\right)^2\right) \\
&- \frac{3}{2} \lambda_1 \left[z_{A A}^{1/2} + z_{h h}^{1/2} + 2 z_{H^+ H^-}^{1/2} + z_{H H}^{1/2} + 2 z_{w^+ w^-}^{1/2} + z_{z z}^{1/2} + \left(z_{H H}^{1/2} - z_{h h}^{1/2}\right) c_{2\alpha} \right. \nn \\
&\left. + \left(2 z_{w^+ w^-}^{1/2} - 2 z_{H^+ H^-}^{1/2} + z_{z z}^{1/2} - z_{A A}^{1/2}\right) c_{2\beta} - \left(z_{H h}^{1/2} + z_{h H}^{1/2}\right) s_{2\alpha} - \left(2 z_{H^+ w^-}^{1/2} + z_{A z}^{1/2}\right) s_{2\beta} \right] \nn
\end{align}
\begin{align}
&\tfrac{1}{\sqrt{2}} (\Phi_2^{\dagger} \Phi_2) \to \tfrac{1}{\sqrt{2}} (\Phi_2^{\dagger} \Phi_2) = - 48 \pi^2 \lambda_2 + 9 \beta_{\lambda_2} + \left(i \pi - 1\right) \left(9 \lambda_2^2 + \left(2 \lambda_3 + \lambda_4\right)^2\right) \\
&- \frac{3}{2} \lambda_2 \left[z_{A A}^{1/2} + z_{h h}^{1/2} + 2 z_{H^+ H^-}^{1/2} + z_{H H}^{1/2} + 2 z_{w^+ w^-}^{1/2} + z_{z z}^{1/2} - \left(z_{H H}^{1/2} - z_{h h}^{1/2}\right) c_{2\alpha} \right. \nn \\
&\left. - \left(2 z_{w^+ w^-}^{1/2} - 2 z_{H^+ H^-}^{1/2} + z_{z z}^{1/2} - z_{A A}^{1/2}\right) c_{2\beta} + \left(z_{H h}^{1/2} + z_{h H}^{1/2}\right) s_{2\alpha} + \left(2 z_{H^+ w^-}^{1/2} + z_{A z}^{1/2}\right) s_{2\beta} \right] \nn
\end{align}
\begin{align}
&\tfrac{1}{\sqrt{2}} (\Phi_1^{\dagger} \Phi_1) \to \tfrac{1}{\sqrt{2}} (\Phi_2^{\dagger} \Phi_2) = - 16 \pi^2 \left(2 \lambda_3 + \lambda_4\right) + 3 \left(i \pi - 1\right) \left(\lambda_1 + \lambda_2\right) \left(2 \lambda_3 + \lambda_4\right)  \\
&+ 3 \left(2 \beta_{\lambda_3} +\beta_{\lambda_4}\right) - \frac{1}{2} \left(2 \lambda_3 + \lambda_4\right) \left[z_{A A}^{1/2} + z_{h h}^{1/2} + 2 z_{H^+ H^-}^{1/2} + z_{H H}^{1/2} + 2 z_{w^+ w^-}^{1/2} + z_{z z}^{1/2}\right] \nn
\end{align}

\subsection{$Y = 0$, $\tau = 0$, $\mathbb{Z}_2$-odd}
\begin{align}
&\tfrac{1}{\sqrt{2}} (\Phi_1^{\dagger} \Phi_2) \to \tfrac{1}{\sqrt{2}} (\Phi_1^{\dagger} \Phi_2) = - 16 \pi^2 \left(\lambda_3 + 2 \lambda_4\right) + \left(i \pi - 1\right) \left(\lambda_3^2 + 4 \lambda_3 \lambda_4 + 4\lambda_4^2 + 9 \lambda_5^2\right) \\
&+ 3 \left(\beta_{\lambda_3} + 2 \beta_{\lambda_4}\right) - \frac{1}{2} \left(\lambda_3 + \lambda_4 + \lambda_5\right) \left[z_{A A}^{1/2} + z_{h h}^{1/2} + 2 z_{H^+ H^-}^{1/2} + z_{H H}^{1/2} + 2 z_{w^+ w^-}^{1/2} + z_{z z}^{1/2}\right] \nn
\end{align}
\begin{align}
\tfrac{1}{\sqrt{2}} (\Phi_2^{\dagger} \Phi_1) \to \tfrac{1}{\sqrt{2}} (\Phi_2^{\dagger} \Phi_1) &= \tfrac{1}{\sqrt{2}} (\Phi_1^{\dagger} \Phi_2) \to \tfrac{1}{\sqrt{2}} (\Phi_1^{\dagger} \Phi_2)\end{align}
\begin{align}
\tfrac{1}{\sqrt{2}} (\Phi_1^{\dagger} \Phi_2) \to \tfrac{1}{\sqrt{2}} (\Phi_2^{\dagger} \Phi_1) &= - 48 \pi^2 \lambda_5 + 9 \beta_{\lambda_5} + 6 \left(i \pi -1\right) \left(\lambda_3 + 2 \lambda_4\right) \lambda_5 \\
&- \frac{1}{2} \left(\lambda_4 + 2 \lambda_5\right) \left[z_{A A}^{1/2} + z_{h h}^{1/2} + 2 z_{H^+ H^-}^{1/2} + z_{H H}^{1/2} + 2 z_{w^+ w^-}^{1/2} + z_{z z}^{1/2}\right] \nn
\end{align}

\subsection{$Y = 0$, $\tau = 1$, $\mathbb{Z}_2$-even}
Neutral Initial States:
\begin{align}
&\tfrac{1}{\sqrt{2}} (\Phi_1^{\dagger} \tau^3 \Phi_1) \to \tfrac{1}{\sqrt{2}} (\Phi_1^{\dagger} \tau^3 \Phi_1) = - 16 \pi^2 \lambda_1 + 3 \beta_{\lambda_1} + \left(i \pi -1\right) \left(\lambda_1^2 + \lambda_4^2\right) \\
&- \frac{1}{2} \lambda_1 \left[z_{A A}^{1/2} + z_{h h}^{1/2} + 2 z_{H^+ H^-}^{1/2} + z_{H H}^{1/2} + 2 z_{w^+ w^-}^{1/2} + z_{z z}^{1/2} + \left(z_{H H}^{1/2} - z_{h h}^{1/2}\right) c_{2\alpha} \right. \nn \\
&\left. + \left(2 z_{w^+ w^-}^{1/2} - 2 z_{H^+ H^-}^{1/2} + z_{z z}^{1/2} - z_{A A}^{1/2}\right) c_{2\beta} - \left(z_{H h}^{1/2} + z_{h H}^{1/2}\right) s_{2\alpha} - \left(2 z_{H^+ w^-}^{1/2} + z_{A z}^{1/2}\right) s_{2\beta} \right] \nn
\end{align}
\begin{align}
&\tfrac{1}{\sqrt{2}} (\Phi_2^{\dagger} \tau^3 \Phi_2) \to \tfrac{1}{\sqrt{2}} (\Phi_2^{\dagger} \tau^3 \Phi_2) = - 16 \pi^2 \lambda_2 + 3 \beta_{\lambda_2} + \left(i \pi -1\right) \left(\lambda_2^2 + \lambda_4^2\right) \\
&- \frac{1}{2} \lambda_2 \left[z_{A A}^{1/2} + z_{h h}^{1/2} + 2 z_{H^+ H^-}^{1/2} + z_{H H}^{1/2} + 2 z_{w^+ w^-}^{1/2} + z_{z z}^{1/2} - \left(z_{H H}^{1/2} - z_{h h}^{1/2}\right) c_{2\alpha} \right. \nn \\
&\left. - \left(2 z_{w^+ w^-}^{1/2} - 2 z_{H^+ H^-}^{1/2} + z_{z z}^{1/2} - z_{A A}^{1/2}\right) c_{2\beta} + \left(z_{H h}^{1/2} + z_{h H}^{1/2}\right) s_{2\alpha} + \left(2 z_{H^+ w^-}^{1/2} + z_{A z}^{1/2}\right) s_{2\beta} \right] \nn
\end{align}
\begin{align}
\tfrac{1}{\sqrt{2}} (\Phi_1^{\dagger} \tau^3 \Phi_1) \to \tfrac{1}{\sqrt{2}} (\Phi_2^{\dagger} \tau^3 \Phi_2) &= - 16 \pi^2 \lambda_4 + 3 \beta_{\lambda_4} + \left(i \pi -1\right) \left(\lambda_1 + \lambda_2\right) \lambda_4 \\
&- \frac{1}{2} \lambda_4 \left[z_{A A}^{1/2} + z_{h h}^{1/2} + 2 z_{H^+ H^-}^{1/2} + z_{H H}^{1/2} + 2 z_{w^+ w^-}^{1/2} + z_{z z}^{1/2}\right] \nn
\end{align}
Singly-Charged Initial States:
\begin{align}
\tfrac{1}{\sqrt{2}} (\Phi_1^{\dagger} \tau^- \Phi_1) \to \tfrac{1}{\sqrt{2}} (\Phi_1^{\dagger} \tau^- \Phi_1) &= \tfrac{1}{\sqrt{2}} (\Phi_1^{\dagger} \tau^3 \Phi_1) \to \tfrac{1}{\sqrt{2}} (\Phi_1^{\dagger} \tau^3 \Phi_1)
\end{align}
\begin{align}
\tfrac{1}{\sqrt{2}} (\Phi_2^{\dagger} \tau^- \Phi_2) \to \tfrac{1}{\sqrt{2}} (\Phi_2^{\dagger} \tau^- \Phi_2) &= \tfrac{1}{\sqrt{2}} (\Phi_2^{\dagger} \tau^3 \Phi_2) \to \tfrac{1}{\sqrt{2}} (\Phi_2^{\dagger} \tau^3 \Phi_2)
\end{align}
\begin{align}
\tfrac{1}{\sqrt{2}} (\Phi_1^{\dagger} \tau^- \Phi_1) \to \tfrac{1}{\sqrt{2}} (\Phi_2^{\dagger} \tau^- \Phi_2) &= - 16 \pi^2 \lambda_4 + 3 \beta_{\lambda_4} + \left(i \pi -1\right) \left(\lambda_1 + \lambda_2\right) \lambda_4 \\
&- \frac{1}{2} \lambda_5 \left[z_{A A}^{1/2} + z_{h h}^{1/2} + 2 z_{H^+ H^-}^{1/2} + z_{H H}^{1/2} + 2 z_{w^+ w^-}^{1/2} + z_{z z}^{1/2}\right] \nn
\end{align}

\subsection{$Y = 0$, $\tau = 1$, $\mathbb{Z}_2$-odd}
Neutral Initial States:
\begin{align}
&\tfrac{1}{\sqrt{2}} (\Phi_1^{\dagger} \tau^3 \Phi_2) \to \tfrac{1}{\sqrt{2}} (\Phi_1^{\dagger} \tau^3 \Phi_2) = - 16 \pi^2 \lambda_3 +3 \beta_{\lambda_3} + \left(i \pi -1\right)\left(\lambda_3^2 + \lambda_5^2\right) \\
&- \frac{1}{2} \left(\lambda_3 + \lambda_4 - \lambda_5\right) \left[z_{A A}^{1/2} + z_{h h}^{1/2} + 2 z_{H^+ H^-}^{1/2} + z_{H H}^{1/2} + 2 z_{w^+ w^-}^{1/2} + z_{z z}^{1/2} \right] \nn
\end{align}
\begin{align}
\tfrac{1}{\sqrt{2}} (\Phi_2^{\dagger} \tau^3 \Phi_1) \to \tfrac{1}{\sqrt{2}} (\Phi_2^{\dagger} \tau^3 \Phi_1) &= \tfrac{1}{\sqrt{2}} (\Phi_1^{\dagger} \tau^3 \Phi_2) \to \tfrac{1}{\sqrt{2}} (\Phi_1^{\dagger} \tau^3 \Phi_2)\end{align}
\begin{align}
&\tfrac{1}{\sqrt{2}} (\Phi_1^{\dagger} \tau^3 \Phi_2) \to \tfrac{1}{\sqrt{2}} (\Phi_2^{\dagger} \tau^3 \Phi_1) = - 16 \pi^2 \lambda_5 +3 \beta_{\lambda_5} + 2 \left(i \pi - 1\right) \lambda_3 \lambda_5 \\
&- \frac{1}{2} \left(\lambda_4 - 2 \lambda_5\right) \left[z_{A A}^{1/2} + z_{h h}^{1/2} + 2 z_{H^+ H^-}^{1/2} + z_{H H}^{1/2} + 2 z_{w^+ w^-}^{1/2} + z_{z z}^{1/2}\right] \nn
\end{align}
Singly-Charged Initial States:
\begin{align}
&\tfrac{1}{\sqrt{2}} (\Phi_1^{\dagger} \tau^- \Phi_2) \to \tfrac{1}{\sqrt{2}} (\Phi_1^{\dagger} \tau^- \Phi_2) = - 16 \pi^2 \lambda_3 +3 \beta_{\lambda_3} + \left(i \pi -1\right)\left(\lambda_3^2 + \lambda_5^2\right) \\
&- \frac{1}{2} \lambda_3 \left[z_{A A}^{1/2} + z_{h h}^{1/2} + 2 z_{H^+ H^-}^{1/2} + z_{H H}^{1/2} + 2 z_{w^+ w^-}^{1/2} + z_{z z}^{1/2} + \left(z_{H H}^{1/2} - z_{h h}^{1/2}\right) c_{2\alpha} \right. \nn \\
&\left. + \left(- 2 z_{w^+ w^-}^{1/2} + 2 z_{H^+ H^-}^{1/2} + z_{z z}^{1/2} - z_{A A}^{1/2}\right) c_{2\beta} - \left(z_{H h}^{1/2} + z_{h H}^{1/2}\right) s_{2\alpha} - \left(- 2 z_{H^+ w^-}^{1/2} + z_{A z}^{1/2}\right) s_{2\beta} \right] \nn
\end{align}
\begin{align}
\tfrac{1}{\sqrt{2}} (\Phi_2^{\dagger} \tau^- \Phi_1) \to \tfrac{1}{\sqrt{2}} (\Phi_2^{\dagger} \tau^- \Phi_1) &= \tfrac{1}{\sqrt{2}} (\Phi_1^{\dagger} \tau^- \Phi_2) \to \tfrac{1}{\sqrt{2}} (\Phi_1^{\dagger} \tau^- \Phi_2)
\end{align}
\begin{align}
\tfrac{1}{\sqrt{2}} (\Phi_1^{\dagger} \tau^- \Phi_2) \to \tfrac{1}{\sqrt{2}} (\Phi_2^{\dagger} \tau^- \Phi_1) &= - 16 \pi^2 \lambda_5 +3 \beta_{\lambda_5} + 2 \left(i \pi - 1\right) \lambda_3 \lambda_5 \\
&- \frac{1}{2} \lambda_4 \left[z_{A A}^{1/2} + z_{h h}^{1/2} + 2 z_{H^+ H^-}^{1/2} + z_{H H}^{1/2} + 2 z_{w^+ w^-}^{1/2} + z_{z z}^{1/2}\right] \nn
\end{align}

\subsection{$Y = 1$, $\tau = 0$, $\mathbb{Z}_2$-odd}
\begin{align}
\label{eq:110}
\tfrac{1}{\sqrt{2}} (\tilde{\Phi}_1 \Phi_2) \to \tfrac{1}{\sqrt{2}} (\tilde{\Phi}_1 \Phi_2) &= - 16 \pi^2 \left(\lambda_3 - \lambda_4\right) + 3 \left(\beta_{\lambda_3} - \beta_{\lambda_4}\right) + \left(i \pi -1\right)\left(\lambda_3 - \lambda_4\right)^2 \\
&- \frac{1}{2} \left(\lambda_3 - \lambda_5\right) \left[z_{A A}^{1/2} + z_{h h}^{1/2} + 2 z_{H^+ H^-}^{1/2} + z_{H H}^{1/2} + 2 z_{w^+ w^-}^{1/2} + z_{z z}^{1/2}\right] \nn
\end{align}

\subsection{$Y = 1$, $\tau = 1$, $\mathbb{Z}_2$-even}
Neutral Initial States:
\begin{align}
\tfrac{1}{2} (\tilde{\Phi}_1 \tau^+ \Phi_1) \to \tfrac{1}{2} (\tilde{\Phi}_1 \tau^+ \Phi_1) &= - 16 \pi^2 \lambda_1 + 3 \beta_{\lambda_1} + \left(i \pi -1\right) \left(\lambda_1^2 + \lambda_5^2\right) \\
&- \lambda_1 \left[z_{A A}^{1/2} + z_{h h}^{1/2} + z_{H H}^{1/2} + z_{z z}^{1/2} + \left(z_{H H}^{1/2} - z_{h h}^{1/2}\right) c_{2\alpha} \right. \nn \\
&\left. + \left(z_{z z}^{1/2} - z_{A A}^{1/2}\right) c_{2\beta} - \left(z_{H h}^{1/2} + z_{h H}^{1/2}\right) s_{2\alpha} - z_{A z}^{1/2} s_{2\beta} \right] \nn
\end{align}
\begin{align}
\tfrac{1}{2} (\tilde{\Phi}_2 \tau^+ \Phi_2) \to \tfrac{1}{2} (\tilde{\Phi}_2 \tau^+ \Phi_2) &= - 16 \pi^2 \lambda_2 + 3 \beta_{\lambda_2} + \left(i \pi -1\right) \left(\lambda_2^2 + \lambda_5^2\right) \\
&- \lambda_2 \left[z_{A A}^{1/2} + z_{h h}^{1/2} + z_{H H}^{1/2} + z_{z z}^{1/2} - \left(z_{H H}^{1/2} - z_{h h}^{1/2}\right) c_{2\alpha} \right. \nn \\
&\left. - \left(z_{z z}^{1/2} - z_{A A}^{1/2}\right) c_{2\beta} + \left(z_{H h}^{1/2} + z_{h H}^{1/2}\right) s_{2\alpha} + z_{A z}^{1/2} s_{2\beta} \right] \nn
\end{align}
\begin{align}
\tfrac{1}{2} (\tilde{\Phi}_1 \tau^+ \Phi_1) \to \tfrac{1}{2} (\tilde{\Phi}_2 \tau^+ \Phi_2) &= - 16 \pi^2 \lambda_5 + 3 \beta_{\lambda_5} + \left(i \pi -1\right)\left(\lambda_1 + \lambda_2\right) \lambda_5 \\
&- \lambda_5 \left[z_{A A}^{1/2} + z_{h h}^{1/2} + z_{H H}^{1/2} + z_{z z}^{1/2}\right] \nn
\end{align}
\begin{equation}
\tfrac{1}{2} (\tilde{\Phi}_i^{*} \tau^- \Phi_i^{*}) \to \tfrac{1}{2} (\tilde{\Phi}_j^{*} \tau^- \Phi_j^{*}) = \tfrac{1}{2} (\tilde{\Phi}_i \tau^+ \Phi_i) \to \tfrac{1}{2} (\tilde{\Phi}_j \tau^+ \Phi_j)
\end{equation}
Singly-Charged Initial States:
\begin{align}
&\tfrac{1}{2} (\tilde{\Phi}_1 \tau^3 \Phi_1) \to \tfrac{1}{2} (\tilde{\Phi}_1 \tau^3 \Phi_1) = - 16 \pi^2 \lambda_1 + 3 \beta_{\lambda_1} + \left(i \pi -1\right) \left(\lambda_1^2 + \lambda_5^2\right) \\
&- \frac{1}{2} \lambda_1 \left[z_{A A}^{1/2} + z_{h h}^{1/2} + 2 z_{H^+ H^-}^{1/2} + z_{H H}^{1/2} + 2 z_{w^+ w^-}^{1/2} + z_{z z}^{1/2} + \left(z_{H H}^{1/2} - z_{h h}^{1/2}\right) c_{2\alpha} \right. \nn \\
&\left. + \left(2 z_{w^+ w^-}^{1/2} - 2 z_{H^+ H^-}^{1/2} + z_{z z}^{1/2} - z_{A A}^{1/2}\right) c_{2\beta} - \left(z_{H h}^{1/2} + z_{h H}^{1/2}\right) s_{2\alpha} - \left(2 z_{H^+ w^-}^{1/2} + z_{A z}^{1/2}\right) s_{2\beta} \right] \nn
\end{align}
\begin{align}
&\tfrac{1}{2} (\tilde{\Phi}_2 \tau^3 \Phi_2) \to \tfrac{1}{2} (\tilde{\Phi}_2 \tau^3 \Phi_2) = - 16 \pi^2 \lambda_2 + 3 \beta_{\lambda_2} + \left(i \pi -1\right) \left(\lambda_2^2 + \lambda_5^2\right) \\
&- \frac{1}{2} \lambda_2 \left[z_{A A}^{1/2} + z_{h h}^{1/2} + 2 z_{H^+ H^-}^{1/2} + z_{H H}^{1/2} + 2 z_{w^+ w^-}^{1/2} + z_{z z}^{1/2} - \left(z_{H H}^{1/2} - z_{h h}^{1/2}\right) c_{2\alpha} \right. \nn \\
&\left. - \left(2 z_{w^+ w^-}^{1/2} - 2 z_{H^+ H^-}^{1/2} + z_{z z}^{1/2} - z_{A A}^{1/2}\right) c_{2\beta} + \left(z_{H h}^{1/2} + z_{h H}^{1/2}\right) s_{2\alpha} + \left(2 z_{H^+ w^-}^{1/2} + z_{A z}^{1/2}\right) s_{2\beta} \right] \nn
\end{align}
\begin{align}
\tfrac{1}{2} (\tilde{\Phi}_1 \tau^3 \Phi_1) \to \tfrac{1}{2} (\tilde{\Phi}_2 \tau^3 \Phi_2) &= - 16 \pi^2 \lambda_5 + 3 \beta_{\lambda_5} + \left(i \pi -1\right)\left(\lambda_1 + \lambda_2\right) \lambda_5 \\
&- \frac{1}{2} \lambda_4 \left[z_{A A}^{1/2} + z_{h h}^{1/2} + 2 z_{H^+ H^-}^{1/2} + z_{H H}^{1/2} + 2 z_{w^+ w^-}^{1/2} + z_{z z}^{1/2}\right] \nn
\end{align}
Doubly-Charged Initial States:
\begin{align}
\tfrac{1}{2} (\tilde{\Phi}_1 \tau^- \Phi_1) \to \tfrac{1}{2} (\tilde{\Phi}_1 \tau^- \Phi_1) &= - 16 \pi^2 \lambda_1 + 3 \beta_{\lambda_1} + \left(i \pi - 1\right) \left(\lambda_1^2 + \lambda_5^2\right) \\
&- 2 \lambda_1 \left[z_{H^+ H^-}^{1/2} + z_{w^+ w^-}^{1/2} + \left(z_{w^+ w^-}^{1/2} - z_{H^+ H^-}^{1/2}\right) c_{2\beta} - z_{H^+ w^-}^{1/2} s_{2\beta}\right] \nn
\end{align}
\begin{align}
\tfrac{1}{2} (\tilde{\Phi}_2 \tau^- \Phi_2) \to \tfrac{1}{2} (\tilde{\Phi}_2 \tau^- \Phi_2) &= - 16 \pi^2 \lambda_2 + 3 \beta_{\lambda_2} + \left(i \pi - 1\right) \left(\lambda_2^2 + \lambda_5^2\right) \\
&- 2 \lambda_2 \left[z_{H^+ H^-}^{1/2} + z_{w^+ w^-}^{1/2} - \left(z_{w^+ w^-}^{1/2} - z_{H^+ H^-}^{1/2}\right) c_{2\beta} + z_{H^+ w^-}^{1/2} s_{2\beta}\right] \nn
\end{align}
\begin{align}
\tfrac{1}{2} (\tilde{\Phi}_1 \tau^- \Phi_1) \to \tfrac{1}{2} (\tilde{\Phi}_2 \tau^- \Phi_2) &= - 16 \pi^2 \lambda_5 + 3 \beta_{\lambda_5} + \left(i \pi -1\right) \left(\lambda_1 + \lambda_2\right) \lambda_5 \\
&- 2 \lambda_5 \left[z_{H^+ H^-}^{1/2} + z_{w^+ w^-}^{1/2} \right] \nn
\end{align}

\subsection{$Y = 1$, $\tau = 1$, $\mathbb{Z}_2$-odd}
Neutral Initial States:
\begin{align}
\tfrac{1}{\sqrt{2}} (\tilde{\Phi}_1 \tau^+ \Phi_2) \to \tfrac{1}{\sqrt{2}} (\tilde{\Phi}_1 \tau^+ \Phi_2) &= - 16 \pi^2 \left(\lambda_3 + \lambda_4\right) + 3 \left(\beta_{\lambda_3} + \beta_{\lambda_4}\right)  \\
&+ \left(i \pi -1\right)\left(\lambda_3 + \lambda_4\right)^2 - \left(\lambda_3 + \lambda_4\right) \left[z_{A A}^{1/2} + z_{h h}^{1/2} + z_{H H}^{1/2} + z_{z z}^{1/2}\right] \nn
\end{align}
\begin{equation}
\tfrac{1}{\sqrt{2}} (\tilde{\Phi}_1^{\dagger} \tau^+ \Phi_2^{\dagger}) \to \tfrac{1}{\sqrt{2}} (\tilde{\Phi}_1^{*} \tau^- \Phi_2^{*}) = \tfrac{1}{\sqrt{2}} (\tilde{\Phi}_1 \tau^+ \Phi_2) \to \tfrac{1}{\sqrt{2}} (\tilde{\Phi}_1 \tau^+ \Phi_2)
\end{equation}
Singly-Charged Initial States:
\begin{align}
&\tfrac{1}{\sqrt{2}} (\tilde{\Phi}_1 \tau^3 \Phi_2) \to \tfrac{1}{\sqrt{2}} (\tilde{\Phi}_1 \tau^3 \Phi_2) = - 16 \pi^2 \left(\lambda_3 + \lambda_4\right) + 3 \left(\beta_{\lambda_3} + \beta_{\lambda_4}\right) \\
&+ \left(i \pi -1\right)\left(\lambda_3 + \lambda_4\right)^2 - \frac{1}{2} \left(\lambda_3 + \lambda_5\right) \left[z_{A A}^{1/2} + z_{h h}^{1/2} + 2 z_{H^+ H^-}^{1/2} + z_{H H}^{1/2} + 2 z_{w^+ w^-}^{1/2} + z_{z z}^{1/2}\right] \nn
\end{align}
Doubly-Charged Initial States:
\begin{align}
\tfrac{1}{\sqrt{2}} (\tilde{\Phi}_1 \tau^- \Phi_2) \to \tfrac{1}{\sqrt{2}} (\tilde{\Phi}_1 \tau^- \Phi_2) &= - 16 \pi^2 \left(\lambda_3 + \lambda_4\right) + 3 \left(\beta_{\lambda_3} + \beta_{\lambda_4}\right) \\
&+ \left(i \pi - 1\right) \left(\lambda_3 + \lambda_4\right)^2 - 2 \left(\lambda_3 + \lambda_4\right) \left[z_{H^+ H^-}^{1/2} + z_{w^+ w^-}^{1/2} \right] \nn
\end{align}

\section{Results for Scattering Amplitudes II: Off-Block Diagonal Elements of $\mathbf{a}_0$}
\label{sec:RSA2}
As in Appendix~\ref{sec:RSA}, each amplitude $ i \to f$ given in Appendix~\ref{sec:RSA2} corresponds to $256 \pi^3 (\mathbf{a}_0)_{i, f}$. The off-block diagonal elements of $\mathbf{a}_0$ are given in Appendix~\ref{sec:RSA2}, all of which vanish at tree level. 

\subsection{$Q = 0$, $\mathbb{Z}_2\text{-even} \to \mathbb{Z}_2\text{-even}$}
\begin{align}
\tfrac{1}{\sqrt{2}} (\Phi_1^{\dagger} \Phi_1) \to \tfrac{1}{\sqrt{2}} (\Phi_1^{\dagger} \tau^3 \Phi_1) &= \lambda_1 \left[ z_{AA}^{1/2} + z_{hh}^{1/2} - 2 z_{H^+H^-}^{1/2} + z_{HH}^{1/2} - 2 z_{w^+w^-}^{1/2} + z_{zz}^{1/2} \right. \\
&\left. + \left(z_{HH}^{1/2} - z_{hh}^{1/2}\right) c_{2\alpha} - \left(z_{AA}^{1/2} - 2 z_{H^+H^-}^{1/2} + 2 z_{w^+w^-}^{1/2} - z_{zz}^{1/2}\right) c_{2\beta} \right. \nn \\
&\left.- \left(z_{hH}^{1/2} + z_{Hh}^{1/2}\right) s_{2\alpha} - \left(z_{Az}^{1/2} - 2 z_{H^+w^-}^{1/2}\right) s_{2\beta}\right] \nn
\end{align}
\begin{align}
\tfrac{1}{\sqrt{2}} (\Phi_2^{\dagger} \Phi_2) \to \tfrac{1}{\sqrt{2}} (\Phi_2^{\dagger} \tau^3 \Phi_2) &= \lambda_2 \left[ z_{AA}^{1/2} + z_{hh}^{1/2} - 2 z_{H^+H^-}^{1/2} + z_{HH}^{1/2} - 2 z_{w^+w^-}^{1/2} + z_{zz}^{1/2} \right. \\
&\left. - \left(z_{HH}^{1/2} - z_{hh}^{1/2}\right) c_{2\alpha} + \left(z_{AA}^{1/2} - 2 z_{H^+H^-}^{1/2} + 2 z_{w^+w^-}^{1/2} - z_{zz}^{1/2}\right) c_{2\beta} \right. \nn \\
&\left.+ \left(z_{hH}^{1/2} + z_{Hh}^{1/2}\right) s_{2\alpha} + \left(z_{Az}^{1/2} - 2 z_{H^+w^-}^{1/2}\right) s_{2\beta}\right] \nn
\end{align}
\begin{align}
&\tfrac{1}{\sqrt{2}} (\Phi_1^{\dagger} \Phi_1) \to \tfrac{1}{\sqrt{2}} (\Phi_2^{\dagger} \tau^3 \Phi_2) = \frac{1}{2} \left(\lambda_3 + \lambda_4\right) \left(z_{AA}^{1/2} + z_{hh}^{1/2} - 2 z_{H^+H^-}^{1/2} + z_{HH}^{1/2} \right. \\
&\left. - 2 z_{w^+w^-}^{1/2} + z_{zz}^{1/2} \right) + \frac{1}{2} \lambda_3 \left[ - \left(z_{HH}^{1/2} - z_{hh}^{1/2}\right) c_{2\alpha} + \left(z_{AA}^{1/2} - 2 z_{H^+H^-}^{1/2} + 2 z_{w^+w^-}^{1/2} - z_{zz}^{1/2}\right) c_{2\beta} \right. \nn \\
&\left. + \left(z_{hH}^{1/2} + z_{Hh}^{1/2}\right) s_{2\alpha} + \left(z_{Az}^{1/2} - 2 z_{H^+w^-}^{1/2}\right) s_{2\beta} \right] \nn
\end{align}
\begin{align}
&\tfrac{1}{\sqrt{2}} (\Phi_2^{\dagger} \Phi_2) \to \tfrac{1}{\sqrt{2}} (\Phi_1^{\dagger} \tau^3 \Phi_1) = \frac{1}{2} \left(\lambda_3 + \lambda_4\right) \left(z_{AA}^{1/2} + z_{hh}^{1/2} - 2 z_{H^+H^-}^{1/2} + z_{HH}^{1/2} \right. \\
&\left. - 2 z_{w^+w^-}^{1/2} + z_{zz}^{1/2} \right) + \frac{1}{2} \lambda_3 \left[\left(z_{HH}^{1/2} - z_{hh}^{1/2}\right) c_{2\alpha} - \left(z_{AA}^{1/2} - 2 z_{H^+H^-}^{1/2} + 2 z_{w^+w^-}^{1/2} - z_{zz}^{1/2}\right) c_{2\beta} \right. \nn \\
&\left. + \left(z_{hH}^{1/2} + z_{Hh}^{1/2}\right) s_{2\alpha} + \left(z_{Az}^{1/2} - 2 z_{H^+w^-}^{1/2}\right) s_{2\beta} \right] \nn
\end{align}
\begin{align}
\tfrac{1}{\sqrt{2}} (\Phi_1^{\dagger} \Phi_1) \to \tfrac{1}{2} (\tilde{\Phi}_1 \tau^+ \Phi_1) &= \lambda_1 \left[ z_{AA}^{1/2} - z_{hh}^{1/2} - z_{HH}^{1/2} + z_{zz}^{1/2} - \left(z_{HH}^{1/2} - z_{hh}^{1/2}\right) c_{2\alpha} \right. \\
&\left. - \left(z_{AA}^{1/2} - z_{zz}^{1/2}\right) c_{2\beta} + \left(z_{hH}^{1/2} + z_{Hh}^{1/2}\right) s_{2\alpha} - z_{Az}^{1/2} s_{2\beta}\right] \nn
\end{align}
\begin{align}
\tfrac{1}{\sqrt{2}} (\Phi_2^{\dagger} \Phi_2) \to \tfrac{1}{2} (\tilde{\Phi}_2 \tau^+ \Phi_2) &= \lambda_2 \left[ z_{AA}^{1/2} - z_{hh}^{1/2} - z_{HH}^{1/2} + z_{zz}^{1/2} + \left(z_{HH}^{1/2} - z_{hh}^{1/2}\right) c_{2\alpha} \right. \\
&\left. + \left(z_{AA}^{1/2} - z_{zz}^{1/2}\right) c_{2\beta} - \left(z_{hH}^{1/2} + z_{Hh}^{1/2}\right) s_{2\alpha} + z_{Az}^{1/2} s_{2\beta}\right] \nn
\end{align}
\begin{align}
\tfrac{1}{\sqrt{2}} (\Phi_1^{\dagger} \Phi_1) \to \tfrac{1}{2} (\tilde{\Phi}_2 \tau^+ \Phi_2) &= \frac{1}{4} \left(2 \lambda_3 + \lambda_4 + \lambda_5\right) \left(z_{AA}^{1/2} - z_{hh}^{1/2} - z_{HH}^{1/2} + z_{zz}^{1/2}\right) \\
&- \frac{1}{4} \left(2 \lambda_3 + \lambda_4 - \lambda_5\right) \left[ \left(z_{hh}^{1/2} - z_{HH}^{1/2}\right) c_{2\alpha} + \left(z_{zz}^{1/2} - z_{AA}^{1/2}\right) c_{2\beta} \right. \nn \\
&\left. + \left(z_{hH}^{12} + z_{Hh}^{1/2}\right) s_{2\alpha} - z_{Az}^{1/2} s_{2\beta}\right] \nn
\end{align}
\begin{align}
\tfrac{1}{\sqrt{2}} (\Phi_2^{\dagger} \Phi_2) \to \tfrac{1}{2} (\tilde{\Phi}_1 \tau^+ \Phi_1) &= \frac{1}{4} \left(2 \lambda_3 + \lambda_4 + \lambda_5\right) \left(z_{AA}^{1/2} - z_{hh}^{1/2} - z_{HH}^{1/2} + z_{zz}^{1/2}\right) \\
&+ \frac{1}{4} \left(2 \lambda_3 + \lambda_4 - \lambda_5\right) \left[ \left(z_{hh}^{1/2} - z_{HH}^{1/2}\right) c_{2\alpha} + \left(z_{zz}^{1/2} - z_{AA}^{1/2}\right) c_{2\beta} \right. \nn \\
&\left. + \left(z_{hH}^{12} + z_{Hh}^{1/2}\right) s_{2\alpha} - z_{Az}^{1/2} s_{2\beta}\right] \nn
\end{align}
\begin{equation}
\tfrac{1}{\sqrt{2}} (\Phi_1^{\dagger} \Phi_1) \to \tfrac{1}{2} (\tilde{\Phi}_1^{*} \tau^- \Phi_1^{*}) = \tfrac{1}{\sqrt{2}} (\Phi_1^{\dagger} \Phi_1) \to \tfrac{1}{2} (\tilde{\Phi}_1 \tau^+ \Phi_1)
\end{equation}
\begin{equation}
\tfrac{1}{\sqrt{2}} (\Phi_2^{\dagger} \Phi_2) \to \tfrac{1}{2} (\tilde{\Phi}_2^{*} \tau^- \Phi_2^{*}) = \tfrac{1}{\sqrt{2}} (\Phi_2^{\dagger} \Phi_2) \to \tfrac{1}{2} (\tilde{\Phi}_2 \tau^+ \Phi_2)
\end{equation}
\begin{equation}
\tfrac{1}{\sqrt{2}} (\Phi_1^{\dagger} \Phi_1) \to \tfrac{1}{2} (\tilde{\Phi}_2^{*} \tau^- \Phi_2^{*}) = \tfrac{1}{\sqrt{2}} (\Phi_1^{\dagger} \Phi_1) \to \tfrac{1}{2} (\tilde{\Phi}_2 \tau^+ \Phi_2)
\end{equation}
\begin{equation}
\tfrac{1}{\sqrt{2}} (\Phi_2^{\dagger} \Phi_2) \to \tfrac{1}{2} (\tilde{\Phi}_2^{*} \tau^- \Phi_2^{*}) = \tfrac{1}{\sqrt{2}} (\Phi_1^{\dagger} \Phi_1) \to \tfrac{1}{2} (\tilde{\Phi}_1 \tau^+ \Phi_1)
\end{equation}
\begin{equation}
\tfrac{1}{\sqrt{2}} (\Phi_1^{\dagger} \tau^3 \Phi_1) \to \tfrac{1}{2} (\tilde{\Phi}_1 \tau^+ \Phi_1) = - \left[\tfrac{1}{\sqrt{2}} (\Phi_1^{\dagger} \Phi_1) \to \tfrac{1}{2} (\tilde{\Phi}_1 \tau^+ \Phi_1)\right]
\end{equation}
\begin{equation}
\tfrac{1}{\sqrt{2}} (\Phi_2^{\dagger} \tau^3 \Phi_2) \to \tfrac{1}{2} (\tilde{\Phi}_2 \tau^+ \Phi_2) = - \left[\tfrac{1}{\sqrt{2}} (\Phi_2^{\dagger} \Phi_2) \to \tfrac{1}{2} (\tilde{\Phi}_2 \tau^+ \Phi_2) \right]
\end{equation}
\begin{align}
\tfrac{1}{\sqrt{2}} (\Phi_1^{\dagger} \tau^3 \Phi_1) \to \tfrac{1}{2} (\tilde{\Phi}_2 \tau^+ \Phi_2) &= - \frac{1}{4} \left(\lambda_4 + \lambda_5\right) \left(z_{AA}^{1/2} - z_{hh}^{1/2} - z_{HH}^{1/2} + z_{zz}^{1/2}\right) \\
&+ \frac{1}{4} \left(\lambda_4 - \lambda_5\right) \left[ \left(z_{hh}^{1/2} - z_{HH}^{1/2}\right) c_{2\alpha} + \left(z_{zz}^{1/2} - z_{AA}^{1/2}\right) c_{2\beta} \right. \nn \\
&\left. + \left(z_{hH}^{12} + z_{Hh}^{1/2}\right) s_{2\alpha} - z_{Az}^{1/2} s_{2\beta}\right] \nn
\end{align}
\begin{align}
\tfrac{1}{\sqrt{2}} (\Phi_2^{\dagger} \tau^3 \Phi_2) \to \tfrac{1}{2} (\tilde{\Phi}_1 \tau^+ \Phi_1) &= - \frac{1}{4} \left(\lambda_4 + \lambda_5\right) \left(z_{AA}^{1/2} - z_{hh}^{1/2} - z_{HH}^{1/2} + z_{zz}^{1/2}\right) \\
&- \frac{1}{4} \left(\lambda_4 - \lambda_5\right) \left[ \left(z_{hh}^{1/2} - z_{HH}^{1/2}\right) c_{2\alpha} + \left(z_{zz}^{1/2} - z_{AA}^{1/2}\right) c_{2\beta} \right. \nn \\
&\left. + \left(z_{hH}^{12} + z_{Hh}^{1/2}\right) s_{2\alpha} - z_{Az}^{1/2} s_{2\beta}\right] \nn
\end{align}
\begin{equation}
\tfrac{1}{\sqrt{2}} (\Phi_1^{\dagger} \tau^3 \Phi_1) \to \tfrac{1}{2} (\tilde{\Phi}_1^{*} \tau^- \Phi_1^{*}) = \tfrac{1}{\sqrt{2}} (\Phi_1^{\dagger} \tau^3 \Phi_1) \to \tfrac{1}{2} (\tilde{\Phi}_1 \tau^+ \Phi_1)
\end{equation}
\begin{equation}
\tfrac{1}{\sqrt{2}} (\Phi_2^{\dagger} \tau^3 \Phi_2) \to \tfrac{1}{2} (\tilde{\Phi}_2^{*} \tau^- \Phi_2^{*}) = \tfrac{1}{\sqrt{2}} (\Phi_2^{\dagger} \tau^3 \Phi_2) \to \tfrac{1}{2} (\tilde{\Phi}_2 \tau^+ \Phi_2)
\end{equation}
\begin{equation}
\tfrac{1}{\sqrt{2}} (\Phi_1^{\dagger} \tau^3 \Phi_1) \to \tfrac{1}{2} (\tilde{\Phi}_2^{*} \tau^- \Phi_2^{*}) = \tfrac{1}{\sqrt{2}} (\Phi_1^{\dagger} \tau^3 \Phi_1) \to \tfrac{1}{2} (\tilde{\Phi}_2 \tau^+ \Phi_2)
\end{equation}
\begin{equation}
\tfrac{1}{\sqrt{2}} (\Phi_2^{\dagger} \tau^3 \Phi_2) \to \tfrac{1}{2} (\tilde{\Phi}_2^{*} \tau^- \Phi_2^{*}) = \tfrac{1}{\sqrt{2}} (\Phi_1^{\dagger} \tau^3 \Phi_1) \to \tfrac{1}{2} (\tilde{\Phi}_1 \tau^+ \Phi_1)
\end{equation}
\begin{equation}
\tfrac{1}{2} (\tilde{\Phi}_1 \tau^+ \Phi_1)
 \to \tfrac{1}{2} (\tilde{\Phi}_1^{*} \tau^- \Phi_1^{*}) = 0
 \end{equation}
\begin{equation}
\tfrac{1}{2} (\tilde{\Phi}_2 \tau^+ \Phi_2)
 \to \tfrac{1}{2} (\tilde{\Phi}_2^{*} \tau^- \Phi_2^{*}) = 0
 \end{equation}
\begin{equation}
\tfrac{1}{2} (\tilde{\Phi}_1 \tau^+ \Phi_1)
 \to \tfrac{1}{2} (\tilde{\Phi}_2^{*} \tau^- \Phi_2^{*}) = 0
 \end{equation}
\begin{equation}
\tfrac{1}{2} (\tilde{\Phi}_2 \tau^+ \Phi_2)
 \to \tfrac{1}{2} (\tilde{\Phi}_1^{*} \tau^- \Phi_1^{*}) = 0
 \end{equation}

\subsection{$Q = 0$, $\mathbb{Z}_2\text{-odd} \to \mathbb{Z}_2\text{-odd}$}
\begin{align}
\tfrac{1}{\sqrt{2}} (\Phi_1^{\dagger} \Phi_2) \to \tfrac{1}{\sqrt{2}} (\Phi_1^{\dagger} \tau^3 \Phi_2) &= \frac{1}{2} \left(\lambda_3 + \lambda_4\right) \left(z_{AA}^{1/2} + z_{hh}^{1/2} - 2 z_{H^+H^-}^{1/2} + z_{HH}^{1/2} \right. \\
&\left. - 2 z_{w^+w^-}^{1/2} + z_{zz}^{1/2} \right) \nn
\end{align}
\begin{equation}
\tfrac{1}{\sqrt{2}} (\Phi_2^{\dagger} \Phi_1) \to \tfrac{1}{\sqrt{2}} (\Phi_2^{\dagger} \tau^3 \Phi_1) = \tfrac{1}{\sqrt{2}} (\Phi_1^{\dagger} \Phi_2) \to \tfrac{1}{\sqrt{2}} (\Phi_1^{\dagger} \tau^3 \Phi_2)
\end{equation}
\begin{equation}
\tfrac{1}{\sqrt{2}} (\Phi_1^{\dagger} \Phi_2) \to \tfrac{1}{\sqrt{2}} (\Phi_2^{\dagger} \tau^3 \Phi_1) = \lambda_5 \left(z_{AA}^{1/2} + z_{hh}^{1/2} - 2 z_{H^+H^-}^{1/2} + z_{HH}^{1/2}  - 2 z_{w^+w^-}^{1/2} + z_{zz}^{1/2} \right)
\end{equation}
\begin{equation}
\tfrac{1}{\sqrt{2}} (\Phi_2^{\dagger} \Phi_1) \to \tfrac{1}{\sqrt{2}} (\Phi_1^{\dagger} \tau^3 \Phi_2) = \tfrac{1}{\sqrt{2}} (\Phi_1^{\dagger} \Phi_2) \to \tfrac{1}{\sqrt{2}} (\Phi_2^{\dagger} \tau^3 \Phi_1)
\end{equation}
\begin{align}
\tfrac{1}{\sqrt{2}} (\Phi_1^{\dagger} \Phi_2) \to \tfrac{1}{\sqrt{2}} (\tilde{\Phi}_1 \tau^+ \Phi_2) &= \frac{\sqrt{2}}{8} \left(2 \lambda_3 + 3 \lambda_4 + 3 \lambda_5\right) \left(z_{AA}^{1/2} - z_{hh}^{1/2} - z_{HH}^{1/2} + z_{zz}^{1/2}\right) \\
& + \frac{\sqrt{2}}{8} \left(2 \lambda_3 + \lambda_4 - \lambda_5\right) \left[\left(z_{hh}^{1/2} - z_{HH}^{1/2}\right) c_{2\alpha} + \left(z_{zz}^{1/2} - z_{AA}^{1/2}\right) c_{2\beta} \right. \nn \\
&\left. + \left(z_{hH}^{1/2} + z_{Hh}^{1/2}\right) s_{2\alpha} - z_{Az}^{1/2} s_{2\beta}\right] \nn
\end{align}
\begin{align}
\tfrac{1}{\sqrt{2}} (\Phi_2^{\dagger} \Phi_1) \to \tfrac{1}{\sqrt{2}} (\tilde{\Phi}_1 \tau^+ \Phi_2) &= \frac{\sqrt{2}}{8} \left(2 \lambda_3 + 3 \lambda_4 + 3 \lambda_5\right) \left(z_{AA}^{1/2} - z_{hh}^{1/2} - z_{HH}^{1/2} + z_{zz}^{1/2}\right) \\
& - \frac{\sqrt{2}}{8} \left(2 \lambda_3 + \lambda_4 - \lambda_5\right) \left[\left(z_{hh}^{1/2} - z_{HH}^{1/2}\right) c_{2\alpha} + \left(z_{zz}^{1/2} - z_{AA}^{1/2}\right) c_{2\beta} \right. \nn \\
&\left. + \left(z_{hH}^{1/2} + z_{Hh}^{1/2}\right) s_{2\alpha} - z_{Az}^{1/2} s_{2\beta}\right] \nn
\end{align}
\begin{equation}
\tfrac{1}{\sqrt{2}} (\Phi_1^{\dagger} \Phi_2) \to \tfrac{1}{\sqrt{2}} (\tilde{\Phi}_1^{*} \tau^- \Phi_2^{*}) = \tfrac{1}{\sqrt{2}} (\Phi_2^{\dagger} \Phi_1) \to \tfrac{1}{\sqrt{2}} (\tilde{\Phi}_1 \tau^+ \Phi_2) 
\end{equation}
\begin{equation}
\tfrac{1}{\sqrt{2}} (\Phi_2^{\dagger} \Phi_1) \to \tfrac{1}{\sqrt{2}} (\tilde{\Phi}_1^{*} \tau^- \Phi_2^{*}) = \tfrac{1}{\sqrt{2}} (\Phi_1^{\dagger} \Phi_2) \to \tfrac{1}{\sqrt{2}} (\tilde{\Phi}_1 \tau^+ \Phi_2) 
\end{equation}
\begin{align}
\tfrac{1}{\sqrt{2}} (\Phi_1^{\dagger} \tau^3 \Phi_2) \to \tfrac{1}{\sqrt{2}} (\tilde{\Phi}_1 \tau^+ \Phi_2) &= - \frac{\sqrt{2}}{8} \left(2 \lambda_3 + \lambda_4 + \lambda_5\right) \left(z_{AA}^{1/2} - z_{hh}^{1/2} - z_{HH}^{1/2} + z_{zz}^{1/2}\right) \\
& - \frac{\sqrt{2}}{8} \left(2 \lambda_3 + 3 \lambda_4 - 3 \lambda_5\right) \left[\left(z_{hh}^{1/2} - z_{HH}^{1/2}\right) c_{2\alpha} \right. \nn \\
&\left. + \left(z_{zz}^{1/2} - z_{AA}^{1/2}\right) c_{2\beta} + \left(z_{hH}^{1/2} + z_{Hh}^{1/2}\right) s_{2\alpha} - z_{Az}^{1/2} s_{2\beta}\right] \nn
\end{align}
\begin{align}
\tfrac{1}{\sqrt{2}} (\Phi_2^{\dagger} \tau^3 \Phi_1) \to \tfrac{1}{\sqrt{2}} (\tilde{\Phi}_1 \tau^+ \Phi_2) &= - \frac{\sqrt{2}}{8} \left(2 \lambda_3 + \lambda_4 + \lambda_5\right) \left(z_{AA}^{1/2} - z_{hh}^{1/2} - z_{HH}^{1/2} + z_{zz}^{1/2}\right) \\
& + \frac{\sqrt{2}}{8} \left(2 \lambda_3 + 3 \lambda_4 - 3 \lambda_5\right) \left[\left(z_{hh}^{1/2} - z_{HH}^{1/2}\right) c_{2\alpha} \right. \nn \\
&\left. + \left(z_{zz}^{1/2} - z_{AA}^{1/2}\right) c_{2\beta} + \left(z_{hH}^{1/2} + z_{Hh}^{1/2}\right) s_{2\alpha} - z_{Az}^{1/2} s_{2\beta}\right] \nn
\end{align}
\begin{equation}
\tfrac{1}{\sqrt{2}} (\Phi_1^{\dagger} \tau^3 \Phi_2) \to \tfrac{1}{\sqrt{2}} (\tilde{\Phi}_1^{*} \tau^- \Phi_2^{*}) = \tfrac{1}{\sqrt{2}} (\Phi_2^{\dagger} \tau^3 \Phi_1) \to \tfrac{1}{\sqrt{2}} (\tilde{\Phi}_1 \tau^+ \Phi_2) 
\end{equation}
\begin{equation}
\tfrac{1}{\sqrt{2}} (\Phi_2^{\dagger} \tau^3 \Phi_1) \to \tfrac{1}{\sqrt{2}} (\tilde{\Phi}_1^{*} \tau^- \Phi_2^{*}) = \tfrac{1}{\sqrt{2}} (\Phi_1^{\dagger} \tau^3 \Phi_2) \to \tfrac{1}{\sqrt{2}} (\tilde{\Phi}_1 \tau^+ \Phi_2) 
\end{equation}
\begin{equation}
\tfrac{1}{\sqrt{2}} (\tilde{\Phi}_1 \tau^+ \Phi_2) \to  \tfrac{1}{\sqrt{2}} (\tilde{\Phi}_1^{*} \tau^- \Phi_2^{*}) = 0
\end{equation}

\subsection{$Q = 0$, $\mathbb{Z}_2\text{-even} \to \mathbb{Z}_2\text{-odd}$}
\begin{align}
\tfrac{1}{\sqrt{2}} (\Phi_1^{\dagger} \Phi_1) \to \tfrac{1}{\sqrt{2}} (\Phi_1^{\dagger} \Phi_2) &= - \frac{3}{8} \left(\lambda_1 - \lambda_{345}\right) \left(z_{Az}^{1/2} + z_{hH}^{1/2} + 2 z_{H^+w^-}^{1/2} - z_{Hh}^{1/2}\right) \\
&+ \frac{3}{8} \left(\lambda_1 + \lambda_{345}\right) \left[- \left(z_{hH}^{1/2} + z_{Hh}^{1/2}\right) c_{2\alpha} - \left(z_{Az}^{1/2} + 2 z_{H^+W^-}^{1/2}\right) c_{2\beta}\right. \nn \\
&\left. + \left(z_{hh}^{1/2} - z_{HH}^{1/2}\right) s_{2\alpha} + \left(z_{AA}^{1/2} + 2 z_{H^+H^-}^{1/2} - 2 z_{w^+w^-}^{1/2} - z_{zz}^{1/2}\right) s_{2\beta}\right] \nn
\end{align}
\begin{align}
\tfrac{1}{\sqrt{2}} (\Phi_2^{\dagger} \Phi_2) \to \tfrac{1}{\sqrt{2}} (\Phi_2^{\dagger} \Phi_1) &= \frac{3}{8} \left(\lambda_2 - \lambda_{345}\right) \left(z_{Az}^{1/2} + z_{hH}^{1/2} + 2 z_{H^+w^-}^{1/2} - z_{Hh}^{1/2}\right) \\
&+ \frac{3}{8} \left(\lambda_2 + \lambda_{345}\right) \left[- \left(z_{hH}^{1/2} + z_{Hh}^{1/2}\right) c_{2\alpha} - \left(z_{Az}^{1/2} + 2 z_{H^+W^-}^{1/2}\right) c_{2\beta}\right. \nn \\
&\left. + \left(z_{hh}^{1/2} - z_{HH}^{1/2}\right) s_{2\alpha} + \left(z_{AA}^{1/2} + 2 z_{H^+H^-}^{1/2} - 2 z_{w^+w^-}^{1/2} - z_{zz}^{1/2}\right) s_{2\beta}\right] \nn
\end{align}
\begin{equation}
\tfrac{1}{\sqrt{2}} (\Phi_1^{\dagger} \Phi_1) \to \tfrac{1}{\sqrt{2}} (\Phi_2^{\dagger} \Phi_1) = \tfrac{1}{\sqrt{2}} (\Phi_1^{\dagger} \Phi_1) \to \tfrac{1}{\sqrt{2}} (\Phi_1^{\dagger} \Phi_2)
\end{equation}
\begin{equation}
\tfrac{1}{\sqrt{2}} (\Phi_2^{\dagger} \Phi_2) \to \tfrac{1}{\sqrt{2}} (\Phi_1^{\dagger} \Phi_2) = \tfrac{1}{\sqrt{2}} (\Phi_2^{\dagger} \Phi_2) \to \tfrac{1}{\sqrt{2}} (\Phi_2^{\dagger} \Phi_1)
\end{equation}
\begin{align}
&\tfrac{1}{\sqrt{2}} (\Phi_1^{\dagger} \Phi_1) \to \tfrac{1}{\sqrt{2}} (\Phi_1^{\dagger} \tau^3 \Phi_2) = \frac{1}{8} \left(3 \lambda_1 - 2 \lambda_3 - \lambda_{345}\right) \left(z_{Az}^{1/2} + z_{hH}^{1/2} - 2 z_{H^+w^-}^{1/2} - z_{Hh}^{1/2}\right) \\
&+ \frac{1}{8} \left(3 \lambda_1 + 2 \lambda_3 + \lambda_{345}\right) \left[\left(z_{hH}^{1/2} + z_{Hh}^{1/2}\right) c_{2\alpha} + \left(z_{Az}^{1/2} - 2 z_{H^+W^-}^{1/2}\right) c_{2\beta} - \left(z_{hh}^{1/2} - z_{HH}^{1/2}\right) s_{2\alpha} \right. \nn \\
&\left. + \left(- z_{AA}^{1/2} + 2 z_{H^+H^-}^{1/2} - 2 z_{w^+w^-}^{1/2} + z_{zz}^{1/2}\right) s_{2\beta}\right] \nn
\end{align}
\begin{align}
&\tfrac{1}{\sqrt{2}} (\Phi_2^{\dagger} \Phi_2) \to \tfrac{1}{\sqrt{2}} (\Phi_2^{\dagger} \tau^3 \Phi_1) = - \frac{1}{8} \left(3 \lambda_2 - 2 \lambda_3 - \lambda_{345}\right) \left(z_{Az}^{1/2} + z_{hH}^{1/2} \right. \\
&\left.- 2 z_{H^+w^-}^{1/2} - z_{Hh}^{1/2}\right) + \frac{1}{8} \left(3 \lambda_2 + 2 \lambda_3 + \lambda_{345}\right) \left[\left(z_{hH}^{1/2} + z_{Hh}^{1/2}\right) c_{2\alpha} + \left(z_{Az}^{1/2} - 2 z_{H^+W^-}^{1/2}\right) c_{2\beta} \right. \nn \\
&\left.  - \left(z_{hh}^{1/2} - z_{HH}^{1/2}\right) s_{2\alpha} + \left(- z_{AA}^{1/2} + 2 z_{H^+H^-}^{1/2} - 2 z_{w^+w^-}^{1/2} + z_{zz}^{1/2}\right) s_{2\beta}\right] \nn
\end{align}
\begin{equation}
\tfrac{1}{\sqrt{2}} (\Phi_1^{\dagger} \Phi_1) \to \tfrac{1}{\sqrt{2}} (\Phi_2^{\dagger} \tau^3 \Phi_1) = \tfrac{1}{\sqrt{2}} (\Phi_1^{\dagger} \Phi_1) \to \tfrac{1}{\sqrt{2}} (\Phi_1^{\dagger} \tau^3 \Phi_2)
\end{equation}
\begin{equation}
\tfrac{1}{\sqrt{2}} (\Phi_2^{\dagger} \Phi_2) \to \tfrac{1}{\sqrt{2}} (\Phi_1^{\dagger} \tau^3 \Phi_2) = \tfrac{1}{\sqrt{2}} (\Phi_2^{\dagger} \Phi_2) \to \tfrac{1}{\sqrt{2}} (\Phi_2^{\dagger} \tau^3 \Phi_1)
\end{equation}
\begin{align}
\tfrac{1}{\sqrt{2}} (\Phi_1^{\dagger} \Phi_1) \to \tfrac{1}{\sqrt{2}} (\tilde{\Phi}_1 \tau^+ \Phi_2) &= \frac{\sqrt{2}}{8} \left(3 \lambda_1 - 3 \lambda_3 - 2 \lambda_4\right) \left(z_{Az}^{1/2} - z_{hH}^{1/2} + z_{Hh}^{1/2}\right) \\
& - \frac{\sqrt{2}}{8} \left(3 \lambda_1 + 3 \lambda_3 + 2 \lambda_4\right) \left[\left(z_{hH}^{1/2} + z_{Hh}^{1/2}\right) c_{2\alpha}  - z_{Az}^{1/2} c_{2\beta} \right. \nn \\
&\left. + \left(z_{HH}^{1/2} - z_{hh}^{1/2}\right) s_{2\alpha} - \left(z_{AA}^{1/2} - z_{zz}^{1/2}\right) s_{2\beta}\right] \nn
\end{align}
\begin{align}
\tfrac{1}{\sqrt{2}} (\Phi_2^{\dagger} \Phi_2) \to \tfrac{1}{\sqrt{2}} (\tilde{\Phi}_1 \tau^+ \Phi_2) &= - \frac{\sqrt{2}}{8} \left(3 \lambda_2 - 3 \lambda_3 - 2 \lambda_4\right) \left(z_{Az}^{1/2} - z_{hH}^{1/2} + z_{Hh}^{1/2}\right) \\
& - \frac{\sqrt{2}}{8} \left(3 \lambda_2 + 3 \lambda_3 + 2 \lambda_4\right) \left[\left(z_{hH}^{1/2} + z_{Hh}^{1/2}\right) c_{2\alpha}  - z_{Az}^{1/2} c_{2\beta} \right. \nn \\
&\left. + \left(z_{HH}^{1/2} - z_{hh}^{1/2}\right) s_{2\alpha} - \left(z_{AA}^{1/2} - z_{zz}^{1/2}\right) s_{2\beta}\right] \nn
\end{align}
\begin{equation}
\tfrac{1}{\sqrt{2}} (\Phi_1^{\dagger} \Phi_1) \to \tfrac{1}{\sqrt{2}} (\tilde{\Phi}_1^{*} \tau^- \Phi_2^{*}) = \tfrac{1}{\sqrt{2}} (\Phi_1^{\dagger} \Phi_1) \to \tfrac{1}{\sqrt{2}} (\tilde{\Phi}_1 \tau^+ \Phi_2)
\end{equation}
\begin{equation}
\tfrac{1}{\sqrt{2}} (\Phi_2^{\dagger} \Phi_2) \to \tfrac{1}{\sqrt{2}} (\tilde{\Phi}_1^{*} \tau^- \Phi_2^{*}) = \tfrac{1}{\sqrt{2}} (\Phi_2^{\dagger} \Phi_2) \to \tfrac{1}{\sqrt{2}} (\tilde{\Phi}_1 \tau^+ \Phi_2)
\end{equation}
\begin{align}
&\tfrac{1}{\sqrt{2}} (\Phi_1^{\dagger} \tau^3 \Phi_1) \to \tfrac{1}{\sqrt{2}} (\Phi_1^{\dagger} \Phi_2) = \frac{1}{8} \left(\lambda_1 + 2 \lambda_3 - 3 \lambda_{345}\right) \left(z_{Az}^{1/2} + z_{hH}^{1/2} - 2 z_{H^+w^-}^{1/2} - z_{Hh}^{1/2}\right) \\
&+ \frac{1}{8} \left(\lambda_1 - 2 \lambda_3 + 3 \lambda_{345}\right) \left[\left(z_{hH}^{1/2} + z_{Hh}^{1/2}\right) c_{2\alpha} + \left(z_{Az}^{1/2} - 2 z_{H^+W^-}^{1/2}\right) c_{2\beta} - \left(z_{hh}^{1/2} - z_{HH}^{1/2}\right) s_{2\alpha} \right. \nn \\
&\left. + \left(- z_{AA}^{1/2} + 2 z_{H^+H^-}^{1/2} - 2 z_{w^+w^-}^{1/2} + z_{zz}^{1/2}\right) s_{2\beta}\right] \nn
\end{align}
\begin{align}
&\tfrac{1}{\sqrt{2}} (\Phi_2^{\dagger} \tau^3 \Phi_2) \to \tfrac{1}{\sqrt{2}} (\Phi_2^{\dagger} \Phi_1) = - \frac{1}{8} \left(\lambda_2 + 2 \lambda_3 - 3 \lambda_{345}\right) \left(z_{Az}^{1/2} + z_{hH}^{1/2} \right. \\
&\left. - 2 z_{H^+w^-}^{1/2} - z_{Hh}^{1/2}\right) + \frac{1}{8} \left(\lambda_2 - 2 \lambda_3 + 3 \lambda_{345}\right) \left[\left(z_{hH}^{1/2} + z_{Hh}^{1/2}\right) c_{2\alpha} + \left(z_{Az}^{1/2} - 2 z_{H^+W^-}^{1/2}\right) c_{2\beta}  \right. \nn \\
&\left. - \left(z_{hh}^{1/2} - z_{HH}^{1/2}\right) s_{2\alpha} + \left(- z_{AA}^{1/2} + 2 z_{H^+H^-}^{1/2} - 2 z_{w^+w^-}^{1/2} + z_{zz}^{1/2}\right) s_{2\beta}\right] \nn
\end{align}
\begin{equation}
\tfrac{1}{\sqrt{2}} (\Phi_1^{\dagger} \tau^3 \Phi_1) \to \tfrac{1}{\sqrt{2}} (\Phi_2^{\dagger} \Phi_1) = \tfrac{1}{\sqrt{2}} (\Phi_1^{\dagger} \tau^3 \Phi_1) \to \tfrac{1}{\sqrt{2}} (\Phi_1^{\dagger} \Phi_2)
\end{equation}
\begin{equation}
\tfrac{1}{\sqrt{2}} (\Phi_2^{\dagger} \tau^3 \Phi_2) \to \tfrac{1}{\sqrt{2}} (\Phi_1^{\dagger} \Phi_2) = \tfrac{1}{\sqrt{2}} (\Phi_2^{\dagger} \tau^3 \Phi_2) \to \tfrac{1}{\sqrt{2}} (\Phi_2^{\dagger} \Phi_1)
\end{equation}
\begin{align}
&\tfrac{1}{\sqrt{2}} (\Phi_1^{\dagger} \tau^3 \Phi_1) \to \tfrac{1}{\sqrt{2}} (\Phi_1^{\dagger} \tau^3 \Phi_2) = - \frac{1}{8} \left(\lambda_1 - \lambda_{345}\right) \left(z_{Az}^{1/2} + z_{hH}^{1/2} + 2 z_{H^+w^-}^{1/2} - z_{Hh}^{1/2}\right) \\
&+ \frac{1}{8} \left(\lambda_1 + \lambda_{345}\right) \left[- \left(z_{hH}^{1/2} + z_{Hh}^{1/2}\right) c_{2\alpha} - \left(z_{Az}^{1/2} + 2 z_{H^+W^-}^{1/2}\right) c_{2\beta} + \left(z_{hh}^{1/2} - z_{HH}^{1/2}\right) s_{2\alpha} \right. \nn \\
&\left. + \left(z_{AA}^{1/2} + 2 z_{H^+H^-}^{1/2} - 2 z_{w^+w^-}^{1/2} - z_{zz}^{1/2}\right) s_{2\beta}\right] \nn
\end{align}
\begin{align}
&\tfrac{1}{\sqrt{2}} (\Phi_2^{\dagger} \tau^3 \Phi_2) \to \tfrac{1}{\sqrt{2}} (\Phi_2^{\dagger} \tau^3 \Phi_1) = \frac{1}{8} \left(\lambda_2 - \lambda_{345}\right) \left(z_{Az}^{1/2} + z_{hH}^{1/2} + 2 z_{H^+w^-}^{1/2} - z_{Hh}^{1/2}\right) \\
&+ \frac{1}{8} \left(\lambda_2 + \lambda_{345}\right) \left[- \left(z_{hH}^{1/2} + z_{Hh}^{1/2}\right) c_{2\alpha} - \left(z_{Az}^{1/2} + 2 z_{H^+W^-}^{1/2}\right) c_{2\beta} + \left(z_{hh}^{1/2} - z_{HH}^{1/2}\right) s_{2\alpha} \right. \nn \\
&\left. + \left(z_{AA}^{1/2} + 2 z_{H^+H^-}^{1/2} - 2 z_{w^+w^-}^{1/2} - z_{zz}^{1/2}\right) s_{2\beta}\right] \nn
\end{align}
\begin{equation}
\tfrac{1}{\sqrt{2}} (\Phi_1^{\dagger} \tau^3 \Phi_1) \to \tfrac{1}{\sqrt{2}} (\Phi_2^{\dagger} \tau^3 \Phi_1) = \tfrac{1}{\sqrt{2}} (\Phi_1^{\dagger} \tau^3 \Phi_1) \to \tfrac{1}{\sqrt{2}} (\Phi_1^{\dagger} \tau^3 \Phi_2)
\end{equation}
\begin{equation}
\tfrac{1}{\sqrt{2}} (\Phi_2^{\dagger} \tau^3 \Phi_2) \to \tfrac{1}{\sqrt{2}} (\Phi_1^{\dagger} \tau^3 \Phi_2) = \tfrac{1}{\sqrt{2}} (\Phi_2^{\dagger} \tau^3 \Phi_2) \to \tfrac{1}{\sqrt{2}} (\Phi_2^{\dagger} \tau^3 \Phi_1)
\end{equation}
\begin{align}
\tfrac{1}{\sqrt{2}} (\Phi_1^{\dagger} \tau^3 \Phi_1) \to \tfrac{1}{\sqrt{2}} (\tilde{\Phi}_1 \tau^+ \Phi_2) &= - \frac{\sqrt{2}}{8} \left(\lambda_1 - \lambda_3 - 2 \lambda_4\right) \left(z_{Az}^{1/2} - z_{hH}^{1/2} + z_{Hh}^{1/2}\right) \\
& + \frac{\sqrt{2}}{8} \left(\lambda_1 + \lambda_3 + 2 \lambda_4\right) \left[\left(z_{hH}^{1/2} + z_{Hh}^{1/2}\right) c_{2\alpha}  - z_{Az}^{1/2} c_{2\beta} \right. \nn \\
&\left. + \left(z_{HH}^{1/2} - z_{hh}^{1/2}\right) s_{2\alpha} - \left(z_{AA}^{1/2} - z_{zz}^{1/2}\right) s_{2\beta}\right] \nn
\end{align}
\begin{align}
\tfrac{1}{\sqrt{2}} (\Phi_2^{\dagger} \tau^3 \Phi_2) \to \tfrac{1}{\sqrt{2}} (\tilde{\Phi}_1 \tau^+ \Phi_2) &= \frac{\sqrt{2}}{8} \left(\lambda_2 - \lambda_3 - 2 \lambda_4\right) \left(z_{Az}^{1/2} - z_{hH}^{1/2} + z_{Hh}^{1/2}\right) \\
& + \frac{\sqrt{2}}{8} \left(\lambda_2 + \lambda_3 + 2 \lambda_4\right) \left[\left(z_{hH}^{1/2} + z_{Hh}^{1/2}\right) c_{2\alpha}  - z_{Az}^{1/2} c_{2\beta} \right. \nn \\
&\left. + \left(z_{HH}^{1/2} - z_{hh}^{1/2}\right) s_{2\alpha} - \left(z_{AA}^{1/2} - z_{zz}^{1/2}\right) s_{2\beta}\right] \nn
\end{align}
\begin{equation}
\tfrac{1}{\sqrt{2}} (\Phi_1^{\dagger} \tau^3 \Phi_1) \to \tfrac{1}{\sqrt{2}} (\tilde{\Phi}_1^{*} \tau^- \Phi_2^{*}) = \tfrac{1}{\sqrt{2}} (\Phi_1^{\dagger} \Phi_1) \to \tfrac{1}{\sqrt{2}} (\tilde{\Phi}_1 \tau^+ \Phi_2)
\end{equation}
\begin{equation}
\tfrac{1}{\sqrt{2}} (\Phi_2^{\dagger} \tau^3 \Phi_2) \to \tfrac{1}{\sqrt{2}} (\tilde{\Phi}_1^{*} \tau^- \Phi_2^{*}) = \tfrac{1}{\sqrt{2}} (\Phi_2^{\dagger} \Phi_2) \to \tfrac{1}{\sqrt{2}} (\tilde{\Phi}_1 \tau^+ \Phi_2)
\end{equation}
\begin{align}
&\tfrac{1}{2} (\tilde{\Phi}_1 \tau^+ \Phi_1) \to \tfrac{1}{\sqrt{2}} (\Phi_1^{\dagger} \Phi_2) = \frac{1}{4} \left(\lambda_4 + 3 \lambda_5\right) \left[ - z_{Az}^{1/2} + z _{hH}^{1/2} - z_{Hh}^{1/2} \right. \\
&\left. - \left(z_{hH}^{1/2} + z_{Hh}^{1/2}\right) c_{2\alpha} + z_{Az}^{1/2} c_{2\beta} + \left(z_{hh}^{1/2} - z_{HH}^{1/2}\right) s_{2\alpha} + \left(z_{zz}^{1/2} - z_{AA}^{1/2}\right) s_{2\beta}\right] \nn
\end{align}
\begin{align}
&\tfrac{1}{2} (\tilde{\Phi}_2 \tau^+ \Phi_2) \to \tfrac{1}{\sqrt{2}} (\Phi_2^{\dagger} \Phi_1) = \frac{1}{4} \left(\lambda_4 + 3 \lambda_5\right) \left[z_{Az}^{1/2} - z _{hH}^{1/2} + z_{Hh}^{1/2} \right. \\
&\left. - \left(z_{hH}^{1/2} + z_{Hh}^{1/2}\right) c_{2\alpha} + z_{Az}^{1/2} c_{2\beta} + \left(z_{hh}^{1/2} - z_{HH}^{1/2}\right) s_{2\alpha} + \left(z_{zz}^{1/2} - z_{AA}^{1/2}\right) s_{2\beta}\right] \nn
\end{align}
\begin{align}
&\tfrac{1}{2} (\tilde{\Phi}_1 \tau^+ \Phi_1) \to \tfrac{1}{\sqrt{2}} (\Phi_2^{\dagger} \Phi_1) = \frac{1}{4} \left(\lambda_1 - \lambda_{345}\right) \left(z_{Az}^{1/2} - z _{hH}^{1/2} + z_{Hh}^{1/2} \right) \\
&- \frac{1}{4}\left(\lambda_1 + \lambda_{345}\right) \left[\left(z_{hH}^{1/2} + z_{Hh}^{1/2}\right) c_{2\alpha} - z_{Az}^{1/2} c_{2\beta} - \left(z_{hh}^{1/2} - z_{HH}^{1/2}\right) s_{2\alpha} - \left(z_{zz}^{1/2} - z_{AA}^{1/2}\right) s_{2\beta}\right] \nn
\end{align}
\begin{align}
&\tfrac{1}{2} (\tilde{\Phi}_2 \tau^+ \Phi_2) \to \tfrac{1}{\sqrt{2}} (\Phi_1^{\dagger} \Phi_2) = - \frac{1}{4} \left(\lambda_2 - \lambda_{345}\right) \left(z_{Az}^{1/2} - z _{hH}^{1/2} + z_{Hh}^{1/2} \right) \\
&- \frac{1}{4}\left(\lambda_2 + \lambda_{345}\right) \left[\left(z_{hH}^{1/2} + z_{Hh}^{1/2}\right) c_{2\alpha} - z_{Az}^{1/2} c_{2\beta} - \left(z_{hh}^{1/2} - z_{HH}^{1/2}\right) s_{2\alpha} - \left(z_{zz}^{1/2} - z_{AA}^{1/2}\right) s_{2\beta}\right] \nn
\end{align}
\begin{align}
&\tfrac{1}{2} (\tilde{\Phi}_1 \tau^+ \Phi_1) \to \tfrac{1}{\sqrt{2}} (\Phi_1^{\dagger} \tau^3 \Phi_2) = \frac{1}{4} \left(\lambda_4 - 3 \lambda_5\right) \left[ - z_{Az}^{1/2} + z _{hH}^{1/2} - z_{Hh}^{1/2} \right. \\
&\left. - \left(z_{hH}^{1/2} + z_{Hh}^{1/2}\right) c_{2\alpha} + z_{Az}^{1/2} c_{2\beta} + \left(z_{hh}^{1/2} - z_{HH}^{1/2}\right) s_{2\alpha} + \left(z_{zz}^{1/2} - z_{AA}^{1/2}\right) s_{2\beta}\right] \nn
\end{align}
\begin{align}
&\tfrac{1}{2} (\tilde{\Phi}_2 \tau^+ \Phi_2) \to \tfrac{1}{\sqrt{2}} (\Phi_2^{\dagger} \tau^3 \Phi_1) = \frac{1}{4} \left(\lambda_4 - 3 \lambda_5\right) \left[z_{Az}^{1/2} - z _{hH}^{1/2} + z_{Hh}^{1/2} \right. \\
&\left. - \left(z_{hH}^{1/2} + z_{Hh}^{1/2}\right) c_{2\alpha} + z_{Az}^{1/2} c_{2\beta} + \left(z_{hh}^{1/2} - z_{HH}^{1/2}\right) s_{2\alpha} + \left(z_{zz}^{1/2} - z_{AA}^{1/2}\right) s_{2\beta}\right] \nn
\end{align}
\begin{equation}
\tfrac{1}{2} (\tilde{\Phi}_1 \tau^+ \Phi_1) \to \tfrac{1}{\sqrt{2}} (\Phi_2^{\dagger} \tau^3 \Phi_1) = - \left[\tfrac{1}{2} (\tilde{\Phi}_1 \tau^+ \Phi_1) \to \tfrac{1}{\sqrt{2}} (\Phi_2^{\dagger} \Phi_1)\right]
\end{equation}
\begin{equation}
\tfrac{1}{2} (\tilde{\Phi}_2 \tau^+ \Phi_2) \to \tfrac{1}{\sqrt{2}} (\Phi_1^{\dagger} \tau^3 \Phi_2) = - \left[\tfrac{1}{2} (\tilde{\Phi}_2 \tau^+ \Phi_2) \to \tfrac{1}{\sqrt{2}} (\Phi_1^{\dagger} \Phi_2)\right]
\end{equation}
\begin{align}
&\tfrac{1}{2} (\tilde{\Phi}_1 \tau^+ \Phi_1) \to \tfrac{1}{\sqrt{2}} (\tilde{\Phi}_1 \tau^+ \Phi_2) = - \frac{\sqrt{2}}{4} \left(\lambda_1 - \lambda_{345}\right) \left(z_{Az}^{1/2} + z_{hH}^{1/2} - z_{Hh}^{1/2}\right) \\
&- \frac{\sqrt{2}}{4} \left(\lambda_1 + \lambda_{345}\right) \left[\left(z_{hH}^{1/2} + z_{Hh}^{1/2}\right) c_{2\alpha} + z_{Az}^{1/2} c_{2\beta} + \left(z_{HH}^{1/2} - z_{hh}^{1/2}\right) s_{2\alpha} + \left(z_{zz}^{1/2} - z_{AA}^{1/2}\right) s_{2\beta} \right] \nn
\end{align}
\begin{align}
&\tfrac{1}{2} (\tilde{\Phi}_2 \tau^+ \Phi_2) \to \tfrac{1}{\sqrt{2}} (\tilde{\Phi}_1 \tau^+ \Phi_2) = \frac{\sqrt{2}}{4} \left(\lambda_2 - \lambda_{345}\right) \left(z_{Az}^{1/2} + z_{hH}^{1/2} - z_{Hh}^{1/2}\right) \\
&- \frac{\sqrt{2}}{4} \left(\lambda_2 + \lambda_{345}\right) \left[\left(z_{hH}^{1/2} + z_{Hh}^{1/2}\right) c_{2\alpha} + z_{Az}^{1/2} c_{2\beta} + \left(z_{HH}^{1/2} - z_{hh}^{1/2}\right) s_{2\alpha} + \left(z_{zz}^{1/2} - z_{AA}^{1/2}\right) s_{2\beta} \right] \nn
\end{align}
\begin{equation}
\tfrac{1}{2} (\tilde{\Phi}_1 \tau^+ \Phi_1) \to \tfrac{1}{\sqrt{2}} (\tilde{\Phi}_1^{*} \tau^- \Phi_2^{*}) = 0
\end{equation}
\begin{equation}
\tfrac{1}{2} (\tilde{\Phi}_2 \tau^+ \Phi_2) \to \tfrac{1}{\sqrt{2}} (\tilde{\Phi}_1^{*} \tau^- \Phi_2^{*}) = 0
\end{equation}
\begin{equation}
\tfrac{1}{2} (\tilde{\Phi}_1^{*} \tau^- \Phi_1^{*}) \to \tfrac{1}{\sqrt{2}} (\Phi_1^{\dagger} \Phi_2) = \tfrac{1}{2} (\tilde{\Phi}_1 \tau^+ \Phi_1) \to \tfrac{1}{\sqrt{2}} (\Phi_2^{\dagger} \Phi_1)
\end{equation}
\begin{equation}
\tfrac{1}{2} (\tilde{\Phi}_2^{*} \tau^- \Phi_2^{*}) \to \tfrac{1}{\sqrt{2}} (\Phi_2^{\dagger} \Phi_1) = \tfrac{1}{2} (\tilde{\Phi}_2 \tau^+ \Phi_2) \to \tfrac{1}{\sqrt{2}} (\Phi_1^{\dagger} \Phi_2)
\end{equation}
\begin{equation}
\tfrac{1}{2} (\tilde{\Phi}_1^{*} \tau^- \Phi_1^{*}) \to \tfrac{1}{\sqrt{2}} (\Phi_2^{\dagger} \Phi_1) = \tfrac{1}{2} (\tilde{\Phi}_1 \tau^+ \Phi_1) \to \tfrac{1}{\sqrt{2}} (\Phi_1^{\dagger} \Phi_2)
\end{equation}
\begin{equation}
\tfrac{1}{2} (\tilde{\Phi}_2^{*} \tau^- \Phi_2^{*}) \to \tfrac{1}{\sqrt{2}} (\Phi_1^{\dagger} \Phi_2) = \tfrac{1}{2} (\tilde{\Phi}_2 \tau^+ \Phi_2) \to \tfrac{1}{\sqrt{2}} (\Phi_2^{\dagger} \Phi_1)
\end{equation}
\begin{equation}
\tfrac{1}{2} (\tilde{\Phi}_1^{*} \tau^- \Phi_1^{*}) \to \tfrac{1}{\sqrt{2}} (\Phi_1^{\dagger} \tau^3 \Phi_2) = \tfrac{1}{2} (\tilde{\Phi}_1 \tau^+ \Phi_1) \to \tfrac{1}{\sqrt{2}} (\Phi_2^{\dagger} \tau^3 \Phi_1)
\end{equation}
\begin{equation}
\tfrac{1}{2} (\tilde{\Phi}_2^{*} \tau^- \Phi_2^{*}) \to \tfrac{1}{\sqrt{2}} (\Phi_2^{\dagger} \tau^3 \Phi_1) = \tfrac{1}{2} (\tilde{\Phi}_2 \tau^+ \Phi_2) \to \tfrac{1}{\sqrt{2}} (\Phi_1^{\dagger} \tau^3 \Phi_2)
\end{equation}
\begin{equation}
\tfrac{1}{2} (\tilde{\Phi}_1^{*} \tau^- \Phi_1^{*}) \to \tfrac{1}{\sqrt{2}} (\Phi_2^{\dagger} \tau^3 \Phi_1) = \tfrac{1}{2} (\tilde{\Phi}_1 \tau^+ \Phi_1) \to \tfrac{1}{\sqrt{2}} (\Phi_1^{\dagger} \tau^3 \Phi_2)
\end{equation}
\begin{equation}
\tfrac{1}{2} (\tilde{\Phi}_2^{*} \tau^- \Phi_2^{*}) \to \tfrac{1}{\sqrt{2}} (\Phi_1^{\dagger} \tau^3 \Phi_2) = \tfrac{1}{2} (\tilde{\Phi}_2 \tau^+ \Phi_2) \to \tfrac{1}{\sqrt{2}} (\Phi_2^{\dagger} \tau^3 \Phi_1)
\end{equation}
\begin{equation}
\tfrac{1}{2} (\tilde{\Phi}_1^{*} \tau^- \Phi_1^{*}) \to \tfrac{1}{\sqrt{2}} (\tilde{\Phi}_1 \tau^+ \Phi_2) = 0
\end{equation}
\begin{equation}
\tfrac{1}{2} (\tilde{\Phi}_2^{*} \tau^- \Phi_2^{*})  \to \tfrac{1}{\sqrt{2}} (\tilde{\Phi}_1 \tau^+ \Phi_2) = 0
\end{equation}
\begin{equation}
\tfrac{1}{2} (\tilde{\Phi}_1^{*} \tau^- \Phi_1^{*}) \to \tfrac{1}{\sqrt{2}} (\tilde{\Phi}_1^{*} \tau^- \Phi_2^{*}) = \tfrac{1}{2} (\tilde{\Phi}_1 \tau^+ \Phi_1) \to \tfrac{1}{\sqrt{2}} (\tilde{\Phi}_1 \tau^+ \Phi_2)
\end{equation}
\begin{equation}
\tfrac{1}{2} (\tilde{\Phi}_2^{*} \tau^- \Phi_2^{*})  \to \tfrac{1}{\sqrt{2}} (\tilde{\Phi}_1^{*} \tau^- \Phi_2^{*}) = \tfrac{1}{2} (\tilde{\Phi}_2 \tau^+ \Phi_2) \to \tfrac{1}{\sqrt{2}} (\tilde{\Phi}_1 \tau^+ \Phi_2)
\end{equation}

\subsection{$Q = 1$, $\mathbb{Z}_2\text{-even} \to \mathbb{Z}_2\text{-even}$}
\begin{align}
&\tfrac{1}{\sqrt{2}} (\Phi_1^{\dagger} \tau^- \Phi_1) \to \tfrac{1}{2} (\tilde{\Phi}_1 \tau^3 \Phi_1)  = \frac{1}{2} \lambda_1 \left[z_{AA}^{1/2} - z_{hh}^{1/2} - z_{HH}^{1/2} + z_{zz}^{1/2} \right. \\
&\left. + \left(z_{hh}^{1/2} - z_{HH}^{1/2}\right) c_{2\alpha} + \left(z_{zz}^{1/2} - z_{AA}^{1/2}\right) c_{2\beta} + \left(z_{hH}^{1/2} + z_{Hh}^{1/2}\right) s_{2\alpha} - z_{Az}^{1/2} s_{2\beta}\right] \nn
\end{align}
\begin{align}
&\tfrac{1}{\sqrt{2}} (\Phi_2^{\dagger} \tau^- \Phi_2) \to \tfrac{1}{2} (\tilde{\Phi}_2 \tau^3 \Phi_2)  = \frac{1}{2} \lambda_2 \left[z_{AA}^{1/2} - z_{hh}^{1/2} - z_{HH}^{1/2} + z_{zz}^{1/2} \right. \\
&\left. - \left(z_{hh}^{1/2} - z_{HH}^{1/2}\right) c_{2\alpha} - \left(z_{zz}^{1/2} - z_{AA}^{1/2}\right) c_{2\beta} - \left(z_{hH}^{1/2} + z_{Hh}^{1/2}\right) s_{2\alpha} + z_{Az}^{1/2} s_{2\beta}\right] \nn
\end{align}
\begin{align}
&\tfrac{1}{\sqrt{2}} (\Phi_1^{\dagger} \tau^- \Phi_1) \to \tfrac{1}{2} (\tilde{\Phi}_2 \tau^3 \Phi_2) = \frac{1}{4} \left(\lambda_4 + \lambda_5\right) \left(z_{AA}^{1/2} - z_{hh}^{1/2} - z_{HH}^{1/2} + z_{zz}^{1/2}\right) \\
&+ \frac{1}{4} \left(\lambda_4 - \lambda_5\right) \left[\left(z_{hh}^{1/2} - z_{HH}^{1/2}\right) c_{2\alpha} + \left(z_{zz}^{1/2} - z_{AA}^{1/2}\right) c_{2\beta} + \left(z_{hH}^{1/2} + z_{Hh}^{1/2}\right) s_{2\alpha} - z_{Az}^{1/2} s_{2\beta}\right] \nn
\end{align}
\begin{align}
&\tfrac{1}{\sqrt{2}} (\Phi_2^{\dagger} \tau^- \Phi_2) \to \tfrac{1}{2} (\tilde{\Phi}_1 \tau^3 \Phi_1)  = \frac{1}{4} \left(\lambda_4 + \lambda_5\right) \left(z_{AA}^{1/2} - z_{hh}^{1/2} - z_{HH}^{1/2} + z_{zz}^{1/2}\right) \\
&- \frac{1}{4} \left(\lambda_4 - \lambda_5\right) \left[\left(z_{hh}^{1/2} - z_{HH}^{1/2}\right) c_{2\alpha} + \left(z_{zz}^{1/2} - z_{AA}^{1/2}\right) c_{2\beta} + \left(z_{hH}^{1/2} + z_{Hh}^{1/2}\right) s_{2\alpha} - z_{Az}^{1/2} s_{2\beta}\right] \nn
\end{align}

\subsection{$Q = 1$, $\mathbb{Z}_2\text{-odd} \to \mathbb{Z}_2\text{-odd}$}
\begin{align}
\tfrac{1}{\sqrt{2}} (\Phi_1^{\dagger} \tau^- \Phi_2) \to \tfrac{1}{\sqrt{2}} (\tilde{\Phi}_1 \Phi_2) &= \frac{\sqrt{2}}{8} \left(2 \lambda_3 - \lambda_4 - \lambda_5\right) \left(z_{AA}^{1/2} - z_{hh}^{1/2} - z_{HH}^{1/2} + z_{zz}^{1/2}\right) \\
&+ \frac{\sqrt{2}}{8} \left(2 \lambda_3 + \lambda_4 - \lambda_5\right) \left[\left(z_{hh}^{1/2} - z_{HH}^{1/2}\right) c_{2\alpha} + \left(z_{zz}^{1/2} - z_{AA}^{1/2}\right) c_{2\beta} \right. \nn \\
&\left.+ \left(z_{hH}^{1/2} + z_{Hh}^{1/2}\right) s_{2\alpha} - z_{Az}^{1/2} s_{2\beta}\right]\nn
\end{align}
\begin{align}
\tfrac{1}{\sqrt{2}} (\Phi_2^{\dagger} \tau^- \Phi_1) \to \tfrac{1}{\sqrt{2}} (\tilde{\Phi}_1 \Phi_2) &= - \frac{\sqrt{2}}{8} \left(2 \lambda_3 - \lambda_4 - \lambda_5\right) \left(z_{AA}^{1/2} - z_{hh}^{1/2} - z_{HH}^{1/2} + z_{zz}^{1/2}\right) \\
&+ \frac{\sqrt{2}}{8} \left(2 \lambda_3 + \lambda_4 - \lambda_5\right) \left[\left(z_{hh}^{1/2} - z_{HH}^{1/2}\right) c_{2\alpha} + \left(z_{zz}^{1/2} - z_{AA}^{1/2}\right) c_{2\beta} \right. \nn \\
&\left.+ \left(z_{hH}^{1/2} + z_{Hh}^{1/2}\right) s_{2\alpha} - z_{Az}^{1/2} s_{2\beta}\right]\nn
\end{align}
\begin{align}
\tfrac{1}{\sqrt{2}} (\Phi_1^{\dagger} \tau^- \Phi_2) \to \tfrac{1}{\sqrt{2}} (\tilde{\Phi}_1 \tau^3 \Phi_2) &= \frac{\sqrt{2}}{8} \left(2 \lambda_3 + \lambda_4 + \lambda_5\right) \left(z_{AA}^{1/2} - z_{hh}^{1/2} - z_{HH}^{1/2} + z_{zz}^{1/2}\right) \\
&+ \frac{\sqrt{2}}{8} \left(2 \lambda_3 - \lambda_4 + \lambda_5\right) \left[\left(z_{hh}^{1/2} - z_{HH}^{1/2}\right) c_{2\alpha} + \left(z_{zz}^{1/2} - z_{AA}^{1/2}\right) c_{2\beta} \right. \nn \\
&\left.+ \left(z_{hH}^{1/2} + z_{Hh}^{1/2}\right) s_{2\alpha} - z_{Az}^{1/2} s_{2\beta}\right]\nn
\end{align}
\begin{align}
\tfrac{1}{\sqrt{2}} (\Phi_2^{\dagger} \tau^- \Phi_1) \to \tfrac{1}{\sqrt{2}} (\tilde{\Phi}_1 \tau^3 \Phi_2) &= \frac{\sqrt{2}}{8} \left(2 \lambda_3 + \lambda_4 + \lambda_5\right) \left(z_{AA}^{1/2} - z_{hh}^{1/2} - z_{HH}^{1/2} + z_{zz}^{1/2}\right) \\
&- \frac{\sqrt{2}}{8} \left(2 \lambda_3 - \lambda_4 + \lambda_5\right) \left[\left(z_{hh}^{1/2} - z_{HH}^{1/2}\right) c_{2\alpha} + \left(z_{zz}^{1/2} - z_{AA}^{1/2}\right) c_{2\beta} \right. \nn \\
&\left.+ \left(z_{hH}^{1/2} + z_{Hh}^{1/2}\right) s_{2\alpha} - z_{Az}^{1/2} s_{2\beta}\right]\nn
\end{align}
\begin{align}
&\tfrac{1}{\sqrt{2}} (\tilde{\Phi}_1 \Phi_2) \to  \tfrac{1}{\sqrt{2}} (\tilde{\Phi}_1 \tau^3 \Phi_2) = \frac{1}{2} \lambda_3 \left[ \left(z_{hh}^{1/2} - z_{HH}^{1/2}\right) c_{2\alpha} \right. \\
&\left. + \left(z_{AA}^{1/2} - 2 z_{H^+H^-}^{1/2} + 2 z_{w^+w^-}^{1/2} - z_{zz}^{1/2}\right) c_{2\beta} +\left(z_{hH}^{1/2} + z_{Hh}^{1/2}\right) s_{2\alpha} + \left(z_{Az}^{1/2} - 2 z_{H^+w^-}^{1/2}\right) s_{2\beta}\right] \nn
\end{align}

\subsection{$Q = 1$, $\mathbb{Z}_2\text{-even} \to \mathbb{Z}_2\text{-odd}$}
\begin{align}
&\tfrac{1}{\sqrt{2}} (\Phi_1^{\dagger} \tau^- \Phi_1) \to \tfrac{1}{\sqrt{2}} (\Phi_1^{\dagger} \tau^- \Phi_2) = - \frac{1}{2} \left(\lambda_1 - \lambda_3\right) z_{H^+w^-}^{1/2} \\
&+ \frac{1}{2} \left(\lambda_1 + \lambda_3\right) \left[\left(z_{H^+H^-}^{1/2} - z_{w^+w^-}^{1/2}\right) s_{2\beta} - z_{H^+w^-}^{1/2} c_{2\beta}\right] + \frac{1}{4} \left(\lambda_4 + \lambda_5\right) \left[z_{Az}^{1/2} + z_{hH}^{1/2} - z_{Hh}^{1/2} \right. \nn \\
&\left. -\left(z_{hH}^{1/2} + z_{Hh}^{1/2}\right) c_{2\alpha} - z_{Az}^{1/2} c_{2\beta} + \left(z_{hh}^{1/2} - z_{HH}^{1/2}\right) s_{2\alpha} + \left(z_{AA}^{1/2} - z_{zz}^{1/2}\right) s_{2\beta} \right] \nn
\end{align}
\begin{align}
&\tfrac{1}{\sqrt{2}} (\Phi_2^{\dagger} \tau^- \Phi_2) \to \tfrac{1}{\sqrt{2}} (\Phi_2^{\dagger} \tau^- \Phi_1) = \frac{1}{2} \left(\lambda_2 - \lambda_3\right) z_{H^+w^-}^{1/2} \\
&+ \frac{1}{2} \left(\lambda_2 + \lambda_3\right) \left[\left(z_{H^+H^-}^{1/2} - z_{w^+w^-}^{1/2}\right) s_{2\beta} - z_{H^+w^-}^{1/2} c_{2\beta}\right] + \frac{1}{4} \left(\lambda_4 + \lambda_5\right) \left[- z_{Az}^{1/2} - z_{hH}^{1/2} + z_{Hh}^{1/2} \right. \nn \\
&\left. -\left(z_{hH}^{1/2} + z_{Hh}^{1/2}\right) c_{2\alpha} - z_{Az}^{1/2} c_{2\beta} + \left(z_{hh}^{1/2} - z_{HH}^{1/2}\right) s_{2\alpha} + \left(z_{AA}^{1/2} - z_{zz}^{1/2}\right) s_{2\beta} \right] \nn
\end{align}
\begin{align}
&\tfrac{1}{\sqrt{2}} (\Phi_1^{\dagger} \tau^- \Phi_1) \to \tfrac{1}{\sqrt{2}} (\Phi_2^{\dagger} \tau^- \Phi_1) = - \frac{1}{4} \left(\lambda_1 - \lambda_3\right) \left(z_{Az}^{1/2} + z_{hH}^{1/2} - z_{Hh}^{1/2} \right)  \\
&+\frac{1}{2} \left(\lambda_4 + \lambda_5\right) \left[z_{H^+w^-}^{1/2} \left(1+ c_{2\beta}\right)+ \left(z_{H^+H^-}^{1/2} - z_{w^+w^-}^{1/2}\right) s_{2\beta}\right] \nn \\
&+ \frac{1}{4} \left(\lambda_1 + \lambda_3\right) \left[- \left(z_{hH}^{1/2} + z_{Hh}^{1/2}\right) c_{2\alpha} - z_{Az}^{1/2} c_{2\beta} + \left(z_{hh}^{1/2} - z_{HH}^{1/2}\right) s_{2\alpha} + \left(z_{AA}^{1/2} - z_{zz}^{1/2}\right) s_{2\beta} \right] \nn
\end{align}
\begin{align}
&\tfrac{1}{\sqrt{2}} (\Phi_2^{\dagger} \tau^- \Phi_2) \to \tfrac{1}{\sqrt{2}} (\Phi_1^{\dagger} \tau^- \Phi_2) = \frac{1}{4} \left(\lambda_2 - \lambda_3\right) \left(z_{Az}^{1/2} + z_{hH}^{1/2} - z_{Hh}^{1/2} \right) \\
&+ \frac{1}{2} \left(\lambda_4 + \lambda_5\right) \left[z_{H^+w^-}^{1/2} \left(- 1+ c_{2\beta}\right) + \left(z_{H^+H^-}^{1/2} - z_{w^+w^-}^{1/2}\right) s_{2\beta}\right] \nn \\
&+ \frac{1}{4} \left(\lambda_2 + \lambda_3\right) \left[- \left(z_{hH}^{1/2} + z_{Hh}^{1/2}\right) c_{2\alpha} - z_{Az}^{1/2} c_{2\beta} + \left(z_{hh}^{1/2} - z_{HH}^{1/2}\right) s_{2\alpha} + \left(z_{AA}^{1/2} - z_{zz}^{1/2}\right) s_{2\beta} \right] \nn
\end{align}
\begin{align}
\tfrac{1}{\sqrt{2}} (\Phi_1^{\dagger} \tau^- \Phi_1) \to \tfrac{1}{\sqrt{2}} (\tilde{\Phi}_1 \Phi_2) &= - \frac{\sqrt{2}}{8} \left(\lambda_1 - \lambda_3 + 2 \lambda_5\right) \left(z_{Az}^{1/2} - z_{hH}^{1/2} + z_{Hh}^{1/2}\right) \\
&+ \frac{\sqrt{2}}{8} \left(\lambda_1 + \lambda_3 - 2 \lambda_5\right) \left[\left(z_{hH}^{1/2} + z_{Hh}^{1/2}\right) c_{2\alpha} - z_{Az}^{1/2} c_{2\beta} \right. \nn \\
&\left.+ \left(z_{HH}^{1/2} - z_{hh}^{1/2}\right) s_{2\alpha} + \left(z_{AA}^{1/2} - z_{zz}^{1/2}\right) s_{2\beta}\right] \nn
\end{align}
\begin{align}
\tfrac{1}{\sqrt{2}} (\Phi_2^{\dagger} \tau^- \Phi_2) \to \tfrac{1}{\sqrt{2}} (\tilde{\Phi}_1 \Phi_2) &= - \frac{\sqrt{2}}{8} \left(\lambda_2 - \lambda_3 + 2 \lambda_5\right) \left(z_{Az}^{1/2} - z_{hH}^{1/2} + z_{Hh}^{1/2}\right) \\
&- \frac{\sqrt{2}}{8} \left(\lambda_2 + \lambda_3 - 2 \lambda_5\right) \left[\left(z_{hH}^{1/2} + z_{Hh}^{1/2}\right) c_{2\alpha} - z_{Az}^{1/2} c_{2\beta} \right. \nn \\
&\left.+ \left(z_{HH}^{1/2} - z_{hh}^{1/2}\right) s_{2\alpha} + \left(z_{AA}^{1/2} - z_{zz}^{1/2}\right) s_{2\beta}\right] \nn
\end{align}
\begin{align}
\tfrac{1}{\sqrt{2}} (\Phi_1^{\dagger} \tau^- \Phi_1) \to \tfrac{1}{\sqrt{2}} (\tilde{\Phi}_1 \tau^3 \Phi_2) &= \frac{\sqrt{2}}{8} \left(\lambda_1 - \lambda_3 - 2 \lambda_5\right) \left(z_{Az}^{1/2} - z_{hH}^{1/2} + z_{Hh}^{1/2}\right) \\
&- \frac{\sqrt{2}}{8} \left(\lambda_1 + \lambda_3 + 2 \lambda_5\right) \left[\left(z_{hH}^{1/2} + z_{Hh}^{1/2}\right) c_{2\alpha} - z_{Az}^{1/2} c_{2\beta} \right. \nn \\
&\left.+ \left(z_{HH}^{1/2} - z_{hh}^{1/2}\right) s_{2\alpha} + \left(z_{AA}^{1/2} - z_{zz}^{1/2}\right) s_{2\beta}\right] \nn
\end{align}
\begin{align}
\tfrac{1}{\sqrt{2}} (\Phi_2^{\dagger} \tau^- \Phi_2) \to \tfrac{1}{\sqrt{2}} (\tilde{\Phi}_1 \tau^3 \Phi_2) &= - \frac{\sqrt{2}}{8} \left(\lambda_2 - \lambda_3 - 2 \lambda_5\right) \left(z_{Az}^{1/2} - z_{hH}^{1/2} + z_{Hh}^{1/2}\right) \\
&- \frac{\sqrt{2}}{8} \left(\lambda_2 + \lambda_3 + 2 \lambda_5\right) \left[\left(z_{hH}^{1/2} + z_{Hh}^{1/2}\right) c_{2\alpha} - z_{Az}^{1/2} c_{2\beta} \right. \nn \\
&\left.+ \left(z_{HH}^{1/2} - z_{hh}^{1/2}\right) s_{2\alpha} + \left(z_{AA}^{1/2} - z_{zz}^{1/2}\right) s_{2\beta}\right] \nn
\end{align}
\begin{align}
&\tfrac{1}{2} (\tilde{\Phi}_1 \tau^3 \Phi_1) \to \tfrac{1}{\sqrt{2}} (\Phi_1^{\dagger} \tau^- \Phi_2) = \frac{1}{2} \lambda_4 \left[- z_{Az}^{1/2} + z_{hH}^{1/2} - z_{Hh}^{1/2} \right. \\
&\left. - \left(z_{hH}^{1/2} + z_{Hh}^{1/2}\right) c_{2\alpha} + z_{Az}^{1/2} c_{2\beta} + \left(z_{hh}^{1/2} + z_{HH}^{1/2}\right) s_{2\alpha} + \left(z_{zz}^{1/2} - z_{AA}^{1/2}\right) s_{2\beta}\right] \nn
\end{align}
\begin{align}
&\tfrac{1}{2} (\tilde{\Phi}_2 \tau^3 \Phi_2) \to \tfrac{1}{\sqrt{2}} (\Phi_2^{\dagger} \tau^- \Phi_1) = \frac{1}{2} \lambda_4 \left[z_{Az}^{1/2} - z_{hH}^{1/2} + z_{Hh}^{1/2} \right. \\
&\left. - \left(z_{hH}^{1/2} + z_{Hh}^{1/2}\right) c_{2\alpha} + z_{Az}^{1/2} c_{2\beta} + \left(z_{hh}^{1/2} + z_{HH}^{1/2}\right) s_{2\alpha} + \left(z_{zz}^{1/2} - z_{AA}^{1/2}\right) s_{2\beta}\right] \nn
\end{align}
\begin{align}
&\tfrac{1}{2} (\tilde{\Phi}_1 \tau^3 \Phi_1) \to \tfrac{1}{\sqrt{2}} (\Phi_2^{\dagger} \tau^- \Phi_1) = \frac{1}{4} \left(\lambda_1 - \lambda_3\right) \left(z_{Az}^{1/2} - z_{hH}^{1/2} + z_{Hh}^{1/2} \right) \\
&+ \frac{1}{4} \left(\lambda_1 + \lambda_3\right) \left[- \left(z_{hH}^{1/2} + z_{Hh}^{1/2}\right) c_{2\alpha} + z_{Az}^{1/2} c_{2\beta} + \left(z_{hh}^{1/2} + z_{HH}^{1/2}\right) s_{2\alpha} + \left(z_{zz}^{1/2} - z_{AA}^{1/2}\right) s_{2\beta}\right] \nn
\end{align}
\begin{align}
&\tfrac{1}{2} (\tilde{\Phi}_2 \tau^3 \Phi_2) \to \tfrac{1}{\sqrt{2}} (\Phi_1^{\dagger} \tau^- \Phi_2) = - \frac{1}{4} \left(\lambda_2 - \lambda_3\right) \left(z_{Az}^{1/2} - z_{hH}^{1/2} + z_{Hh}^{1/2} \right) \\
&+ \frac{1}{4} \left(\lambda_2 + \lambda_3\right) \left[- \left(z_{hH}^{1/2} + z_{Hh}^{1/2}\right) c_{2\alpha} + z_{Az}^{1/2} c_{2\beta} + \left(z_{hh}^{1/2} + z_{HH}^{1/2}\right) s_{2\alpha} + \left(z_{zz}^{1/2} - z_{AA}^{1/2}\right) s_{2\beta}\right] \nn
\end{align}
\begin{align}
&\tfrac{1}{2} (\tilde{\Phi}_1 \tau^3 \Phi_1) \to \tfrac{1}{\sqrt{2}} (\tilde{\Phi}_1 \Phi_2) = \frac{\sqrt{2}}{8} \left(\lambda_1 - \lambda_3 + \lambda_4 + \lambda_5\right) \left(z_{Az}^{1/2} + z_{hH}^{1/2} \right. \\
&\left. - 2 z_{H^+w^-}^{1/2} - z_{Hh}^{1/2}\right) + \frac{\sqrt{2}}{8} \left(\lambda_1 + \lambda_3 - \lambda_4 - \lambda_5\right) \left[\left(z_{hH}^{1/2} + z_{Hh}^{1/2}\right) c_{2\alpha} + \left(z_{Az}^{1/2} - 2 z_{H^+w^-}^{1/2}\right) c_{2\beta} \right. \nn \\
&\left. + \left(z_{HH}^{1/2} - z_{hh}^{1/2}\right) s_{2\alpha} + \left(2 z_{H^+H^-}^{1/2} - 2 z_{w^+w^-}^{1/2} - z_{AA}^{1/2} + z_{zz}^{1/2}\right) \right] \nn
\end{align}
\begin{align}
&\tfrac{1}{2} (\tilde{\Phi}_2 \tau^3 \Phi_2) \to \tfrac{1}{\sqrt{2}} (\tilde{\Phi}_1 \Phi_2) = \frac{\sqrt{2}}{8} \left(\lambda_2 - \lambda_3 + \lambda_4 + \lambda_5\right) \left(z_{Az}^{1/2} + z_{hH}^{1/2} \right. \\
&\left. - 2 z_{H^+w^-}^{1/2} - z_{Hh}^{1/2}\right) - \frac{\sqrt{2}}{8} \left(\lambda_2 + \lambda_3 - \lambda_4 - \lambda_5\right) \left[\left(z_{hH}^{1/2} + z_{Hh}^{1/2}\right) c_{2\alpha} + \left(z_{Az}^{1/2} - 2 z_{H^+w^-}^{1/2}\right) c_{2\beta} \right. \nn \\
&\left. + \left(z_{HH}^{1/2} - z_{hh}^{1/2}\right) s_{2\alpha} + \left(2 z_{H^+H^-}^{1/2} - 2 z_{w^+w^-}^{1/2} - z_{AA}^{1/2} + z_{zz}^{1/2}\right) \right] \nn
\end{align}
\begin{align}
&\tfrac{1}{2} (\tilde{\Phi}_1 \tau^3 \Phi_1) \to \tfrac{1}{\sqrt{2}} (\tilde{\Phi}_1 \tau^3 \Phi_2) = - \frac{\sqrt{2}}{8} \left(\lambda_1 - \lambda_3 + \lambda_4 + \lambda_5\right) \left(z_{Az}^{1/2} + z_{hH}^{1/2} \right. \\
&\left. + 2 z_{H^+w^-}^{1/2} - z_{Hh}^{1/2}\right) + \frac{\sqrt{2}}{8} \left(\lambda_1 + \lambda_3 - \lambda_4 - \lambda_5\right) \left[- \left(z_{hH}^{1/2} + z_{Hh}^{1/2}\right) c_{2\alpha} - \left(z_{Az}^{1/2} + 2 z_{H^+w^-}^{1/2}\right) c_{2\beta} \right. \nn \\
&\left. - \left(z_{HH}^{1/2} - z_{hh}^{1/2}\right) s_{2\alpha} + \left(2 z_{H^+H^-}^{1/2} - 2 z_{w^+w^-}^{1/2} + z_{AA}^{1/2} - z_{zz}^{1/2}\right) \right] \nn
\end{align}
\begin{align}
&\tfrac{1}{2} (\tilde{\Phi}_2 \tau^3 \Phi_2) \to \tfrac{1}{\sqrt{2}} (\tilde{\Phi}_1 \tau^3 \Phi_2) = \frac{\sqrt{2}}{8} \left(\lambda_2 - \lambda_3 + \lambda_4 + \lambda_5\right) \left(z_{Az}^{1/2} + z_{hH}^{1/2} \right. \\
&\left. + 2 z_{H^+w^-}^{1/2} - z_{Hh}^{1/2}\right) + \frac{\sqrt{2}}{8} \left(\lambda_2 + \lambda_3 - \lambda_4 - \lambda_5\right) \left[- \left(z_{hH}^{1/2} + z_{Hh}^{1/2}\right) c_{2\alpha} - \left(z_{Az}^{1/2} + 2 z_{H^+w^-}^{1/2}\right) c_{2\beta} \right. \nn \\
&\left. - \left(z_{HH}^{1/2} - z_{hh}^{1/2}\right) s_{2\alpha} + \left(2 z_{H^+H^-}^{1/2} - 2 z_{w^+w^-}^{1/2} + z_{AA}^{1/2} - z_{zz}^{1/2}\right) \right] \nn
\end{align}

\subsection{$Q = 2$, $\mathbb{Z}_2\text{-even} \to \mathbb{Z}_2\text{-odd}$}
\begin{align}
\tfrac{1}{2} (\tilde{\Phi}_1 \tau^- \Phi_1) \to \tfrac{1}{\sqrt{2}} (\tilde{\Phi}_1 \tau^- \Phi_2) &= \frac{\sqrt{2}}{2} \left[ - \left(\lambda_1 - \lambda_{345}\right) z_{H^+w^-}^{1/2} \right. \\
&\left. - \left(\lambda_1 + \lambda_{345}\right) \left(z_{H^+w^-}^{1/2} c_{2\beta} + \left(z_{w^+w^-}^{1/2} - z_{H^+H^-}^{1/2}\right) s_{2\beta}\right)\right] \nn
\end{align}
\begin{align}
\tfrac{1}{2} (\tilde{\Phi}_2 \tau^- \Phi_2) \to \tfrac{1}{\sqrt{2}} (\tilde{\Phi}_1 \tau^- \Phi_2) &= \frac{\sqrt{2}}{2} \left[ \left(\lambda_2 - \lambda_{345}\right) z_{H^+w^-}^{1/2} \right. \\
&\left. - \left(\lambda_2 + \lambda_{345}\right) \left(z_{H^+w^-}^{1/2} c_{2\beta} + \left(z_{w^+w^-}^{1/2} - z_{H^+H^-}^{1/2}\right) s_{2\beta}\right)\right] \nn
\end{align}

\bibliography{2HDM1loopUnitarity_v2}
\bibliographystyle{jhep}

\end{document}